\documentclass[11pt]{article}
\usepackage{arxiv}

\usepackage[utf8]{inputenc}
\usepackage{epsfig}
\usepackage{graphicx}
\usepackage{ulem}
\usepackage[letterpaper]{geometry}
\usepackage{multirow}
\usepackage{graphicx}
\usepackage[usenames,dvipsnames]{color}
\usepackage{todonotes}
\usepackage{bbm}

\newcommand{\eps}{\varepsilon}

\newcommand{\R}{{\mathbb R}}
\newcommand{\N}{{\mathbb N}}

\newcommand{\dist}{\mathrm{dist}}
\newcommand{\emb}{\mathcal E}

\usepackage[numbers]{natbib}
\setcitestyle{aysep={}}

\usepackage{graphicx}
\usepackage[colorlinks=true, allcolors=blue]{hyperref}

\usepackage{paralist}
\usepackage{amsmath}
\usepackage{amssymb}
\usepackage{amsthm}
\usepackage{algorithmic}
\usepackage{algorithm}
\usepackage{tikz}
\usetikzlibrary{arrows}
\usetikzlibrary{arrows.meta}

\usepackage[]{todonotes}   
\usepackage{soul}

\DeclareMathOperator*{\argmin}{argmin}

\title{A Multi-purposed Unsupervised Framework for Comparing Embeddings of Undirected and Directed Graphs}

\author{
Bogumi\l{} Kami\'nski\thanks{Decision Analysis and Support Unit, SGH Warsaw School of Economics, Warsaw, Poland; e-mail: \texttt{bkamins@sgh.waw.pl}}
\And
\L{}ukasz Krai\'nski\thanks{Decision Analysis and Support Unit, SGH Warsaw School of Economics, Warsaw, Poland; e-mail: \texttt{lkrain@sgh.waw.pl}}
\And
Pawe\l{}~Pra\l{}at\thanks{Department of Mathematics, Ryerson University, Toronto, ON, Canada; e-mail: \texttt{pralat@ryerson.ca}}
\And
Fran\c{c}ois Th\'eberge\thanks{Tutte Institute for Mathematics and Computing, Ottawa, ON, Canada; email: \texttt{theberge@ieee.org}}
}

\begin{document}

\maketitle

\begin{abstract}
Graph embedding is a transformation of nodes of a network into a set of vectors. A~good embedding should capture the underlying graph topology and structure, node-to-node relationship, and other relevant information about the graph, its subgraphs, and nodes themselves. If these objectives are achieved, an embedding is a meaningful, understandable, and often compressed representation of a network. Unfortunately, selecting the best embedding is a challenging task and very often requires domain experts.

In this paper, we extend the framework for evaluating graph embeddings that was recently introduced in~\cite{Embedding_Complex_Networks}. Now, the framework assigns two scores, local and global, to each embedding that measure the quality of an evaluated embedding for tasks that require good representation of local and, respectively, global properties of the network. The best embedding, if needed, can be selected in an unsupervised way, or the framework can identify a few embeddings that are worth further investigation. The framework is flexible, scalable, and can deal with undirected/directed, weighted/unweighted graphs.
\end{abstract}

\section{Introduction}

Network Geometry is a rapidly developing approach in Network Science~\cite{hoff2002latent} which enriches the system by modelling the nodes of the network as points in a geometric space. There are many successful examples of this approach that include latent space models~\cite{krioukov2016clustering}, and connections between geometry and network clustering and community structure~\cite{zuev2015emergence,gastner2006spatial}. Very often, these geometric embeddings naturally correspond to physical space, such as when modelling wireless networks or when networks are embedded in some geographic space~\cite{expert2011uncovering,janssen2010spatial}. 

Another important application of geometric graphs is in graph embeddings~\cite{aggarwal2021machine}. In order to extract useful structural information from graphs, one might want to try to embed them in a geometric space by assigning coordinates to each node such that nearby nodes are more likely to share an edge than those far from each other. Moreover, the embedding should also capture global structure and topology of the associated network, identify specific roles of nodes, etc. 

Due to their spectacular successes in various applications, graph embedding methods are becoming increasingly popular in the machine learning community. They are widely used for tasks such as node classification, community detection, and link prediction but other applications such as anomaly detection are currently explored. As reported in~\citep{chen2021symbols}, the ratio between the number of papers published in top 3 conferences (ACL, WWW, KDD) closely related to Computational Social Science (CSS) applying symbol-based representations and the number of papers using embeddings decreased from 10 in 2011 to 1/5 in 2020. We expect this trend to continue with embeddings eventually playing a central role in many machine learning tasks.

There are over 100 algorithms proposed in the literature for node embeddings. The techniques and possible approaches to construct the desired embedding can be broadly divided into the following three families: linear algebra algorithms, random walk based algorithms, and deep learning methods~\cite{aggarwal2021machine,book}. All of these algorithms have plenty of parameters to tune, the dimension of the embedding being only one of them but an important one. Moreover, most of them are randomized algorithms which means that even for a given graph, each time we run them we get a different embedding, possibly of different quality. As a result, it is not clear which algorithm and which parameters one should use for a given application at hand. There is no clear winner and, even for the same application, the decision which algorithm to use might depend on the properties of the investigated network~\cite{Arash}. 

\medskip

In the initial version of the framework, as detailed in~\cite{Embedding_Complex_Networks}, only undirected, unweighted graphs were considered. A null model was introduced by generalizing the well-known Chung-Lu model~\cite{CL} and, based on that, a global divergence score was defined. The global score can be computed for each embedding under consideration by comparing the number of edges within and between communities (obtained via some stable graph clustering algorithm) with the corresponding expected values under the null model. This global score measures how well an embedding preserves the global structure of the graph. In order to handle huge graphs, a landmark-based version of the framework was introduced in~\cite{Embedding_Complex_Networks_Scalable}, which can be calibrated to provide a good trade-off between performance and accuracy. 

In this paper, we generalize the original framework in several ways. Firstly, we add the capability of handling both directed and undirected graphs as well as taking edge weights into account. Moreover, we introduce a new, local score which measures how well each embedding preserves local properties related to the presence of edges between pairs of nodes. The global and local scores can be combined in various ways to select good embedding(s). We illustrate the usefulness of our framework for several applications by comparing those two scores, and we show various ways to combine them to select embeddings.
After appropriate adjustments to directed and/or weighted graphs, landmarks can be used the same way as in the original framework to handle huge graphs.

\medskip

The paper is structured as follows. The framework is introduced in Section~\ref{sec:framework}. The Geometric Chung-Lu model that is the heart of the framework is introduced and discussed in Section~\ref{sec:chung-lu}. Section~\ref{sec:experiments} presents experiments justifying the usefulness of the framework and investigating the quality of a scalable approximation algorithm. Finally, some conclusions and future work is outlined in Section~\ref{sec:conclusions}.

Finally, let us mention that standard measures such as various correlations coefficients, the accuracy, and the AMI score are not formally defined in this paper. The reader is directed to any book on data science or machine learning (for example, to~\cite{book}) for definitions and more.

\section{The Framework}\label{sec:framework}

In this section, we introduce the unsupervised framework for comparing graph embeddings, the main contribution of this paper. Subsection~\ref{sec:framework1} is devoted to high level description and intuition behind the two embedding quality scores, local and global one, returned by the framework. The algorithm that computes them is formally defined in Subsection~\ref{sec:framework2}. Looking at the two scores to make an informed decision which embedding to chose is always recommended but if one wants to use the framework to select the best embedding in an unsupervised manner, then one may combine the two scores into a single value. We discuss this process in Subsection~\ref{sec:many_embeddgins}. The description of the scoring algorithm assumes that the graph is unweighted and directed. Generalizing it to undirected or weighted graphs is straightforward and we discuss it in Subsection~\ref{sec:framework4}.

\subsection{Intuition Behind the Algorithm}\label{sec:framework1}

The proposed framework is multi-purposed, that is, it independently evaluates embeddings using two approaches.

The first approach looks at the network and the associated embeddings ``from the distance'', trying to see a ``big picture''. It evaluates the embeddings based on their ability to capture global properties of the network, namely, edge densities. In order to achieve it, the framework compares edge density between and within the communities that are present (and can be easily recovered) in the evaluated graph $G$ with the corresponding expected edge density in the associated random null-model. This score is designed to identify embeddings that should perform well in tasks requiring global knowledge of the graph such as node classification or community detection. We will call this measure a \emph{global score}.

The second approach looks at the network and embeddings ``under the microscope'', trying to see if a local structure of a graph $G$ is well reflected by the associated embedding. The \emph{local score} will be designed in such a way that it is able to evaluate if the embedding is a strong predictor of (directed) adjacency between nodes in the network. In general, this property could be easily tested using any strong supervised machine learning algorithm. However, our objective is to test not only predictive power but also explainability of the embedding (sometimes referred to as interpretability). Namely, we assume that the adjacency probability between nodes should be monotonically linked with their distance in the embedding and their in and out degrees. This approach has the following advantage: embeddings that score well should not only be useful for link prediction but they should perform well in any task that requires a local knowledge of the graph. To achieve this, we use the same random null-model as we use to calculate the global score to estimate the probability of two nodes to be adjacent. This question is the well-known and well-studied problem of link prediction in which one seeks to find the node pairs most likely to be linked by an edge. When computing the local score we calculate a ranking of the node pairs from most to least likely of being linked. A common evaluation method for such problem is to compute the AUC (area under the ROC curve), which can be interpreted as the probability that a randomly chosen positive sample (here, a pair of nodes connected by an edge) is ranked higher that a negative sample (pair of nodes without an edge). It is important to note that the AUC is independent of the ratio between the number of edges and the number of non-edges in the graph. As a result, the it can be easily approximated using random sampling.

Despite the fact that the global and local scores take diametrically different points of view, there is often correlation between the two. Good embeddings tend to capture both global and local properties and, as a result, they score well in both approaches. On the other hand, poor embeddings have problems with capturing any useful properties of graphs and so their corresponding scores are both bad. Having said that, they are certainly not equivalent and we will show examples in which the two scores are different.

\subsection{The Algorithm}\label{sec:framework2}

In this section, and later in the paper, we use $[n]$ to denote the set of natural numbers less than or equal to $n$; that is, $[n] := \{1, \ldots, n\}$. Given a directed graph $G=(V,E)$ (in particular, given its in- and out- degree distributions $\textbf{w}^{in}$ and $\textbf{w}^{out}$ on $V$) and an embedding $\emb : V \to \R^k$ of its nodes in $k$-dimensional space, we perform the steps detailed below to obtain $(\Delta_\emb^g(G),\Delta_\emb^{\ell}(G))$, a pair of respectively global and local \emph{divergence scores} for the embedding. Indeed, as already mentioned the framework is multi-purposed and, depending on the specific application in mind, one might want the selected embeddings that preserve global properties (density based evaluation) and/or pay attention to local properties (link based evaluation). As a result, we independently compute the two corresponding divergence scores, $\Delta^{g}_{\emb_i}(G)$ and $\Delta^{\ell}_{\emb_i}(G)$, to provide the users of the framework with a more complete picture and let them make an informative decision which embedding to use. We typically apply this algorithm to compare several embeddings $\emb_1,\dots,\emb_m$ of the same graph, and select the best one via $\argmin_{i\in[m]} \Delta_{\emb_i}(G)$, where $\Delta_{\emb_i}(G)$ is the \emph{combined divergence score} that takes into account the scores of the competitors and that can be tuned for a given application at hand.

Note that our algorithm is a general framework and some parts have flexibility. We clearly identify these below and provide a specific, default, approach that we applied in our implementation. In the description below, we assume that the graph is directed and unweighted, and we then discuss how the framework deals with undirected and/or weighted graphs.

The code can be accessed at the GitHub repository\footnote{\url{https://github.com/KrainskiL/CGE.jl}}. The first version of the framework (c.f.~\cite{Embedding_Complex_Networks}) was designed for undirected, unweighted graphs, and only used the density based evaluation. Since it was used for experiments reported in various papers, for reproducibility purpose the code can still be accessed on GitHub\footnote{\url{https://github.com/ftheberge/Comparing\_Graph\_Embeddings}}.

\subsubsection{Global (Density Based) Evaluation: $\Delta_\emb^g(G)$}

\medskip \noindent \textbf{Step 1:} Run some stable \emph{graph} clustering algorithm on $G$ to obtain a partition $\textbf{C}$ of the set of nodes $V$ into $\ell$ communities $C_1, \ldots, C_\ell$. \\
\noindent {\it Note:}
In our implementation, we used the ensemble clustering algorithm for unweighted graphs (ECG) which is based on the Louvain algorithm~\cite{louvain} and the concept of consensus clustering~\cite{Poulin2019}, and is shown to have good stability. For weighted graphs, by default we use the Louvain algorithm. \\
\noindent {\it Note:}
In some applications, the desired partition may be provided together with a graph (for example, when nodes contain some natural labelling and so some form of a ground-truth is provided). The framework is flexible and allows for communities to be provided as an input to the framework instead of using a clustering algorithm.

\medskip \noindent \textbf{Step 2:} For each $i \in [\ell]$, let $c_{i}$ be the proportion of directed edges of $G$ with both endpoints in $C_i$. Similarly, for each $1 \le i, j \le \ell$, $i \neq j$, let $c_{i,j}$ be the proportion of directed edges of $G$ from some node in $C_i$ to some node in $C_j$. Let
\begin{eqnarray}
\bar{\textbf{c}} &=& (c_{1,2},c_{2,1},\ldots, c_{1,\ell}, c_{\ell,1}, c_{2,3}, c_{3,2}, \ldots, c_{\ell-1,\ell}, c_{\ell,\ell-1} ), \nonumber \\
\hat{\textbf{c}} &=& (c_1, \ldots, c_\ell), \label{eq:c}
\end{eqnarray}
and let $\textbf{c} = \bar{\textbf{c}} \oplus \hat{\textbf{c}}$ be the concatenation of the two vectors with a total of $2 \binom{\ell}{2} + \ell = \ell^2$ entries which together sum to one. This {\it graph vector} $\textbf{c}$ characterizes the partition $\textbf{C}$ from the perspective of the graph $G$. \\
\noindent{\it Note:}
The embedding $\emb$ does \emph{not} affect the vectors $\bar{\textbf{c}}$ and $\hat{\textbf{c}}$ (and so also $\textbf{c}$).
They are calculated purely based on $G$ and the partition $\textbf{C}$.

\medskip \noindent \textbf{Step 3:} For a given parameter $\alpha \in \R_+$ and the same partition of nodes $\textbf{C}$, we consider the Geometric Chung-Lu Directed Graph model $\mathcal{G}(\textbf{w}^{in}, \textbf{w}^{out}, \emb, \alpha)$ presented in Section~\ref{sec:chung-lu} that can be viewed in this context as the associated null-model. For each $1 \le i, j \le \ell$, $i \neq j$, we compute $b_{i,j}$, the expected proportion of directed edges of $\mathcal{G}(\textbf{w}^{in}, \textbf{w}^{out}, \emb, \alpha)$ from some node in $C_i$ to some node in $C_j$. Similarly, for each $i \in [\ell]$, let $b_i$ be the expected proportion of directed edges within $C_i$. That gives us another two vectors:
\begin{eqnarray}
\bar{\textbf{b}}_\emb(\alpha) &=& (b_{1,2}, b_{2,1}, \ldots, b_{1,\ell}, b_{\ell,1}, b_{2,3}, b_{3,2}, \ldots, b_{\ell-1,\ell}, b_{\ell,\ell-1} ), \nonumber \\
\hat{\textbf{b}}_\emb(\alpha) &=& (b_{1},\ldots,  b_{\ell} ),  \label{eq:b}
\end{eqnarray}
and let $\textbf{b}_\emb(\alpha) = \bar{\textbf{b}}_\emb(\alpha) \oplus \hat{\textbf{b}}_\emb(\alpha)$ be the concatenation of the two vectors with a total of $\ell^2$ entries which together sum to one. This \emph{model vector} $\textbf{b}_\emb(\alpha)$ characterizes the partition $\textbf{C}$ from the perspective of the embedding $\emb$. \\
\noindent {\it Note:}
The structure of graph $G$ does \emph{not} affect the vectors $\bar{\textbf{b}}_\emb(\alpha)$ and $\hat{\textbf{b}}_\emb(\alpha)$; only its degree distribution $\textbf{w}^{in}$, $\textbf{w}^{out}$, and embedding $\emb$ are used.\\
\noindent {\it Note:}
We used the Geometric Chung-Lu Directed Graph model but the framework is flexible. If, for any reason (perhaps there are some restrictions for the maximum edge length; such restrictions are often present in, for example, wireless networks) it makes more sense to use some other model of random geometric graphs, it can be easily implemented here. If the model is too complicated and computing the expected number of edges between two parts is challenging, then it can be approximated easily via simulations.

\medskip \noindent \textbf{Step 4:} Compute the distance $\Delta_\alpha$ between the two vectors, $\textbf{c}$ and $\textbf{b}_\emb(\alpha)$, in order to measure how well the model $\mathcal{G}(\textbf{w}^{in}, \textbf{w}^{out}, \emb, \alpha)$ fits the graph $G$. \\
\noindent {\it Note:}
We used the well-known and widely used Jensen–Shannon divergence (JSD) to measure the dissimilarity between two probability distributions, that is, $\Delta_\alpha = JSD(\textbf{c},\textbf{b}_\emb(\alpha))$. The JSD was originally proposed in~\cite{Lin1991} and can be viewed as a smoothed version of the Kullback-Leibler divergence. \\
\noindent {\it Note:}
Alternatively, one may independently treat internal and external edges to compensate the fact that there are $2 \binom{\ell}{2} = \Theta(\ell^2)$ coefficients related to external densities whereas only $\ell$ ones related to internal ones. Then, for example, after appropriate normalization of the vectors a simple average of the two corresponding distances can be used, that is,
\begin{equation*}
\Delta_\alpha = \frac 12 \cdot \left(
JSD(\bar{\bf c},\bar{\bf b}_\emb(\alpha)) +  JSD(\hat{\bf c},\hat{\bf b}_\emb(\alpha))
\right).
\end{equation*}
Depending on the application at hand, other weighted averages can be used if more weight needs to be put on internal or external edges.

\medskip \noindent \textbf{Step 5:} Select $\hat{\alpha} = \argmin_{\alpha} \Delta_\alpha$, and define the \emph{global (density based) score} for embedding $\emb$ on $G$ as $\Delta_\emb^g(G) = \Delta_{\hat{\alpha}}$. \\
\noindent \textbf{\it Note:}
The parameter $\alpha$ is used to define a distance in the embedding space, as we detail in Section~\ref{sec:chung-lu}.
In our implementation we simply checked values of $\alpha$ from a dense grid, starting from $\alpha=0$ and finishing the search if no improvement is found for 5 consecutive values of $\alpha$. Clearly, there are potentially faster ways to find an optimum value of $\alpha$ but, since our algorithm is fast performing grid search, this approach was chosen as both easy to implement and robust to potential local optima.

\subsubsection{Local (Link Based) Evaluation: $\Delta_\emb^{\ell}(G)$}

\medskip \noindent \textbf{Step 6:} We let
\begin{eqnarray*}
S^+ &=& \{ (u, v) \in V \times V, u \ne v ; uv \in E \}, \\
S^- &=& \{ (u, v) \in V \times V, u \ne v ; uv \notin E \}.
\end{eqnarray*}
For a given parameter $\alpha \in \R_+$, we again consider the Geometric Chung-Lu Directed Graph model $\mathcal{G}(\textbf{w}^{in}, \textbf{w}^{out}, \emb, \alpha)$ detailed in Section~\ref{sec:chung-lu}. Let $p(u,v)$ be the probability of a directed edge $u \rightarrow v$ to be present under this model.
Then, the AUC (the area under the ROC curve) can be expressed as follows:
$$
p_\alpha = \frac{\sum_{(s,t) \in S^+} \sum_{(u,v) \in S^-} \mathbbm{1}{\{ p(s,t) > p(u,v) \}}}{|S^+| \cdot |S^-|}.
$$
The AUC measures how much the model is capable of distinguishing between the two classes, $S^+$ and $S^-$. In other words, it may be viewed as the probability that $p(s,t) > p(u,v)$, provided that a directed edge $s \rightarrow t$ and a directed non-edge $u \not\rightarrow v$ are selected uniformly at random from $S^+$ and, respectively, $S^-$. \\
\noindent {\it Note:}
In practice, there is no need to investigate all $|S^+| \cdot |S^-|$ pairs of nodes. Instead, we can randomly sample (with replacement) $k$ pairs $(s_i,t_i) \in S^+$ and $k$ pairs $(u_i,v_i) \in S^-$ and then compute
$$
\hat{p}_\alpha = \frac{\sum_{i=1}^k \mathbbm{1}{\{ p(s_i,t_i) > p(u_i,v_i) \}}}{k}
$$
to approximate $p_\alpha$. The value of $k$ is adjusted so that the approximate 95\% confidence interval, namely,
$$
\left[ \hat{p}_\alpha - 1.96 \sqrt{\hat{p}_\alpha(1-\hat{p}_\alpha)/k}, \hat{p}_\alpha + 1.96 \sqrt{\hat{p}_\alpha(1-\hat{p}_\alpha)/k} \right]
$$
is shorter that some precision level. In our implementation the default value of $k$ is set to $k=10{,}000$ so that the length of the interval is guaranteed to be at most $0.02$.

\medskip \noindent \textbf{Step 7:}
Select $\hat{\alpha} = \argmin_{\alpha} (1-\hat{p}_\alpha)$, and define the \emph{local (link based) score} for embedding $\emb$ on $G$ as $\Delta_\emb^{\ell}(G) = 1-\hat{p}_{\hat{\alpha}}$. 

\subsection{Combined Divergence Scores for Evaluating Many Embeddings}\label{sec:many_embeddgins}

As already mentioned a few times, the framework is multi-purposed and, depending on the specific application in mind, one might want the selected embeddings that preserve global properties (global, density based evaluation) or pay more attention to local properties (local, link based evaluation). That is the reason, we independently compute the two corresponding divergence scores, $\Delta^{g}_{\emb_i}(G)$ and $\Delta^{\ell}_{\emb_i}(G)$.

In order to compare several embeddings for the same graph $G$, we repeat steps 3--7 above, each time computing the two scores for a given embedding. Let us stress again that steps 1--2 are done only once; that is, we use the same partition of the graph into $\ell$ communities for all embeddings. In order to select (in an unsupervised way) the best embedding to be used, one may simply compute the combined divergence score, a linear combination of the two scores:
\begin{equation}
\Delta_{\emb_i}(G) = q \cdot \frac { (\Delta^{g}_{\emb_i}(G) + \eps) } { \min_{j \in [m]} ( \Delta^{g}_{\emb_j}(G) + \eps) } + (1-q) \cdot \frac { (\Delta^{\ell}_{\emb_i}(G) + \eps) } {\min_{j \in [m]} ( \Delta^{\ell}_{\emb_j}(G) + \eps) }
\label{eq:final_score}
\end{equation}
for a fixed parameter $q \in [0,1]$, carefully selected for a given application at hand, and $\eps=0.01$, introduced to prevent rare but possible numerical issues when one of the scores is close to zero. Note that $\Delta_{\emb_i}(G) \ge 1$ and $\Delta_{\emb_i}(G) = 1$ if and only if a given embedding $\emb_i$ does not have a better competitor in \emph{any} of the two evaluation criteria. Of course, the lower the score, the better the embedding is. The winner $\emb_j$ can easily be identified by taking $j = \argmin_i \Delta_{\emb_i}(G)$.

Let us briefly justify the choice of function~(\ref{eq:final_score}). First note that both $\Delta^{g}_{\emb_i}(G)$ and $\Delta^{\ell}_{\emb_i}(G)$ are in $[0,1]$. However, since they might have different orders of magnitude (and typically they do), the corresponding scores need to be normalized. In decision theory, one typically simply tunes $q$ to properly take this into account. While we allow for the more advanced user to change the value of $q$, we believe that it is preferable to provide a reasonable scaling when the default value of $q$, namely, $q=1/2$ is used. When choosing a normalization by minimum we were guided by the fact that it is not uncommon that most of the embeddings score poorly in both dimensions; if this is so, then they affect for example the average score but they ideally should not influence the selection process. On the other hand, the minimum clearly should not be affected by bad embeddings. In particular, the normalization by the minimum allows us to distinguish the situation in which two embeddings have similar but large scores (indicating that both embeddings are bad) from the situation in which two embeddings have similar but small scores (one of the two corresponding embeddings can still be significantly better).

Finally, let us mention that while having a single score assigned to each embedding is useful, it is always better to look at the composition of the scores,
$$
\left( \frac { (\Delta^{g}_{\emb_i}(G) + \eps) } { \min_{j \in [m]} (\Delta^{g}_{\emb_j}(G) + \eps) }, \frac { (\Delta^{\ell}_{\emb_i}(G) + \eps) } {\min_{j \in [m]} (\Delta^{\ell}_{\emb_j}(G) + \eps) }  \right),
$$
to make a more informative decision. See Subsection~\ref{sec:toy_example} for an example of such a selection process. 

\subsection{Weighted and Undirected Graphs}\label{sec:framework4}

For simplicity, we defined the framework for unweighted but directed graphs. Extending the density based evaluation to weighted graphs can be easily and naturally done by considering the sum of weights of the edges instead of the number of them; for example, $c_i$ is the proportion of the total weight concentrated on the directed edges with both endpoints in $C_i$. For the link based evaluation, we need to adjust the definition of the AUC so that it is equal to the probability that $p(s,t) > p(u,v)$ times the weight of a directed edge $s \rightarrow t$ (scaled appropriately such that the average scaled weight of all edges investigated is equal to one), provided that a directed edge $s \rightarrow t$ is selected from $S^+$ (the set of all edges) and a directed non-edge $u \not\rightarrow v$ is selected from $S^-$ (the set of non-edges), both of them uniformly at random. As before, such quantity may be easily and efficiently approximated by sampling.

On the other hand, clearly undirected graphs can be viewed as directed ones by replacing each undirected edge $uv$ by two directed edges $uv$ and $vu$. Hence, one can transform an undirected graph $G$ to its directed counterpart and run the framework on it. However, the framework is tuned for a faster running-time when undirected graphs are used but, of course, the divergence score remains unaffected.

\section{Geometric Chung-Lu Model}\label{sec:chung-lu}

The heart of the framework is the associated random graph null-model that is used to design both the global and the local score. The Geometric Chung-Lu model, a generalization of the original Chung-Lu model~\cite{CL}, was introduced in~\cite{Embedding_Complex_Networks} to benchmark embeddings of undirected graphs (at that time only from the global perspective). Now, we need to generalize it even further to include directed and weighted graphs. We do it in Subsection~\ref{sec:geom_dir_model}. A scalable implementation in discussed in Subsection~\ref{sec:scalable}.

\subsection{Geometric Directed Model}\label{sec:geom_dir_model}

In the Geometric Chung-Lu Directed Graph Model we are not only given the expected degree distribution of a directed graph $G$
\begin{eqnarray*}
\textbf{w}^{in} &=& (w_1^{in}, \ldots, w_n^{in}) = (\deg^{in}_G(v_1), \ldots, \deg^{in}_G(v_n)) \\
\textbf{w}^{out} &=& (w_1^{out}, \ldots, w_n^{out}) = (\deg^{out}_G(v_1), \ldots, \deg^{out}_G(v_n))
\end{eqnarray*}
but also an embedding $\emb$ of nodes of $G$ in some $k$-dimensional space, $\emb : V \to \R^k$. In particular, for each pair of nodes, $v_i$, $v_j$, we know the distance between them:
$$
d_{i,j} = \dist( \emb(v_i), \emb(v_j)).
$$
It is desired that the probability that nodes $v_i$ and $v_j$ are adjacent to be a function of $d_{i,j}$, that is, to be proportional to $g(d_{i,j})$ for some function $g$. The function $g$ should be a decreasing function as long edges should occur less frequently than short ones. There are many natural choices such as $g(d) = d^{-\beta}$ for some $\beta \in [0, \infty)$ or $g(d) = \exp(-\gamma d)$ for some $\gamma \in [0, \infty)$. We use the following, normalized function $g:[0,\infty) \to [0,1]$: for a fixed $\alpha \in [0,\infty)$, let
$$
g(d) := \left( 1 - \frac{d - d_{\min}}{d_{\max} - d_{\min}} \right)^{\alpha} = \left( \frac{d_{\max} - d}{d_{\max} - d_{\min}} \right)^{\alpha},
$$
where
\begin{eqnarray*}
d_{\min} &=& \min \{ \dist(\emb(v), \emb(w)): v,w \in V,  v\neq w\} \\
d_{\max} &=& \max \{ \dist(\emb(v), \emb(w)): v,w \in V \}
\end{eqnarray*}
are the minimum, and respectively the maximum, distance between nodes in embedding $\emb$. One convenient and desired property of this function is that it is invariant with respect to an affine transformation of the distance measure. Clearly, $g(d_{\min})=1$ and $g(d_{\max})=0$; in the computations, we can use clipping to force $g(d_{\min})<1$
and/or $g(d_{\max})>0$ if required.
Let us also note that if $\alpha = 0$ (that is, $g(d)=1$ for any $d \in [d_{\min},d_{\max})$ with the convention that $g(d_{\max})=0^0=1$), then the pairwise distances are neglected. As a result, in particular, for undirected graphs we recover the original Chung-Lu model. Moreover, the larger parameter $\alpha$ is, the larger the aversion to long edges is. Since this family of functions (for various values of the parameter $\alpha$) captures a wide spectrum of behaviours, it should be enough to concentrate on this choice but one can easily experiment with other functions. So, for now we may assume that the only parameter of the model is $\alpha \in [0,\infty)$.

The \emph{Geometric Chung-Lu Directed Graph} model is the random graph $G(\textbf{w}^{in}, \textbf{w}^{out}, \emb, \alpha)$ on the set of nodes $V = \{ v_1, \ldots, v_n \}$ in which each pair of nodes $v_i, v_j$, independently of other pairs, forms a directed edge from $v_i$ to $v_j$ with probability $p_{i,j}$, where
\begin{equation*}
p_{i,j} = x_i^{out} x_j^{in} g(d_{i,j})
\end{equation*}
for some carefully tuned weights $x_i^{in}, x_i^{out} \in \R_+$. The weights are selected such that the expected in-degree and out-degree of $v_i$ is $w_i^{in}$ and, respectively, $w_i^{out}$; that is, for all $i \in [n]$
\begin{eqnarray*}
w_i^{out} &=& \sum_{j \in [n], j\neq i} p_{i,j} =  x_i^{out} \sum_{j \in [n], j\neq i} x_j^{in} g(d_{i,j}), \\
w_i^{in} &=& \sum_{j \in [n], j\neq i} p_{j,i} =  x_i^{in} \sum_{j \in [n], j\neq i} x_j^{out} g(d_{i,j}).
\end{eqnarray*}
Additionally, we set $p_{i,i}=0$ for $i\in[n]$ which corresponds to the fact that the model does not allow loops.

In Appendix~\ref{apdx:model} we prove that there exists the unique selection of weights, \emph{unless} $G$ has an independent set of size $n-1$, that is, $G$ is a star with one node being part of \emph{every} edge. (Since each connected component of $G$ can be embedded independently, we always assume that $G$ is connected.) This very mild condition is satisfied in practice. Let us mention that in Appendix~\ref{apdx:model} it is assumed that $g(d_{i,j}) > 0$ for all pairs $i,j$. In our case, $g(d_{i,j}) = 0$ for a pair of nodes that are at the maximum distance. It causes no problems in practice but, as mentioned earlier, one can easily scale the outcome of function $g(\cdot)$ to move away from zero without affecting the divergence score in any non-negligible way. 

Finally, note that it is not clear how to find weights explicitly but they can be easily and efficiently approximated numerically to any desired precision. In Appendix~\ref{apdx:model}, we prove that, if the solution exists, which is easy to check, then the set of right hand sides of the equations, considered as a function from $\mathbb{R}^{2n}$ to $\mathbb{R}^{2n}$, is a local diffeomorphism everywhere in its domain. As a result, standard gradient root-finding algorithms should be quite effective in finding the desired weights. In our implementation we use even simpler numerical approximation procedure.

\noindent {\it Note:} The specification of the Geometric Chung-Lu Directed Graph Model implies that the probability of having a directed edge from one node to another one increases with their out- and, respectively, in- degrees and decreases with the distance between them. This is a crucial feature that ensures that the local divergence score is explainable. For instance, potentially one could imagine embeddings where nodes that are far apart are more likely to be connected by an edge. However, under our framework, such embeddings would get a low local score.

\subsection{Approximated but Scalable Implementation}\label{sec:scalable}

The main bottleneck of the algorithm is the process of tuning $2n$ weights $x_i^{out}, x_i^{in} \in \R_+$ ($i \in [n]$) in the Geometric Chung-Lu Graph (both in directed as well as in undirected counterpart). This part requires $\Theta(n^2)$ steps and so it is not feasible for large graphs. Fortunately, one may modify the algorithm slightly to obtain a scalable approximation algorithm that can be efficiently run on large networks. It was done for the framework for undirected graphs in~\cite{Embedding_Complex_Networks_Scalable} to obtain the running time of $O(n \ln n)$ which is practical.

The main idea behind our approximation algorithm is quite simple. The goal is to group together nodes from the same part of the partition $\textbf{C}$ obtained in Step~1 of the algorithm that are close to each other in the embedded space. Once such refinement of partition $\textbf{C}$ is generated, one may simply replace each group by the corresponding auxiliary node (that we call a \emph{landmark}) that is placed in the appropriately weighted center of mass of the group it is associated with. The reader is directed to~\cite{Embedding_Complex_Networks_Scalable} for more details. Minor adjustments are only needed for the density based evaluation to accommodate directed graphs and approximating the link based score is easy. Both issues are discussed in Appendix~\ref{apdx:landmarks}.

The framework, by default, uses the approximation algorithm for networks with 10,000 nodes or more. In Subsection~\ref{sec:landmarks} we show how well the approximation algorithm works in practice.

\section{Experiments}\label{sec:experiments}

In this section we detail some experiments we performed showing that the framework works as desired. In Subsection~\ref{sec:experiment1} we provide descriptions of  both synthetic and real-world networks that we used for the experiments. A few embedding algorithms for directed graphs are introduced in Subsection~\ref{sec:experiments2}. We used two of them for our experiments. In Subsection~\ref{sec:toy_example} we present results of an experiment with a small ``toy example'' to illustrate how one can use the framework to select the best embedding amongst several ones. As explained earlier, in order to have a scalable framework that can be used to evaluate huge graphs, we introduced landmarks to provide approximated scores. We show in Subsection~\ref{sec:landmarks} that this approximation works well. 
Finally, we ran experiments showing the usefulness of the framework. In Subsection~\ref{sec:correlation_with_global_score} we show that the global score may be used to predict how good the evaluated embedding is for algorithms that require good global properties such as node classification or community detection algorithms. Similarly, the correlation between the local score and the ability of embeddings to capture local properties (for example with link prediction algorithm) is investigated in Subsection~\ref{sec:correlation_with_local_score}. A few more experiments showing the differences between the two scores are presented in Subsection~\ref{sec:experiments7}.

\subsection{Synthetic and Real-World Graphs}\label{sec:experiment1}

In order to test the framework, we performed various experiments on both synthetic and real-world networks. 

\subsubsection{Stochastic Block Model}

To model networks with simple community structure, we used the classical Stochastic Block Model (SBM)~\cite{holland1983stochastic} (see~\cite{funke2019stochastic} for an overview of various generalizations). The model takes the following parameters as its input: the number of nodes $n$, a partition of nodes into $\ell$ communities, and an $\ell \times \ell$ matrix $P$ of edge probabilities. Two nodes, $u$ from $i$th community and $v$ from $j$th community, are adjacent with probability $P_{ij}$ and the events associated with different pairs of nodes are independent. The model can be used to generate undirected graphs in which case matrix $P$ should be symmetric but one can also use the model to generate directed graphs. 

For our experiments, we generated directed \textbf{SBM} graphs with $n=10{,}000$ nodes and 30 communities of similar size. Nodes from two different communities are adjacent with probability $P_{ij} = 0.001=10/n$ and nodes from the same community are adjacent with probability $P_{ii} = 0.025=250/n$. As a result, about 54\% of the edges ended up between communities. 

\subsubsection{LFR Model}

The LFR (Lancichinetti, Fortunato, Radicchi) model~\cite{lancichinetti2008benchmark,lancichinetti2009benchmarks} generates networks with communities and at the same time it allows for the heterogeneity in the distributions of both node degrees and of community sizes. As a result, it became a standard and extensively used method for generating artificial networks. The original model~\cite{lancichinetti2008benchmark} generates undirected graphs but it was soon after generalized to directed and weighted graphs~\cite{lancichinetti2009benchmarks}. The model has various parameters: the number of nodes $n$, the mixing parameter $\mu$ that controls the fraction of edges that are between communities, power law exponent $\gamma$ for the degree distribution, power law exponent $\beta$ for the distribution of community sizes, average degree $d$, and the maximum degree $\Delta$. 

For our experiments, we generated two families of directed graphs that we called \textbf{LFR} and \textbf{noisy-LFR}. To generate \textbf{LFR} graphs we used $n=10{,}000$, $\mu=0.2$, $\gamma=3$, $\beta=2$, $d=100$, and $\Delta =  500$ whereas for \textbf{noisy-LFR} we used $n=10{,}000$, $\mu=0.5$, $\gamma=2$, $\beta=1$, $d=100$, and $\Delta=500$.

\subsubsection{ABCD Model}

In order to generate undirected graphs, we used an ``LFR-like'' random graph model, the Artificial Benchmark for Community Detection (ABCD graph)~\cite{kaminski2020artificial} that was recently introduced and implemented\footnote{\url{https://github.com/bkamins/ABCDGraphGenerator.jl/}}, including a fast implementation that uses multiple threads (ABCDe)\footnote{\url{https://github.com/tolcz/ABCDeGraphGenerator.jl/}}. Undirected variant of LFR and ABCD produce graphs with comparable properties but ABCD/ABCDe is faster than LFR and can be easily tuned to allow the user to make a smooth transition between the two extremes: pure (independent) communities and random graph with no community structure.

For our experiments, we generated \textbf{ABCD} graphs on $n=10{,}000$ nodes, $\xi=0.2$ (the counterpart of $\mu \approx 0.194$ in LFR), $\gamma=3$, $\beta=2$, $d=8.3$, and $\Delta =  50$. The number of edges generated was $m=41{,}536$ and the number of communities was $\ell = 64$.

\subsubsection{EU Email Communication Network}

We also used the real-world network that was generated using email data from a large European research institution~\cite{paranjape2017motifs}. The network is made available through Stanford Network Analysis Project~\cite{leskovec2016snap}\footnote{\url{https://snap.stanford.edu/data/email-Eu-core.html}}. Emails are anonymized and there is an edge between $u$ and $v$ if person $u$ sent person $v$ at least one email. The dataset does not contain incoming messages from or outgoing messages to the rest of the world. More importantly, it contains ``ground-truth'' community memberships of the nodes indicating which of 42 departments at the research institute individuals belong to. As a result, this dataset is suitable for experiments aiming to detect communities but we ignore this external knowledge in our experiments. 

The associated \textbf{EMAIL} directed graph consists of $n=1005$ nodes, $m=25{,}571$ edges, and $\ell = 42$ communities.

\subsubsection{College Football Network}

In order to see the framework ``in action'', in Section~\ref{sec:toy_example} we performed an illustrative experiment with the well-known College Football real-world network with known community structures. This graph represents the schedule of United States football games between Division IA colleges during the regular season in Fall 2000~\cite{girvan2002community}. The associated \textbf{FOOTBALL} graph consists of 115 teams (nodes) and 613 games (edges). The teams are divided into conferences containing 8--12 teams each. In general, games are more frequent between members of the same conference than between members of different conferences, with teams playing an average of about seven intra-conference games and four inter-conference games in the 2000 season. There are a few exceptions to this rule, as detailed in~\cite{lu2018community}: one of the conferences is really a group of independent teams, one conference is really broken into two groups, and 3 other teams play mainly against teams from other conferences. We refer to those as outlying nodes.

\subsection{Node Embeddings}\label{sec:experiments2}

As mentioned in the introduction, there are over 100+ node embedding algorithms for undirected graphs. 
There are also some algorithms explicitly designed for directed graphs, or that can handle both types of graphs, but their number is much smaller. 
We selected two of them for our experiments, \textbf{Node2Vec} and \textbf{HOPE}, but there are more to choose from. Embeddings were produced for 16 different dimensions between 2 and 32 (with a step of 2). 

\subsubsection{Node2Vec}

\textbf{Node2Vec}\footnote{\url{https://github.com/eliorc/node2vec}}~\cite{node2vec} is based on random walks performed on the graph, an approach that was successfully used in Natural Language Processing. In this embedding algorithm, biased random walks are defined via two main parameters.
The return parameter ($p$) controls the likelihood of immediately revisiting a node in the random walk. Setting it to a high value ensures that we are less likely to sample an already-visited node in the following two steps.
The in-out parameter ($q$) allows the search to differentiate between inward and outward nodes so we can smoothly interpolate between breadth-first-search (BFS) and depth-first search (DFS) exploration. 
We tested three variants of parameters $p$ and $q$: $(p=1/9,q=9)$, $(p=1,q=1)$, and $(p=9,q=1/9)$.

\subsubsection{HOPE}

\textbf{HOPE} (High-Order Proximity preserved Embedding)~\cite{Ou2016} is based on the notion of asymmetric transitivity; for example, the existence of several short directed paths from node $u$ to node $v$ makes the existence of a directed edge from $u$ to $v$ more plausible. The algorithm learns node embeddings as two concatenated vectors representing the source and the target roles. HOPE can also be used for undirected graphs, in which case the source and target roles are identical, so only one is retained. Four different high order proximity measures can be used within the same framework. For our experiments we used three of them: Katz, Adamic-Adar, and personalized PageRank (denoted in our experiments as \textit{katz}, \textit{aa} and, respectively, \textit{ppr}). 

\subsubsection{A Few Other Ones}

Similarly to HOPE, \textbf{APP} (Asymmetric Proximity Preserving)~\cite{Zhou2017} is also using asymmetric transitivity and is based on directed random walks and preserved rooted page rank proximity between nodes. It learns node embedding as a concatenation of two vectors representing the node's source and target roles.
\textbf{NERD} (Node Embeddings Respecting Directionality)~\cite{Khosla2020}
also learns 2 embeddings for each node as a source or a target,
using alternating random walks with starting nodes used as a source or a target;
this approach can be interpreted as optimizing with respect to first order proximity for 3 graphs: source-target (directed edges), source-source (pointing to common nodes) and target-target (pointed to by common nodes).
Other methods for directed graphs try to learn node embeddings as well as some directional vector field. For example, \textbf{ANSE} (Asymmetric Node Similarity Embedding)~\cite{Dernbach2020}
uses skip-gram-like random walks, namely, forward and reverse random walks, to limit dead-end issues. It also has an option to embed on a hypersphere.
In another method that also try to learn a directional vector field \cite{Perrault2011}, a similarity kernel on some learned manifold is defined from two components: (i) a symmetric component, which depends only on distance, and (ii) an asymmetric one, which depends also on the directional vector field. The algorithm is based on asymptotic results for (directed graph) Laplacians embedding, and the directional vector space can be decoupled.

\subsection{Illustration of the Framework}\label{sec:toy_example}

\begin{figure}[ht]
     \centering
     \includegraphics[width=0.5\textwidth]{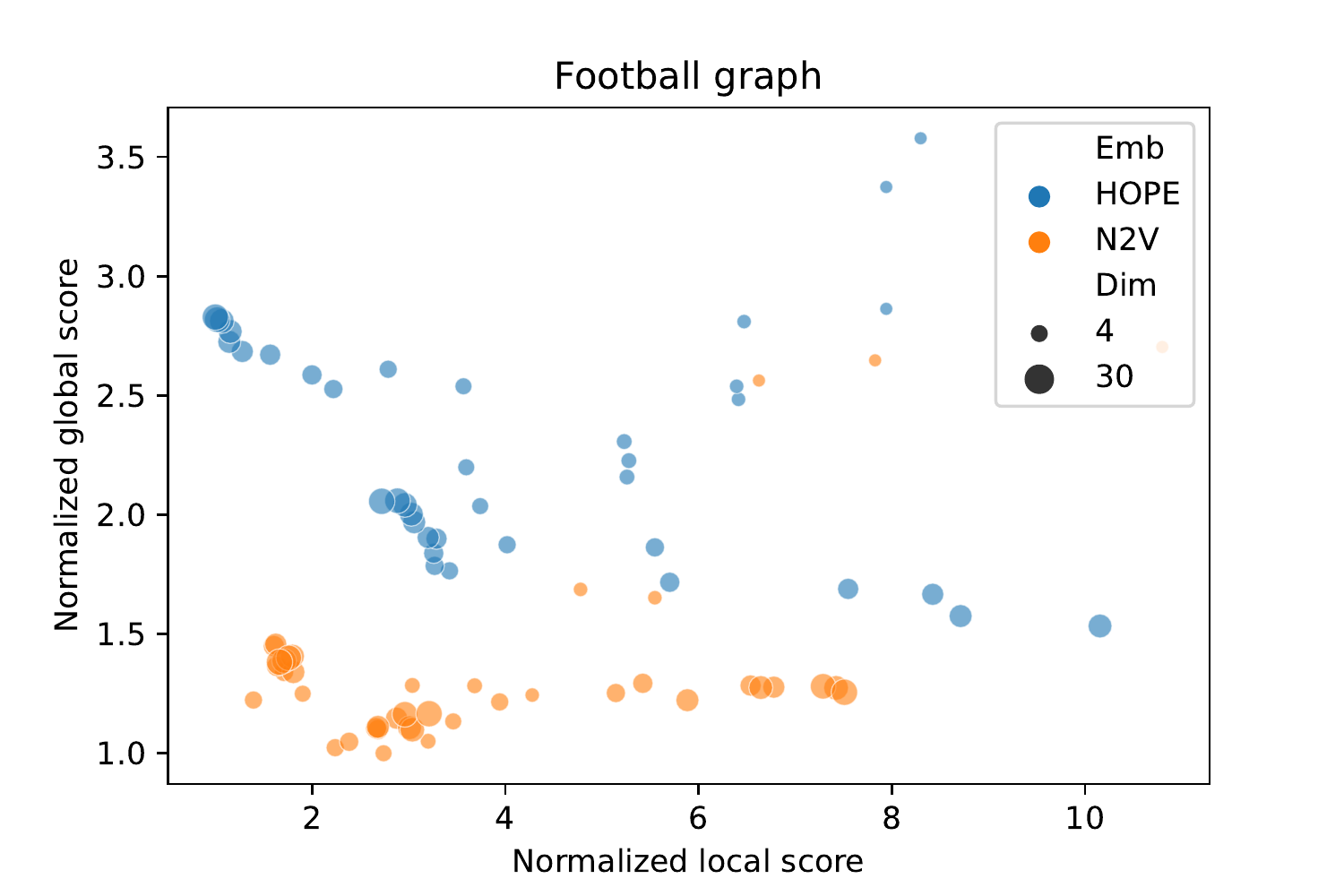}
     \caption{Normalized global and local scores for \textbf{FOOTBALL} graph.}
     \label{fig_paper:toy_normalized}
 \end{figure}

In order to illustrate the application of the framework, we ran the two embedding algorithms (\textbf{HOPE} and \textbf{Node2Vec}) in different dimensions and sets of parameters on the \textbf{FOOTBALL} graph. For each combination of the parameters, an embedding was produced and assessed with our framework. Local and global scores were normalized, as explained in Subsection~\ref{sec:many_embeddgins}, and are presented in Figure~\ref{fig_paper:toy_normalized}. In particular, good embeddings are concentrated around the auxiliary point $(1,1)$ that corresponds to the embedding that is the best from both local and global perspective. (Such embedding might or might not exist.) 

\begin{figure}[ht]
     \centering
     \includegraphics[width=0.31\textwidth]{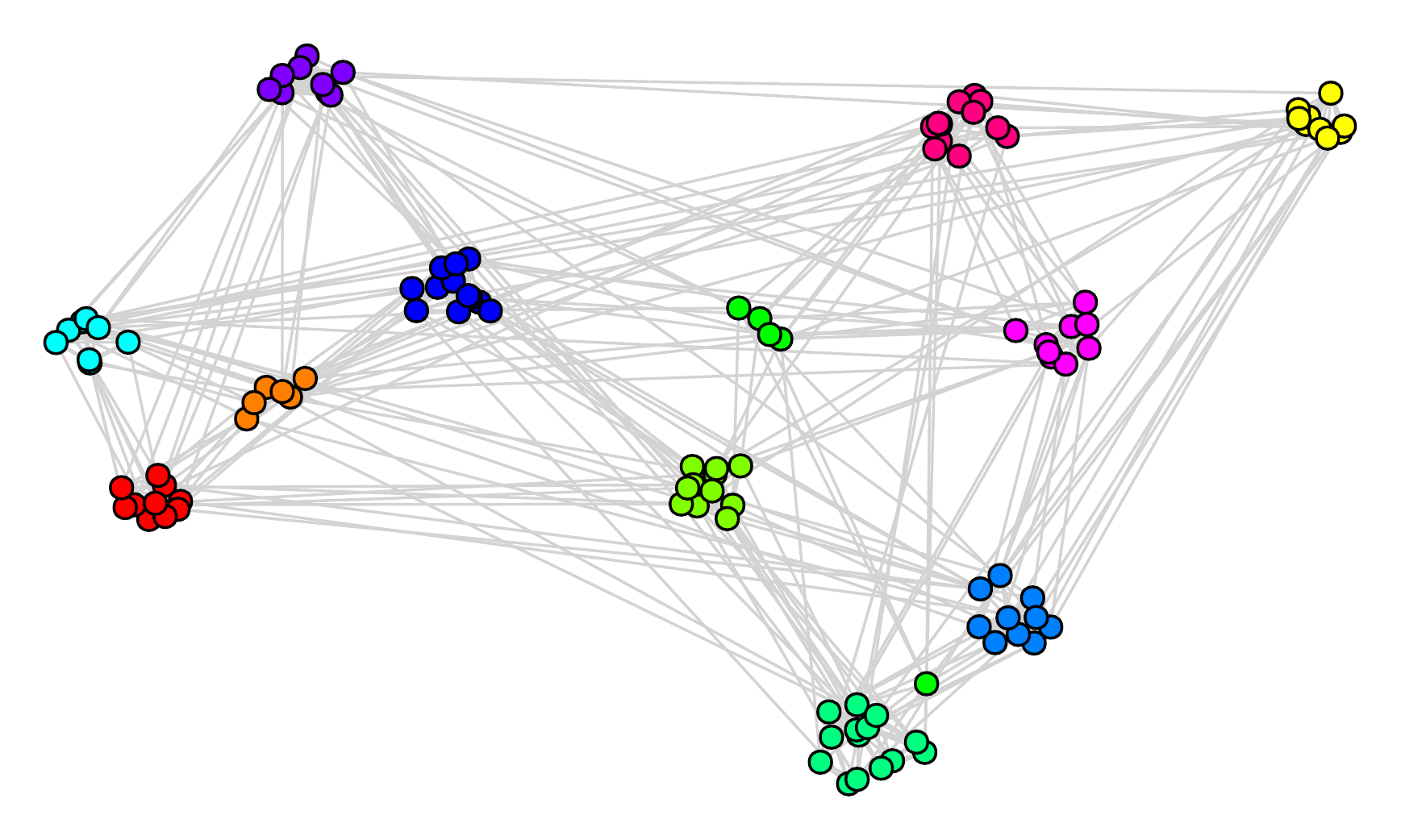}
         \hspace{.1cm}
     \includegraphics[width=0.31\textwidth]{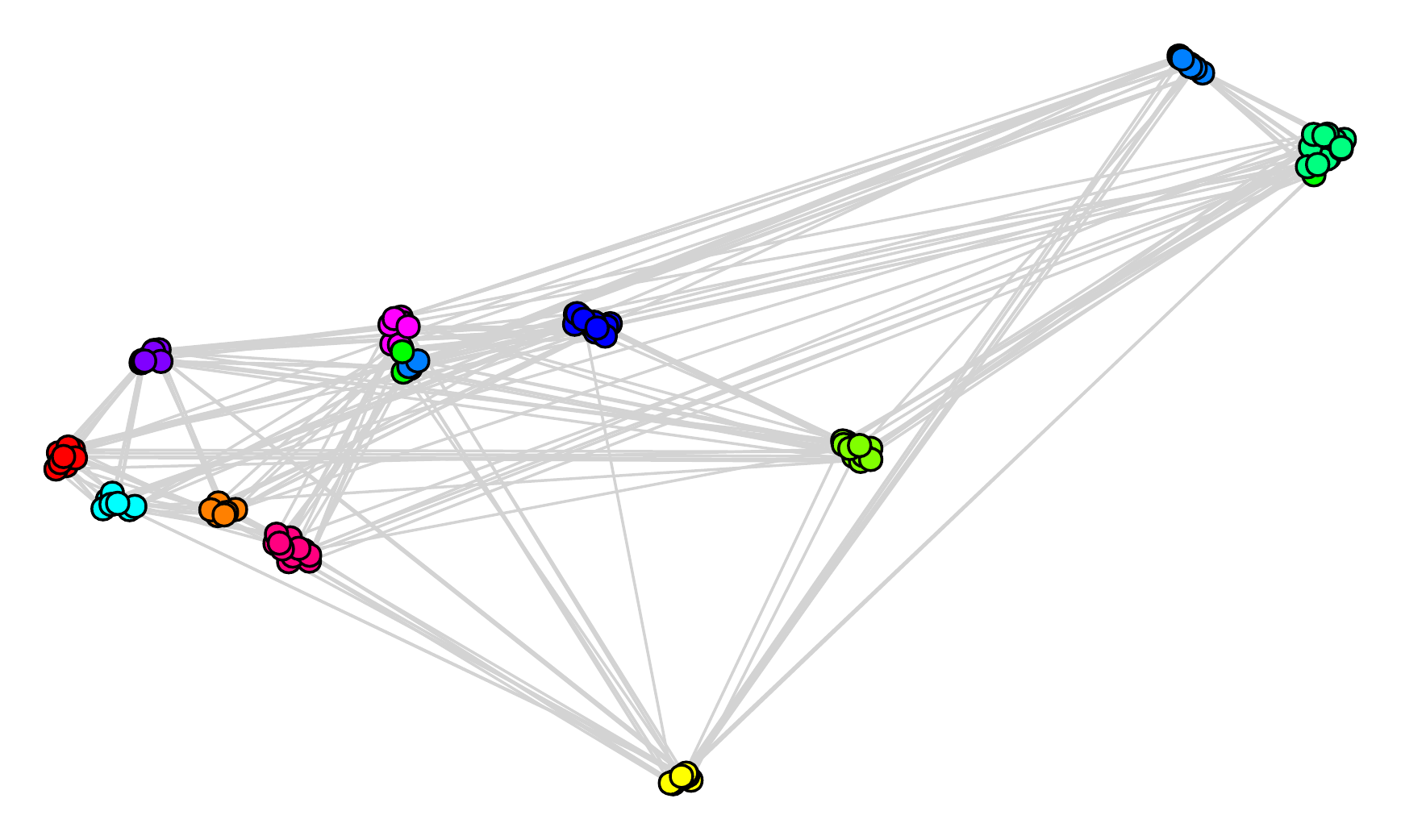}
     \includegraphics[width=0.31\textwidth]{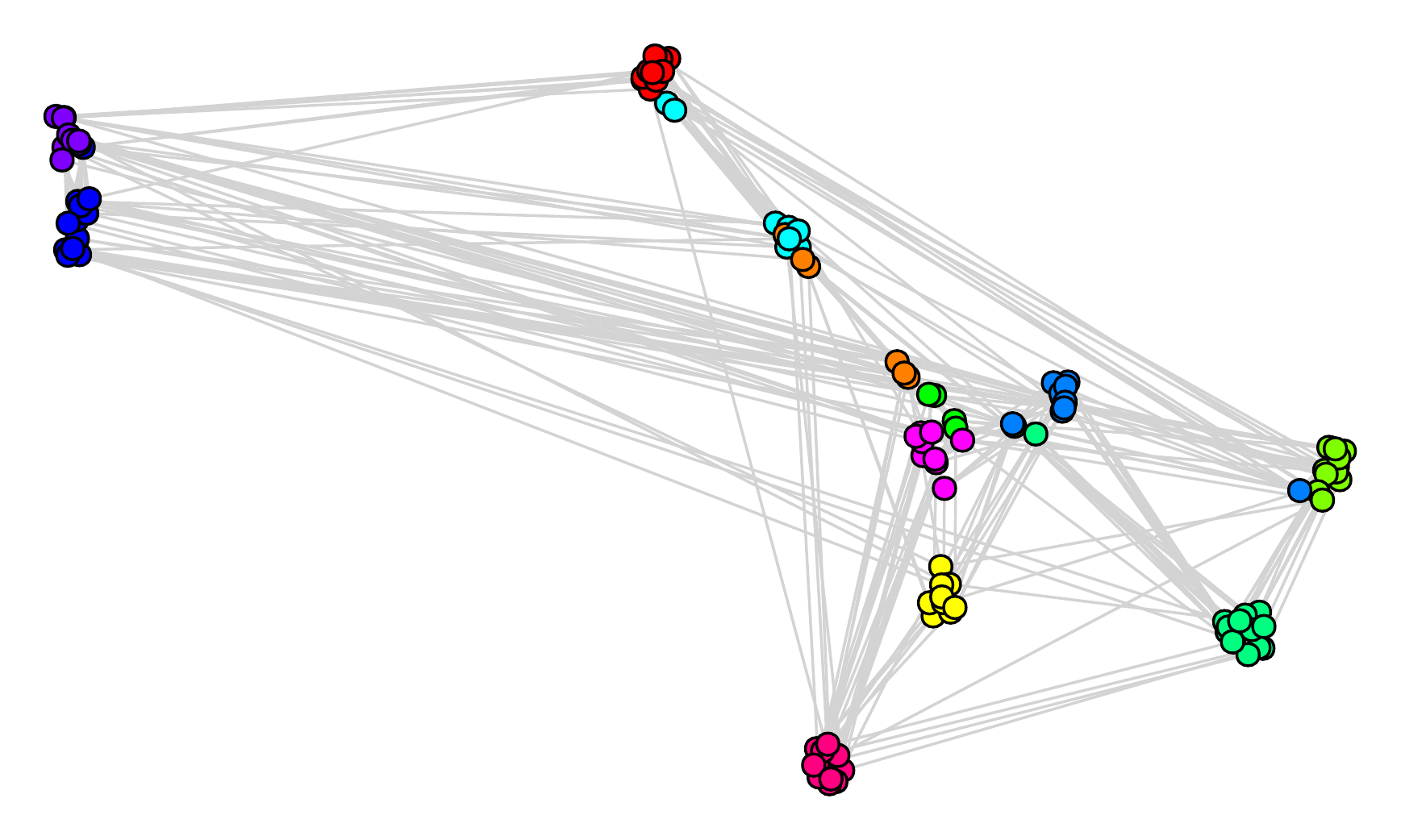}
     \caption{2-dimensional projections of embeddings of \textbf{FOOTBALL} graph according to the framework: the best (left), the worst with respect to the local score (middle), and the worst with respect to the global score (right).}
     \label{fig_paper:toy_best_worst_layout}
 \end{figure}

We recommend to select the best embedding by careful investigation of both scores but, alternatively, one can ask the framework to decide which embedding to use in an unsupervised fashion. In order to do it, one may combine the global and the local scores, as explained in Equation~(\ref{eq:final_score}), to make a decision based on a single combined divergence score. This way we identified the best and the worst embeddings. To visualize the selected embeddings in high dimensions we needed to perform dimension reduction that seeks to produce a low dimensional representation of high dimensional data that preserves relevant structure. For that purpose we used the Uniform Manifold Approximation and Projection (UMAP)\footnote{\url{ https://github.com/lmcinnes/umap}}~\cite{mcinnes2018umap}, a novel manifold learning technique for dimension reduction. UMAP is constructed from a theoretical framework based in Riemannian geometry and algebraic topology; it provides a practical scalable algorithm that applies to real world datasets. The results are presented in Figure~\ref{fig_paper:toy_best_worst_layout}. The embedding presented on the left hand side shows the winner that seems to not only separate nicely the communities but also puts pairs of communities far from each other if there are few edges between them; as a result, both scores are good. The embedding in the middle separates communities but the nodes within communities are clumped together resulting in many edges that are too short. There are also a lot of edges that are too long. As a result, the local score for this embedding is bad. Finally, the embedding on the right has a clear problem with distinguishing communities and some of them are put close to each other resulting in a bad global score.

\subsection{Approximating the Two Scores}\label{sec:landmarks}

Let us start with an experiment testing whether our scalable implementation provides a good approximation of both measures, the 
density based (global) score $\Delta_\emb^g(G)$, and the link based (local) score $\Delta_\emb^{\ell}(G)$. We tested \textbf{SBM}, \textbf{LFR}, and \textbf{noisy-LFR} graphs with the two available embeddings (using the parameters and selected dimensions as described above). The number of landmarks was set to be 5 times larger than the number of communities in each graph, namely, \textbf{SBM} graph used $30\cdot 5 = 150$ landmarks, \textbf{LFR} had $75\cdot 5 = 375$ landmarks, and for \textbf{noisy-LFR} the number of landmarks was set to $54\cdot 5 = 270$. 

 \begin{figure}[htb]
     \centering
     \includegraphics[width=0.4\textwidth]{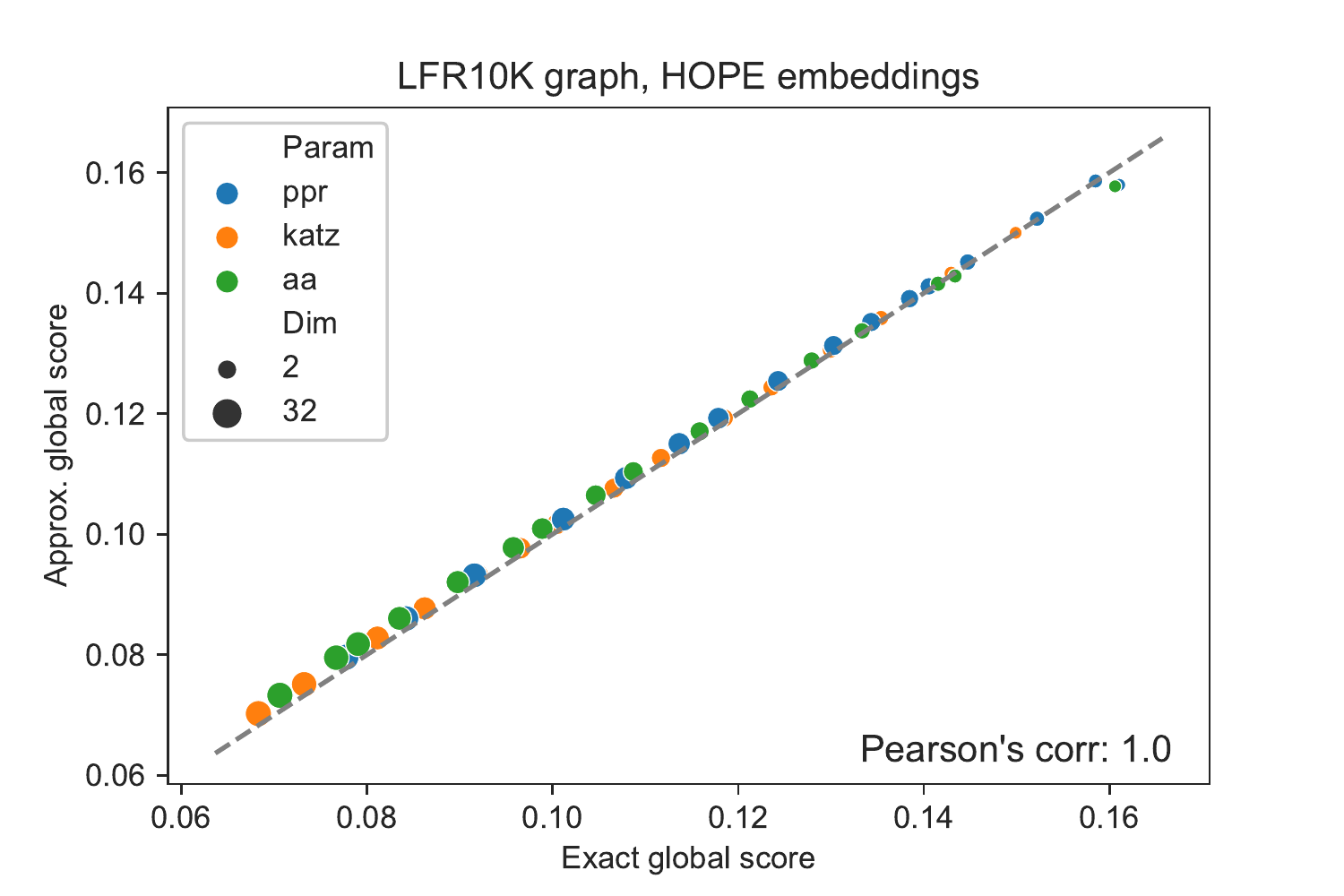}
         \hspace{.1cm}
     \includegraphics[width=0.4\textwidth]{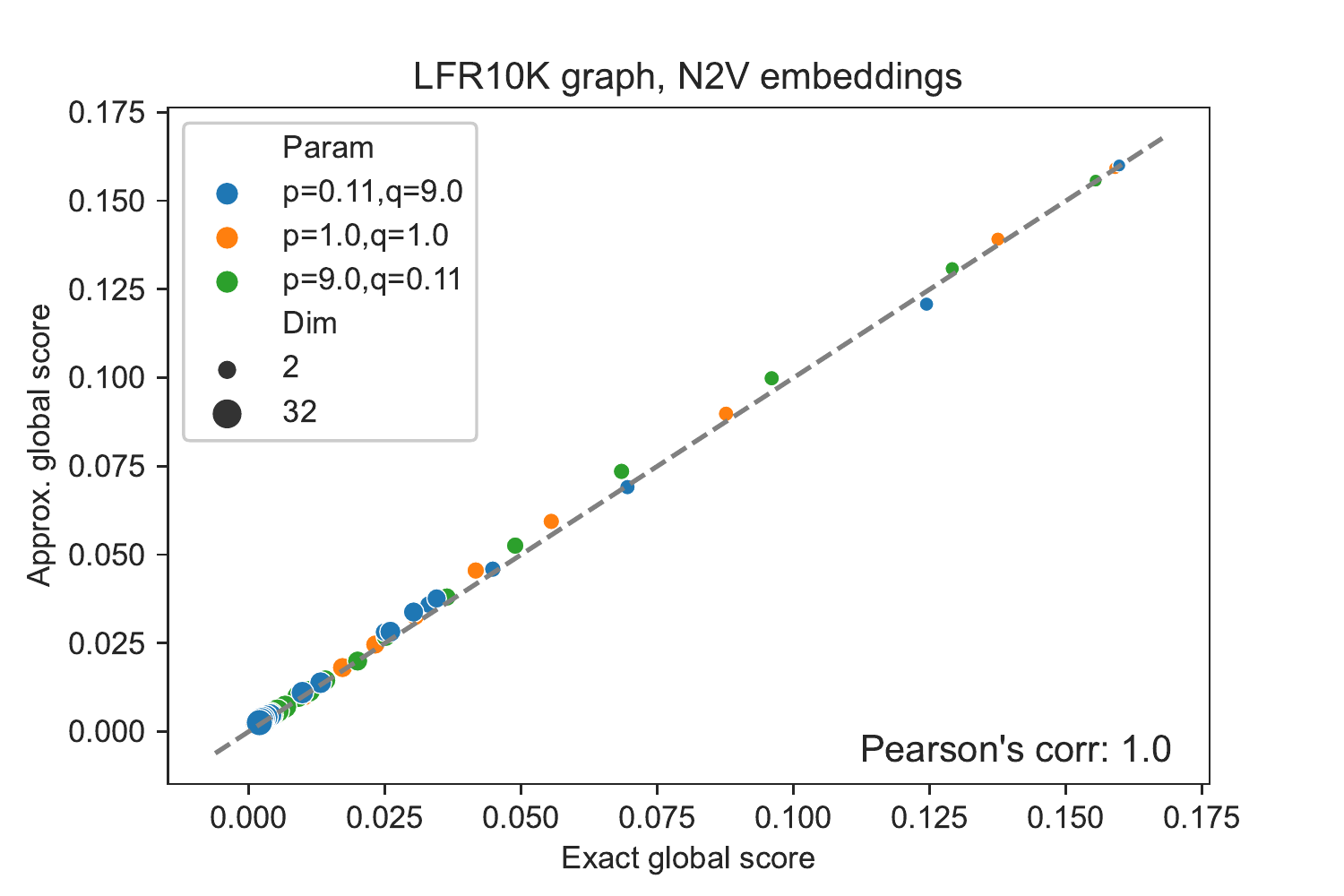}
     \caption{Approximated vs.\ exact global scores for \textbf{LFR} graphs and \textbf{HOPE} (left) and \textbf{Node2Vec} (right) embeddings.}
     \label{fig_paper:10kapprox_exact_div}
 \end{figure}cd ..

 \begin{figure}[htb]
     \centering
     \includegraphics[width=0.4\textwidth]{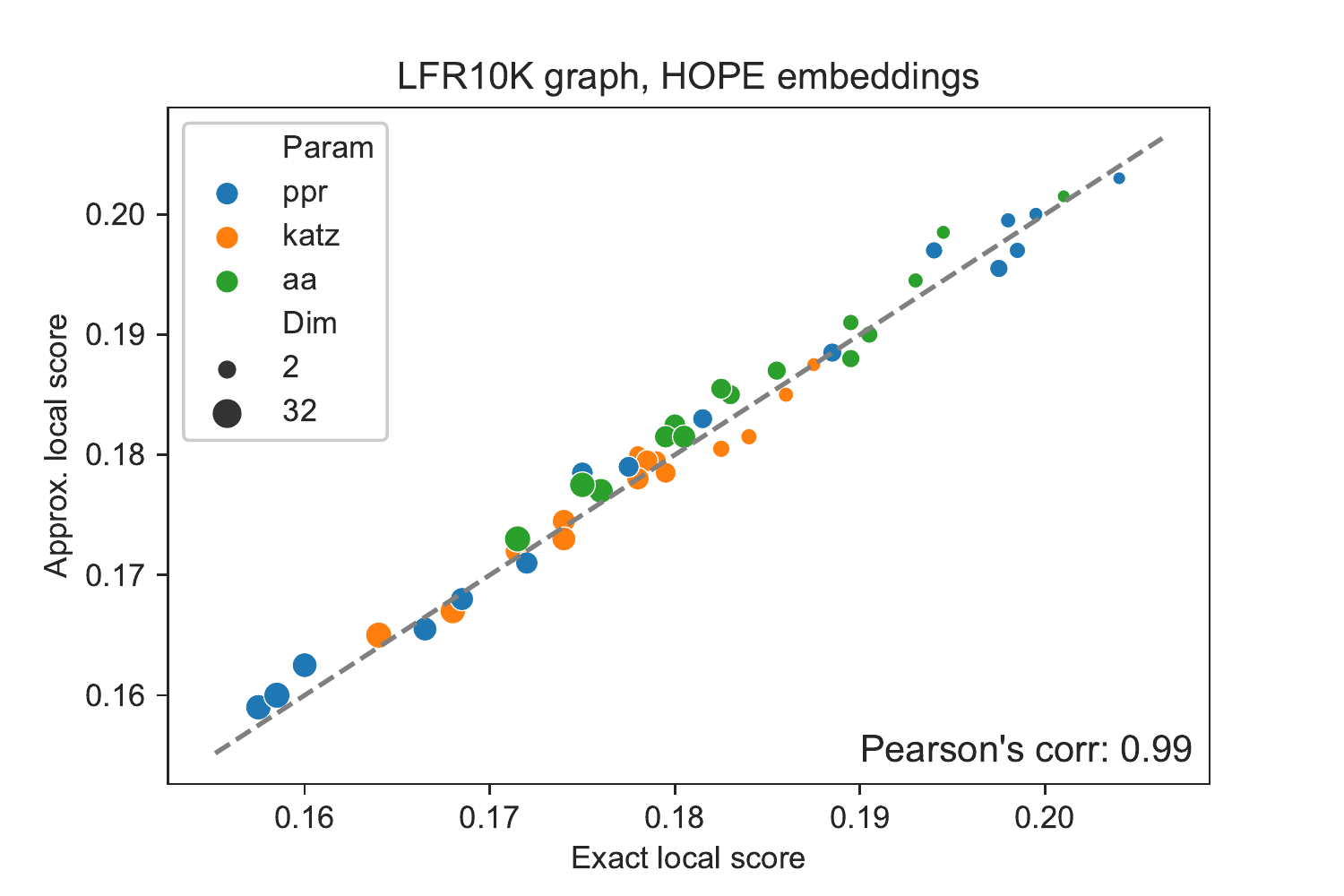}
         \hspace{.1cm}
     \includegraphics[width=0.4\textwidth]{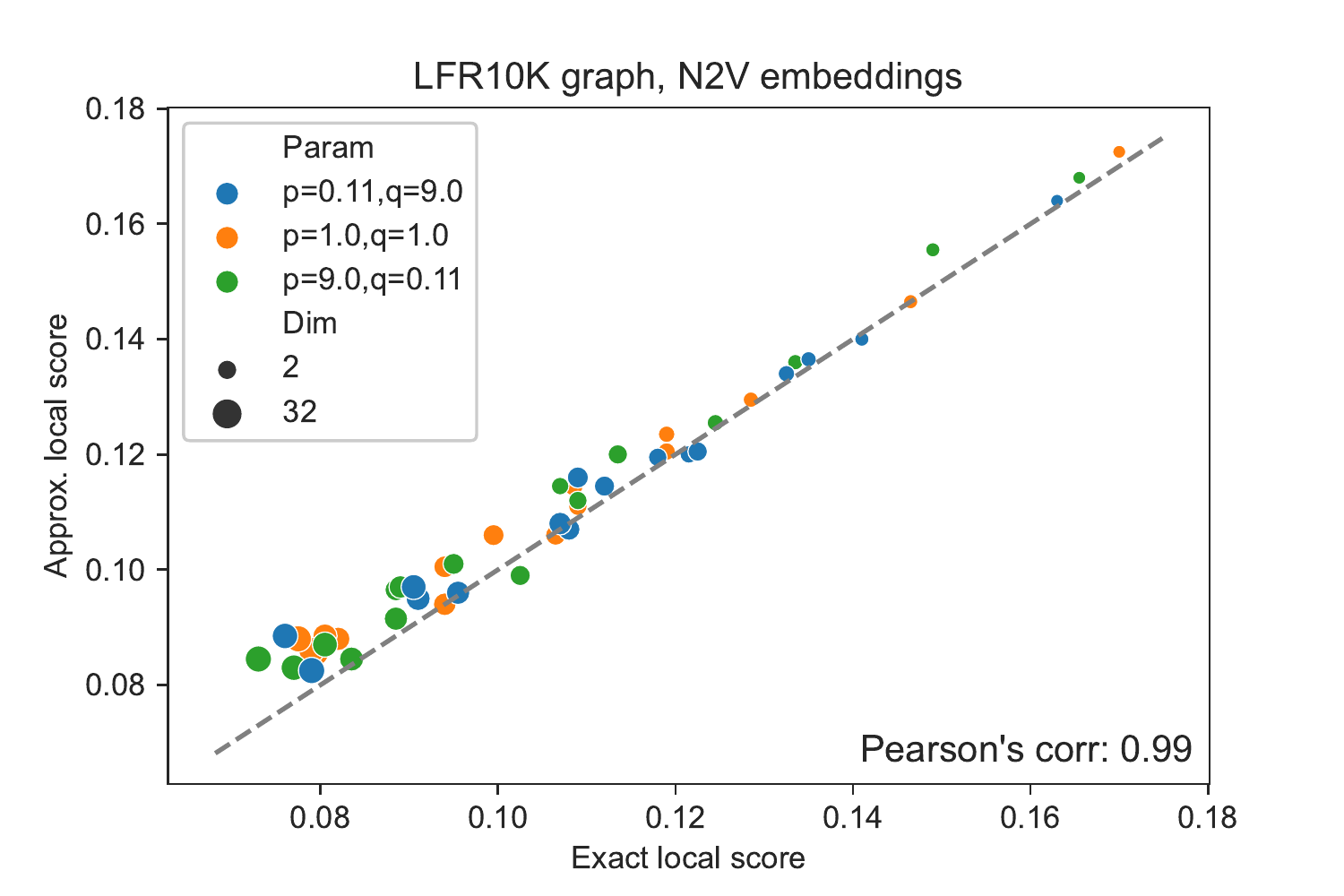}
     \caption{Approximated vs.\ exact local scores for \textbf{LFR} graphs and \textbf{HOPE} (left) and \textbf{Node2Vec} (right) embeddings.}
     \label{fig_paper:10kapprox_exact_auc}
 \end{figure}

Figures~\ref{fig_paper:10kapprox_exact_div} and~\ref{fig_paper:10kapprox_exact_auc} present the results of experiments for \textbf{LFR} graph and both embedding algorithms. (Experiments for other graphs can be found in Appendix~\ref{sec:appendix_experiments}.) The first figure shows how well the global score is approximated by a scalable algorithm and the second figure concentrates on the local score. Recall that graphs on $10{,}000$ nodes are, by default, the smallest graphs for which the framework uses scalable algorithm. The default setting in the framework is to use at least $4$ landmarks per community and at least $4\sqrt{n}$ of them overall, so $400$ landmarks for graphs of size $10{,}000$ and many more for larger graphs. In our experiment, we see good results with even fewer landmarks. This was done in purpose to test approximation precision in a challenging corner case. As expected, global properties reflected by the global score are easier to approximate than local properties investigated by the local score that are more sensitive to small perturbations. Having said that, both scores are approximated to a satisfactory degree. The Pearson correlation coefficients for both \textbf{HOPE} and \textbf{Node2Vec} are close to 1 for the global score and roughly 0.99 for the local score. For other standard correlation measures see Tables~\ref{tab:cor_approx_exact_glob_scores} and~\ref{tab:cor_approx_exact_loc_scores}.

\begin{table}[htb]
\centering
\scalebox{0.7}{
 \begin{tabular}{|l|l|l|l|l}
 \cline{1-4}
 \textbf{Graph-Embedding} & \textbf{Pearson} & \textbf{Spearman} & \textbf{Kendall-Tau} &  \\ \cline{1-4}
 SBM10K-HOPE              & 0.98             & 0.98              & 0.94                 &  \\ \cline{1-4}
 SBM10K-N2V               & 1.0             & 1.0              & 0.97                 &  \\ \cline{1-4}
 LFR10K-HOPE              & 1.0             & 1.0               & 0.99                 &  \\ \cline{1-4}
 LFR10K-N2V               & 1.0              & 1.0               & 0.99                 &  \\ \cline{1-4}
 nLFR10K-HOPE             & 1.0              & 1.0               & 1.0                 &  \\ \cline{1-4}
 nLFR10K-N2V              & 1.0              & 0.96               & 0.87                 &  \\ \cline{1-4}
 \end{tabular}
}
     \caption{Correlation between approximated and (exact) global scores.}
     \label{tab:cor_approx_exact_glob_scores}
 \end{table}


\begin{table}[ht!]
\centering
\scalebox{0.7}{
 \begin{tabular}{|l|l|l|l|l}
 \cline{1-4}
 \textbf{Graph-Embedding} & \textbf{Pearson} & \textbf{Spearman} & \textbf{Kendall-Tau} &  \\ \cline{1-4}
 SBM10K-HOPE              & 0.99             & 0.99               & 0.92                 &  \\ \cline{1-4}
 SBM10K-N2V               & 0.99             & 0.97              & 0.87                 &  \\ \cline{1-4}
 LFR10K-HOPE              & 0.99             & 0.99              & 0.93                 &  \\ \cline{1-4}
 LFR10K-N2V               & 0.99             & 0.98               & 0.90                 &  \\ \cline{1-4}
 nLFR10K-HOPE             & 1.0              & 1.0              & 0.98                 &  \\ \cline{1-4}
 nLFR10K-N2V              & 1.0             & 0.99              & 0.94                 &  \\ \cline{1-4}
 \end{tabular}
 }
     \caption{Correlation between approximated and (exact) local scores.}
     \label{tab:cor_approx_exact_loc_scores}
 \end{table}

\begin{figure}[htb]
    \centering
    \includegraphics[width=0.4\textwidth]{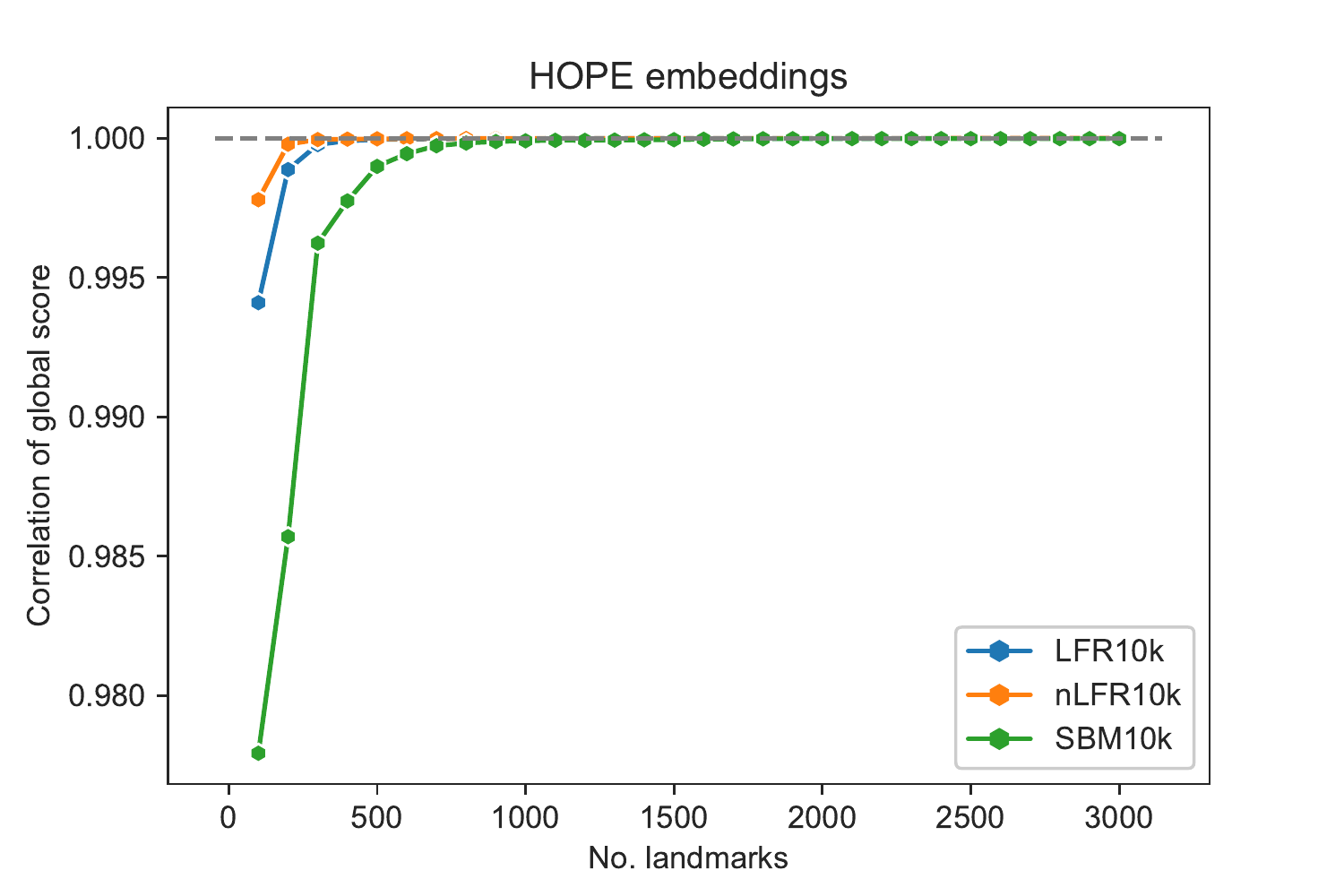}
        \hspace{.1cm}
    \includegraphics[width=0.4\textwidth]{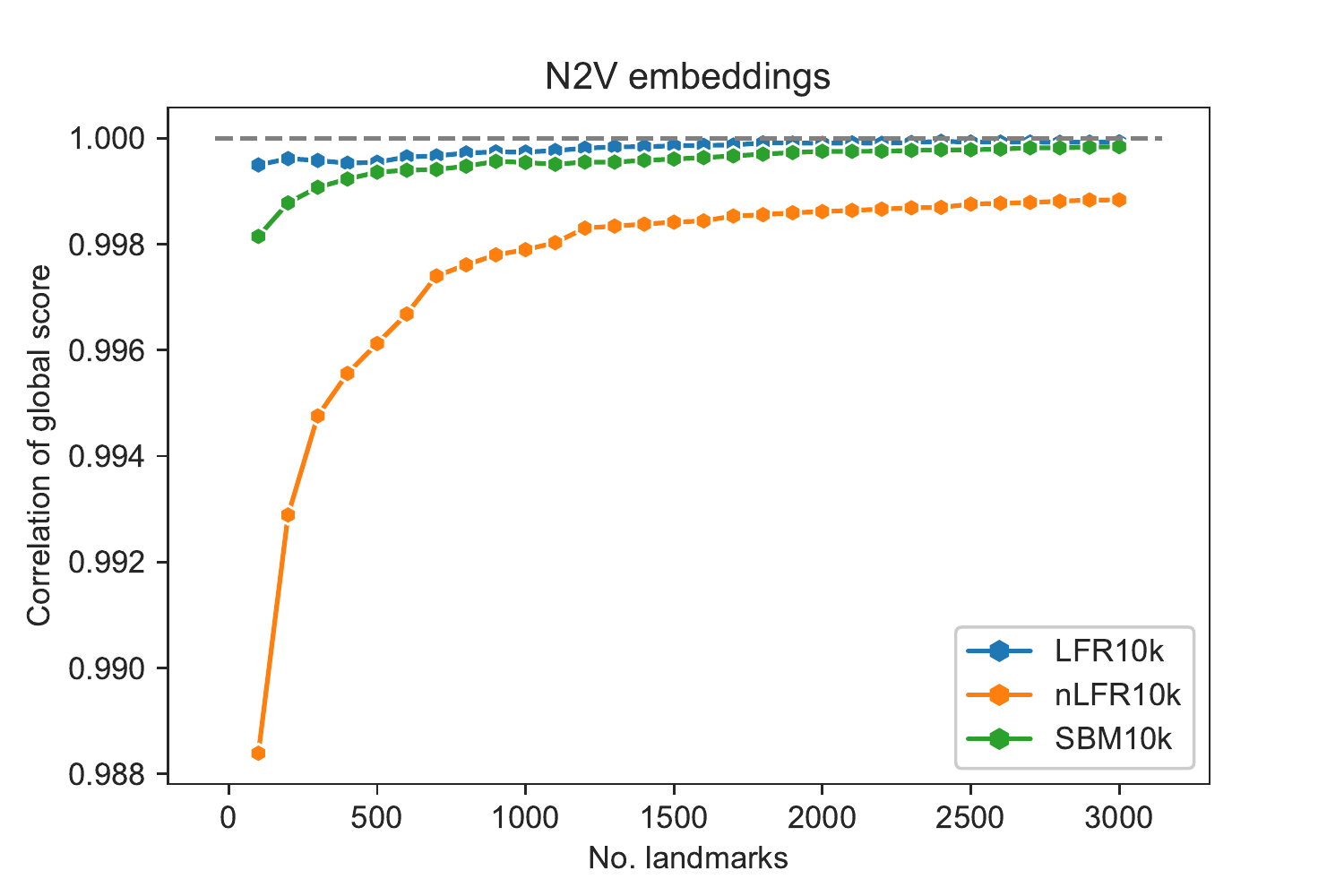}
     \caption{Pearson correlation between approximated and exact global scores for \textbf{SMB}, \textbf{LFR}, \textbf{noisy-LFR} graphs and \textbf{HOPE} (left) and \textbf{Node2Vec} (right) embeddings.}
    \label{fig:10kratioapprox_2measseed}
\end{figure}

\begin{figure}[htb]
    \centering
    \includegraphics[width=0.4\textwidth]{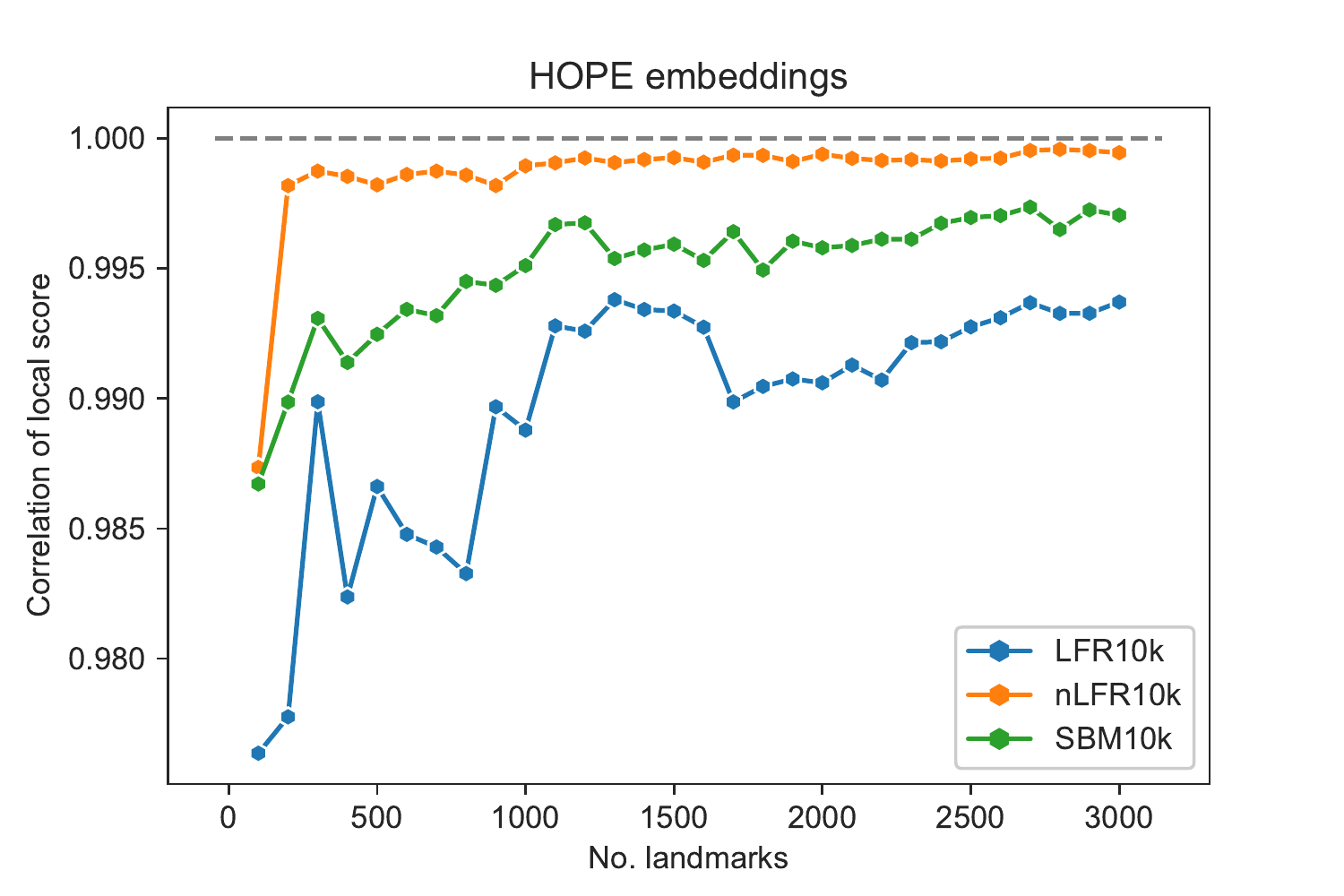}
        \hspace{.1cm}
    \includegraphics[width=0.4\textwidth]{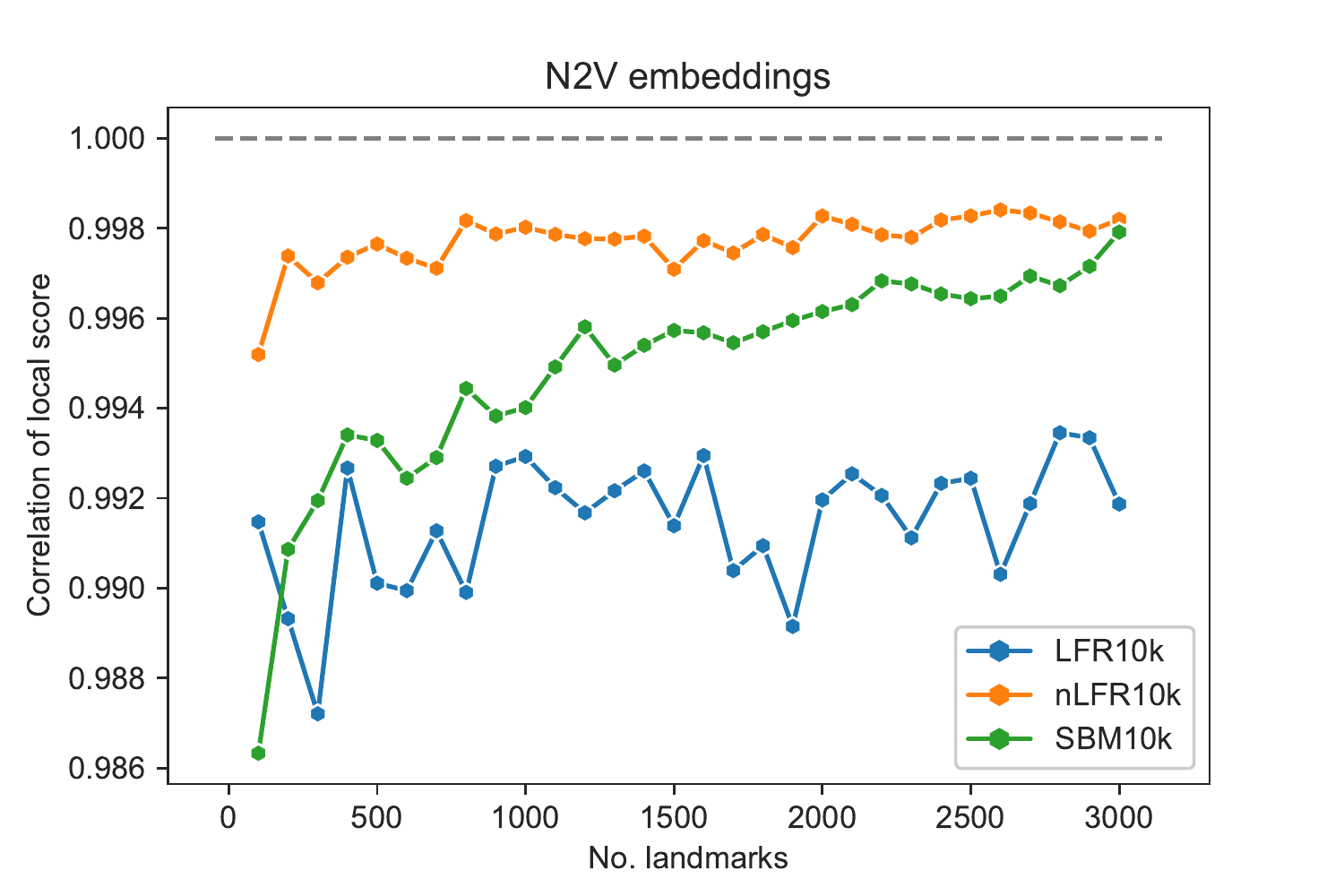}
     \caption{Pearson correlation between approximated and exact local scores for \textbf{SMB}, \textbf{LFR}, \textbf{noisy-LFR}  graphs and \textbf{HOPE} (left) and \textbf{Node2Vec} (right) embeddings.}
    \label{fig:10kratioapprox_exact_aucseed}
\end{figure}

Finally, we checked if the number of landmarks used by the framework as a default value (namely, $4\sqrt{n}$) is a good choice. We see in Figures~\ref{fig:10kratioapprox_2measseed} and~\ref{fig:10kratioapprox_exact_aucseed} that the approximation quickly stabilizes and so there is no need for a large number of landmarks to get the desired approximation. As commented earlier, the local score is more challenging to approximate (see Figure~\ref{fig:10kratioapprox_exact_aucseed} again) and the Pearson correlation between the approximated and the exact local scores does not seem to tend to 1 quickly but it is very close to 1, providing a satisfactory precision.

\subsection{Node Classification and Community Detection vs.\ the Global Score}\label{sec:correlation_with_global_score}

It is expected and desired that the global score measures how useful a given embedding is for any application that requires a good understanding of a global structure of the network. In our next experiments, we tested two such applications: node classification and community detection. 

\medskip

Node classification task aims to train a model to learn to which class a node belongs to. Typically, the goal is to label each node with a categorical class (binary classification or multi-class classification), or to predict a continuous number (regression). The process is supervised in nature, that is, the model is trained using a subset of nodes that have ground-truth labels.

For each graph (\textbf{SBM}, \textbf{LFR}, \textbf{noisy-LFR}, and \textbf{EMAIL}), each of the two embeddings (\textbf{HOPE} and \textbf{Node2Vec}) was used as an input for community detection task based on XGBoost model. XGBoost\footnote{\url{https://github.com/dmlc/xgboost}} is an open-source software library that provides a regularizing gradient boosting framework, the algorithm of choice for many winning teams of machine learning competitions. Having said that, since we aim to have a fair benchmark evaluating if an embedding extracts any useful global properties of the network instead of optimizing the quality of the outcome, we used vanilla XGBoost with default hyper-parameters. For each embedding we conducted 10 independent repetitions of the standard training process. First, nodes with the corresponding embedding was randomly partitioned into a training and a test set (25\% of all observations). Then, XGBoost model was trained on the training set together with the corresponding labels indicated the ground-truth community the nodes belong to. Finally, the model was used to predict the communities on the test set and the accuracy was reported. 

\medskip

Similarly, in order to investigate how much of the community structure got preserved by the evaluated embeddings, each embedding was independently clustered 20 times using the classic $k$-means algorithm, a well-known method that aims to partition $n$ vectors into $k$ clusters in which each vector belongs to the cluster with the nearest mean (center of mass). As before, since we aim for a fair and easy benchmark rather than a carefully tuned specific approach, we simply used the vanilla $k$-means algorithm and the value of $k$ set to the true number of communities. The adjusted mutual information (AMI) score was then calculated based on the two partitions of the set of nodes: the clusters assignment returned by the $k$-means algorithm and the ground-truth communities. This approach is similar to the one used in the recent paper~\cite{tandon2021community} in which the authors investigate if embeddings can be used to detect communities.

\begin{figure}[h!]
    \centering
    \includegraphics[width=0.4\textwidth]{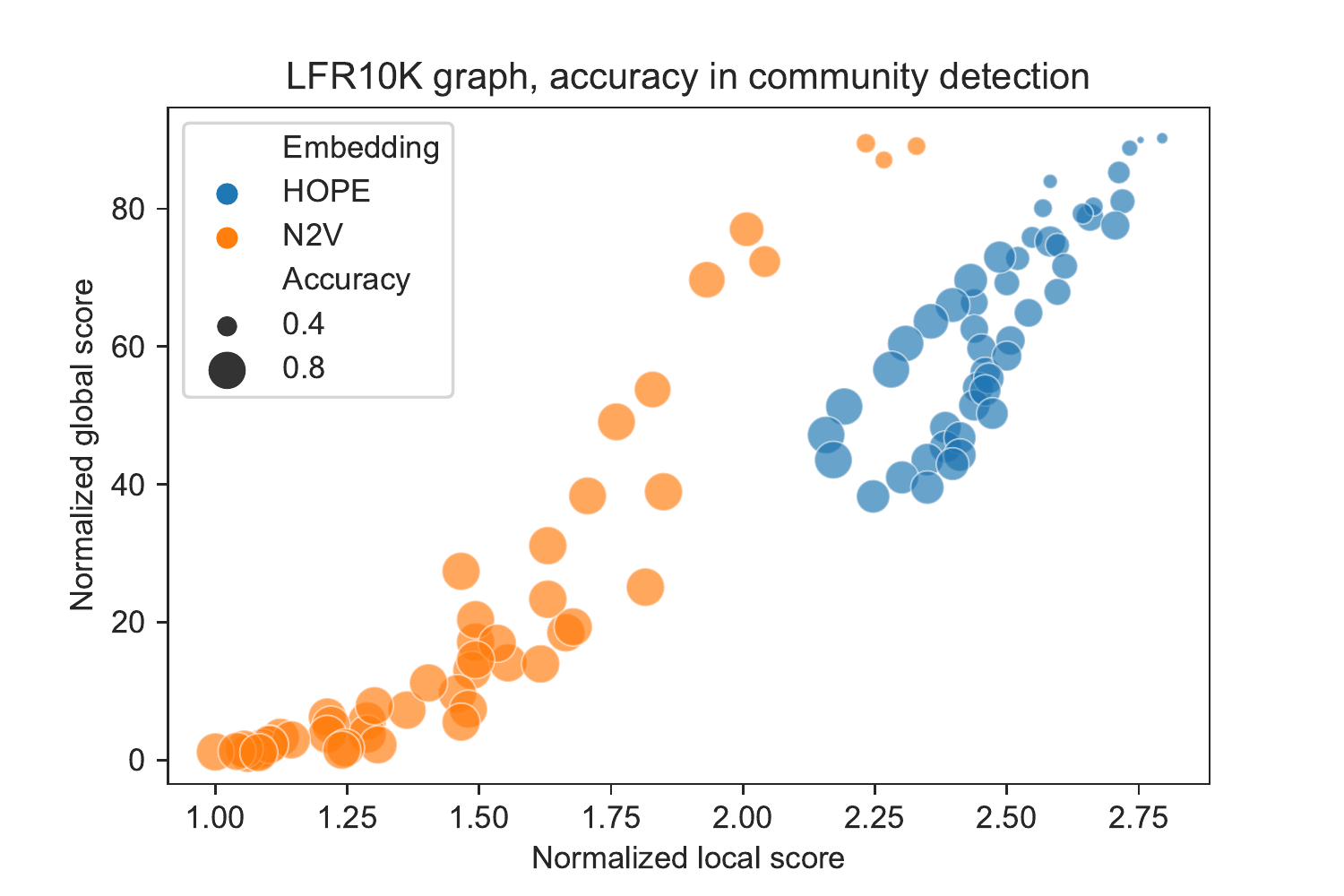}
        \hspace{.1cm}
    \includegraphics[width=0.4\textwidth]{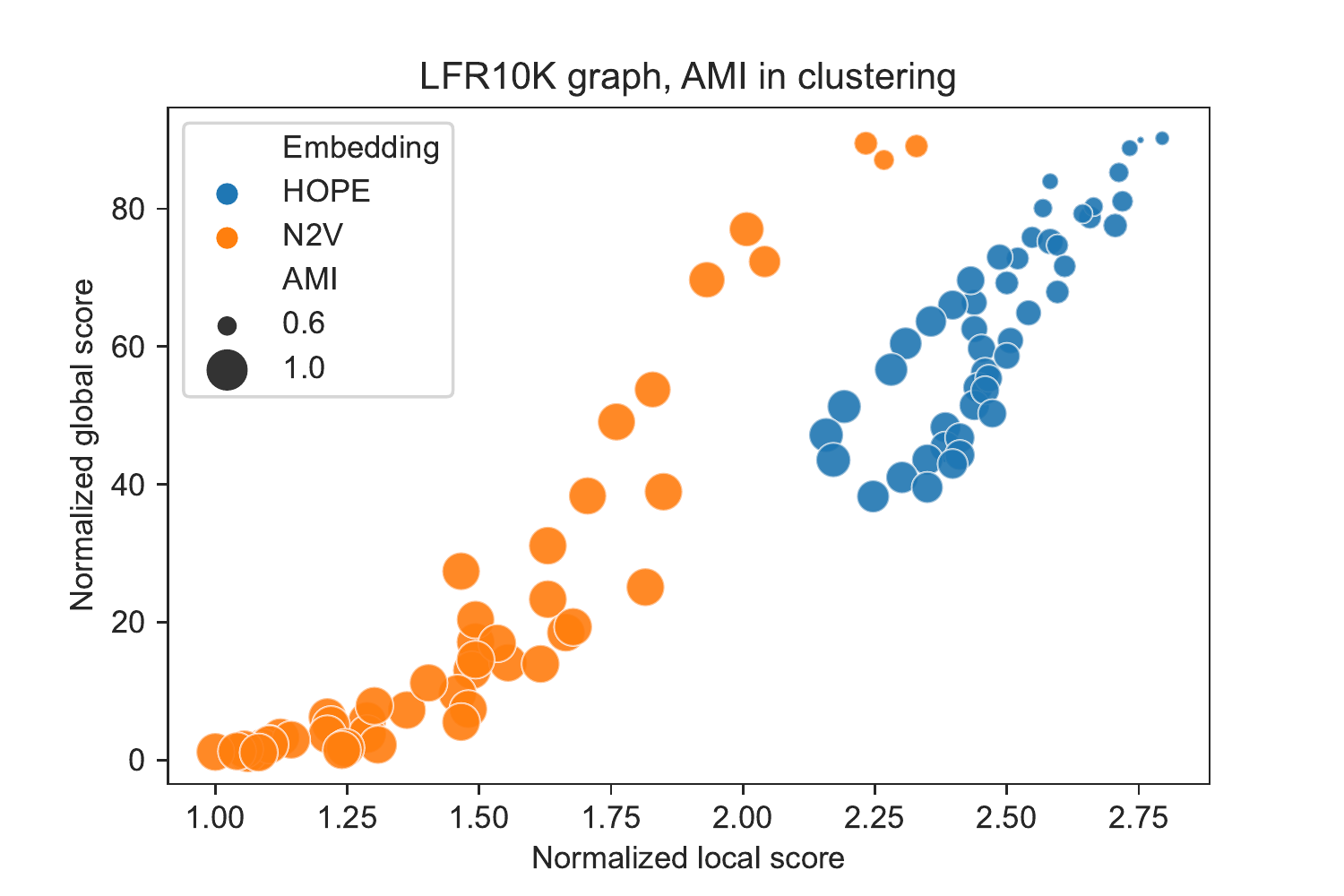}
    \vspace{.1cm}
    \includegraphics[width=0.4\textwidth]{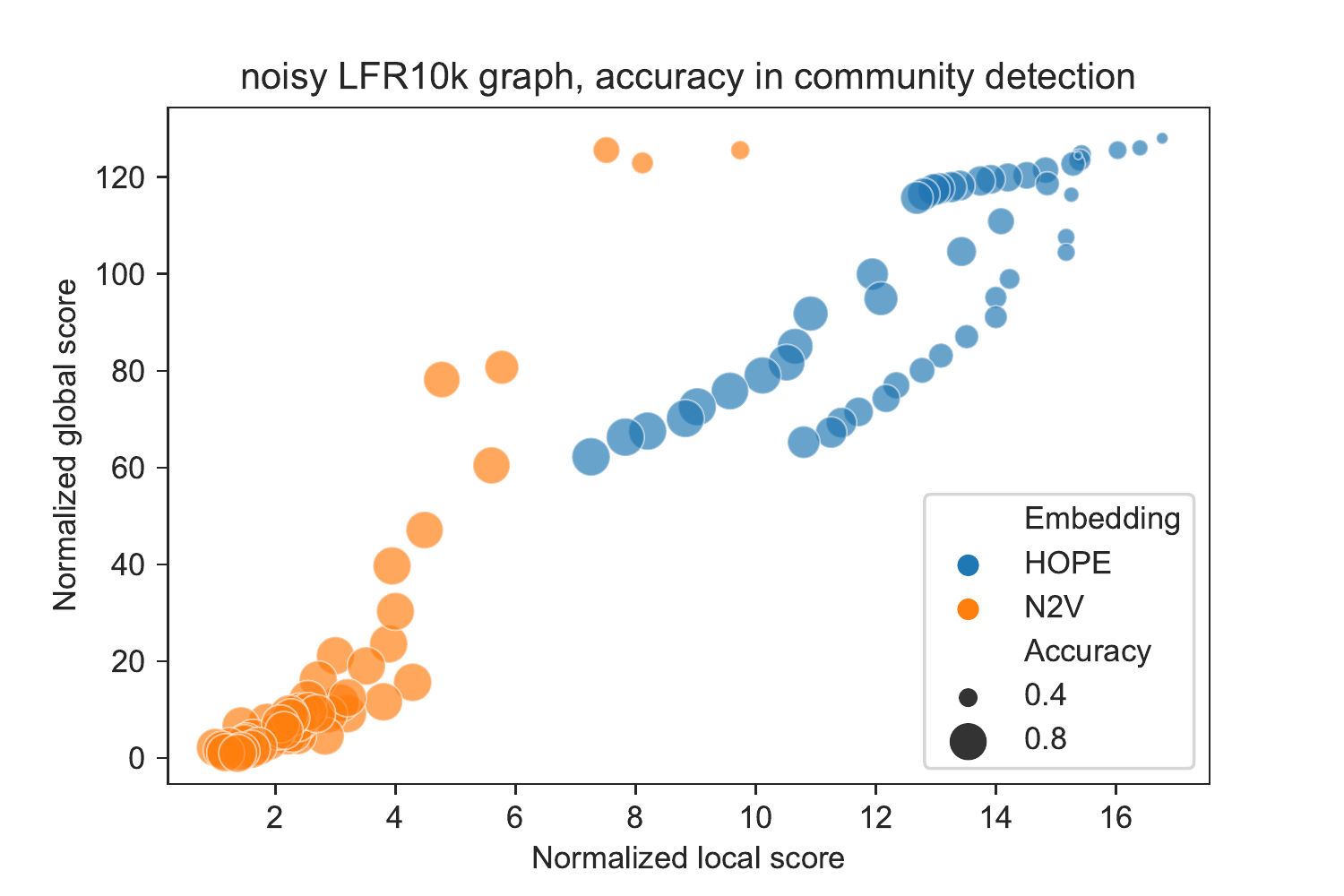}
        \hspace{.1cm}
    \includegraphics[width=0.4\textwidth]{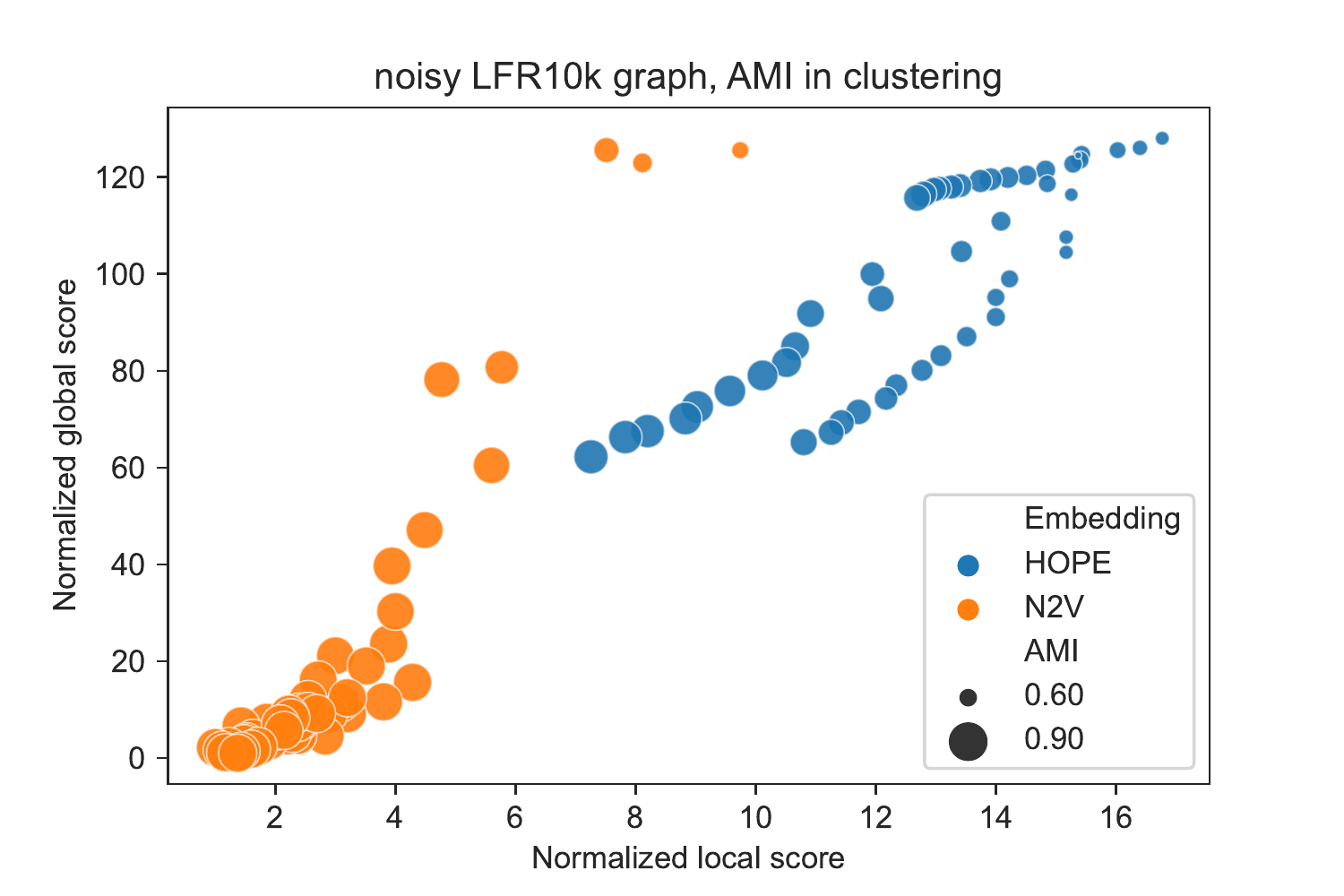}
    \vspace{.1cm}
    \includegraphics[width=0.4\textwidth]{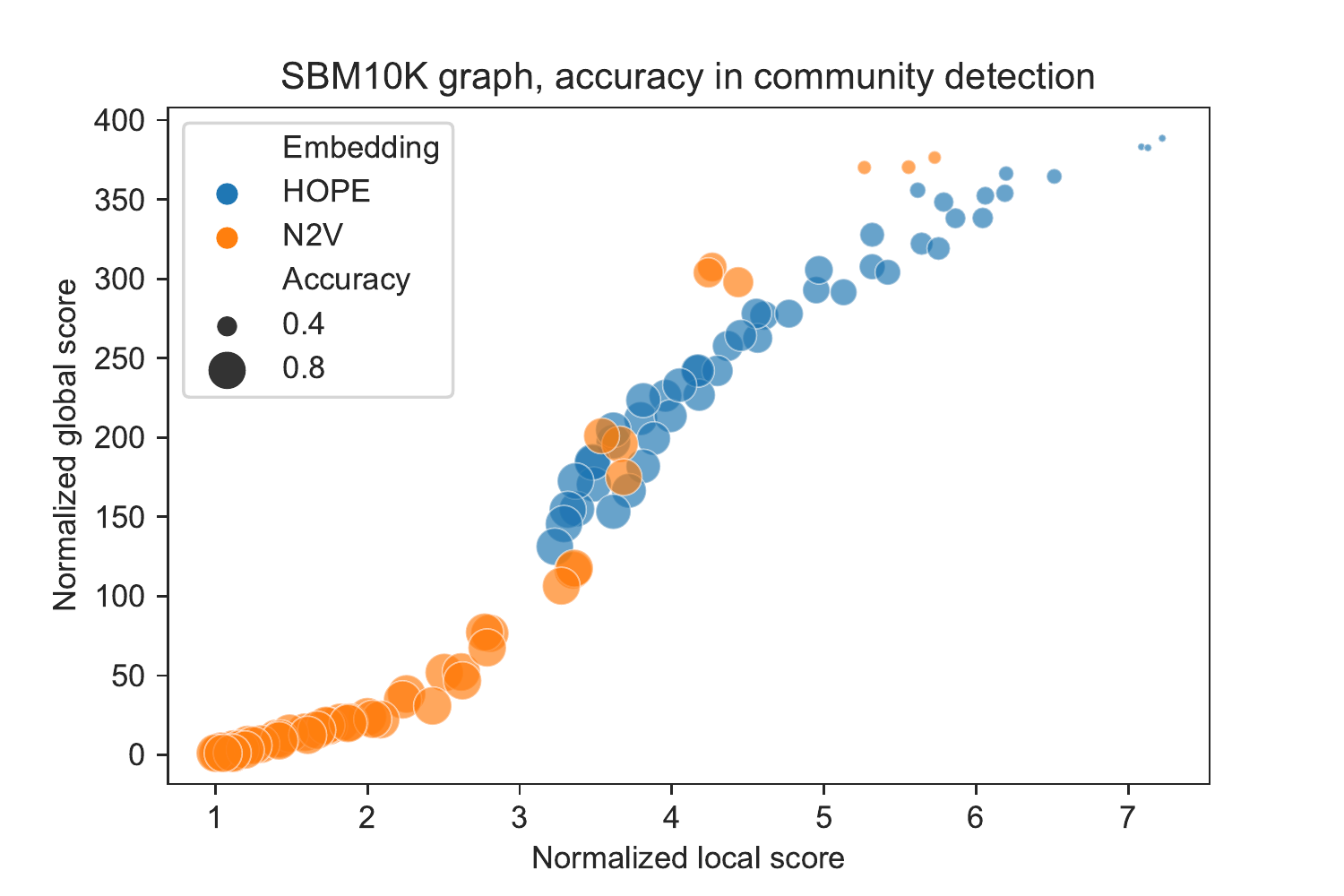}
        \hspace{.1cm}
    \includegraphics[width=0.4\textwidth]{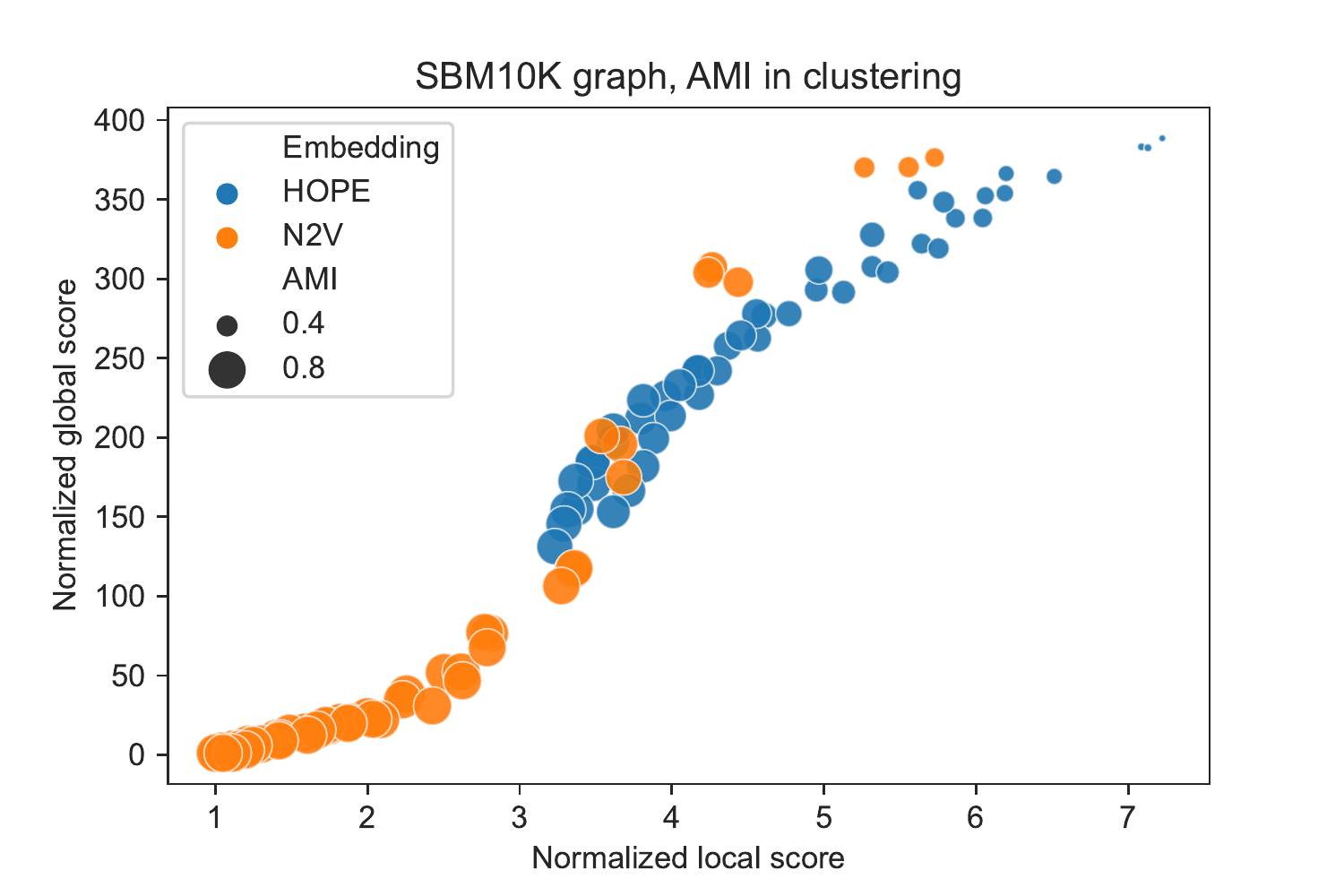}
    \vspace{.1cm}
    \includegraphics[width=0.4\textwidth]{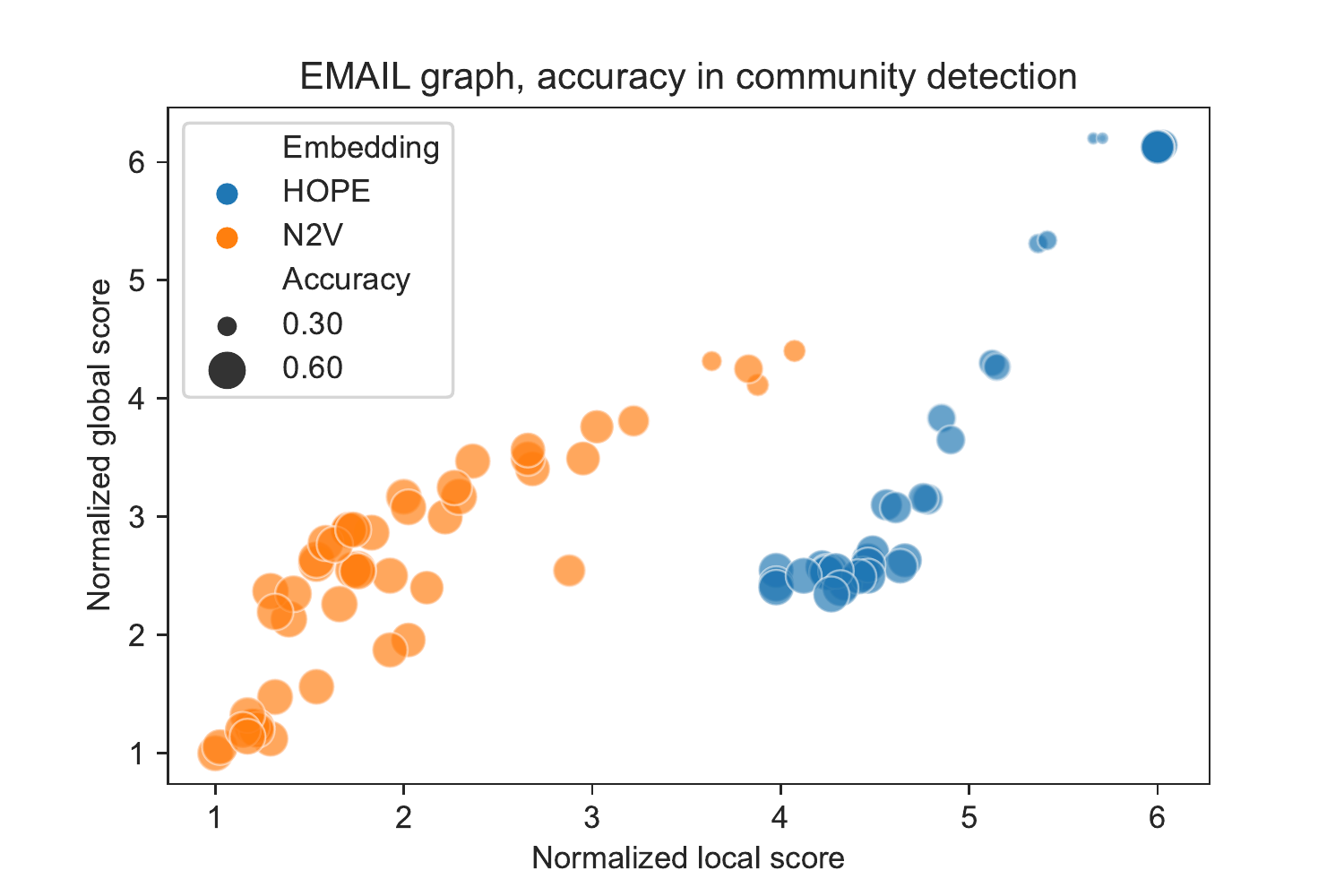}
        \hspace{.1cm}
    \includegraphics[width=0.4\textwidth]{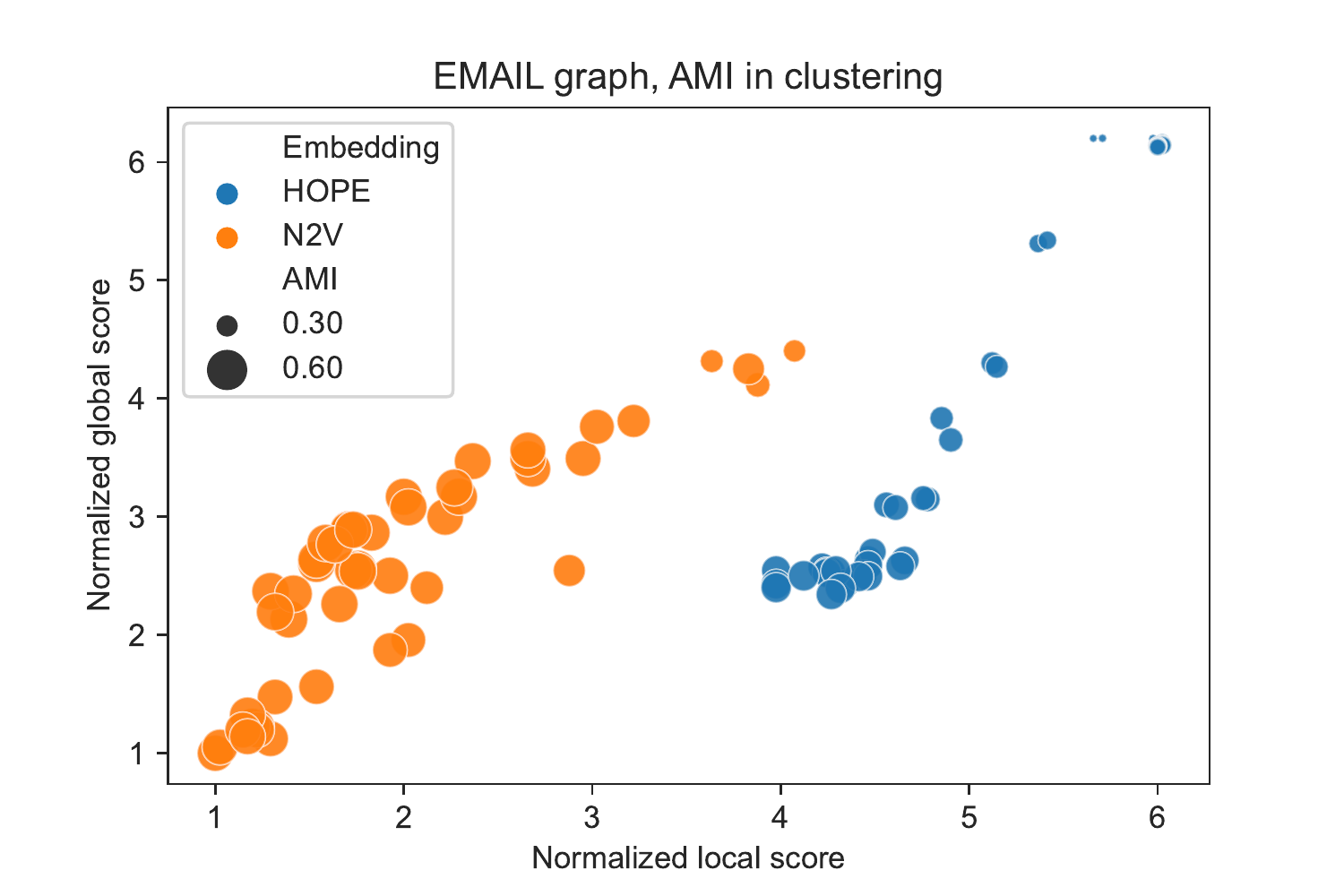}
    \caption{Global and local scores with accuracy (left) and AMI (right) overlay for \textbf{SBM}, \textbf{LFR}, \textbf{noisy-LFR}, \textbf{EMAIL} graphs and \textbf{HOPE}, \textbf{Node2Vec} embeddings.}
    \label{fig:normalized}
\end{figure}

\medskip

The results of both experiments are presented on Figure~\ref{fig:normalized} (the accuracy score for node classification task is presented on the left side whereas the AMI score for community detection task can be found on the right side). The embeddings that are identified by the framework as good (left-bottom corner, close to the auxiliary point $(1,1)$ representing the hypothetical perfect score) tend to perform well in both applications (large balls representing large values of the accuracy/AMI). On the other hand, poorly scored embeddings (top-right corner) perform poorly (small balls). The same conclusion is obtained after more rigorous investigation of the correlation coefficients between the accuracy/AMI and the global/local scores (see Table~\ref{tab:accuracy_loc_glob} and, respectively, Table~\ref{tab:AMI_loc_glob}). Let us note that the rank-based correlations (Spearman and Kendall-Tau) report a smaller correlation for \textbf{Node2Vec} embeddings as in our experiment we generated many embeddings that were comparable and of very good quality. As a result, their rankings with respect to the global/local scores and the accuracy/AMI might not be exactly the same but, in any case, the framework was able to distinguish good embeddings from bad ones. More experiments can be found in Appendix~\ref{sec:appendix_experiments}.

\begin{table}[ht]
\centering
\scalebox{0.7}{
\begin{tabular}{|l|l|l|l|l|l|l|}
\hline
\multicolumn{1}{|c|}{\multirow{2}{*}{\textbf{Graph-Embedding}}} & \multicolumn{2}{l|}{\textbf{Pearson}} & \multicolumn{2}{l|}{\textbf{Spearman}} & \multicolumn{2}{l|}{\textbf{Kendall-Tau}} \\ \cline{2-7} 
\multicolumn{1}{|c|}{}                                          & \textbf{Global}    & \textbf{Local}   & \textbf{Global}    & \textbf{Local}    & \textbf{Global}      & \textbf{Local}     \\ \hline
SBM10K-HOPE                                                     & -0.97              & -0.99            & -1.0               & -1.0              & -0.97                & -0.96              \\ \hline
SBM10K-N2V                                                      & -0.9               & -0.82            & -0.57              & -0.57             & -0.42                & -0.42              \\ \hline
LFR10K-HOPE                                                     & -0.96              & -0.93            & -1.0               & -0.97             & -0.97                & -0.89              \\ \hline
LFR10K-N2V                                                      & -0.8               & -0.71            & -0.25              & -0.23             & -0.16                & -0.14              \\ \hline
NLFR10K-HOPE                                                    & -0.98              & -0.98            & -1.0               & -1.0              & -0.96                & -0.96              \\ \hline
NLFR10K-N2V                                                     & -0.91              & -0.85            & -0.14              & -0.1              & -0.08                & -0.04              \\ \hline
EMAIL-HOPE                                                     & -0.97              & -0.68            & -0.9              & -0.72              & -0.77               & -0.61              \\ \hline
EMAIL-N2V                                                     & -0.86              & -0.87            & -0.69              & -0.69              & -0.54               & -0.54             \\ \hline
\end{tabular}
}
\caption{Correlation coefficients between the accuracy scores in node classification task and global/local scores (averaged over all parameters).}
\label{tab:accuracy_loc_glob}
\end{table}

\begin{table}[ht]
\centering
\scalebox{0.7}{
\begin{tabular}{|l|l|l|l|l|l|l|}
\hline
\multirow{2}{*}{\textbf{Graph-Embedding}} & \multicolumn{2}{l|}{\textbf{Pearson}} & \multicolumn{2}{l|}{\textbf{Spearman}} & \multicolumn{2}{l|}{\textbf{Kendall-Tau}} \\ \cline{2-7} 
                                          & \textbf{Global}    & \textbf{Local}   & \textbf{Global}    & \textbf{Local}    & \textbf{Global}      & \textbf{Local}     \\ \hline
SBM10K-HOPE                               & -0.98              & -0.99            & -1.0               & -1.0              & -0.97                & -0.95              \\ \hline
SBM10K-N2V                                & -0.93              & -0.85            & -0.91              & -0.91             & -0.8                 & -0.81              \\ \hline
LFR10K-HOPE                               & -0.98              & -0.96            & -1.0               & -0.97             & -0.96                & -0.89              \\ \hline
LFR10K-N2V                                & -0.82              & -0.73            & -0.76              & -0.75             & -0.6                 & -0.59              \\ \hline
NLFR10K-HOPE                              & -0.99              & -0.99            & -0.99              & -0.99             & -0.95                & -0.95              \\ \hline
NLFR10K-N2V                               & -0.92              & -0.86            & -0.4               & -0.39             & -0.29                & -0.29              \\ \hline
EMAIL-HOPE                                                     & -0.86             & -0.46            & -0.62              & -0.48              & -0.54               & -0.42              \\ \hline
EMAIL-N2V                                                     & -0.86              & -0.87            & -0.77              & -0.80              & -0.61               & -0.65             \\ \hline
\end{tabular}
}
\caption{Correlation coefficients between the AMI scores in community detection task and global/local scores (averaged over all parameters).}
\label{tab:AMI_loc_glob}
\end{table}

\subsection{Link Prediction vs.\ the Local Score}\label{sec:correlation_with_local_score}

As discussed in the previous section, the global score returned by the framework captures how well the embedding preserves global properties of embedded networks. Similarly, it is expected and desired that the local score evaluates the power of embeddings to encapsulate local properties. In the next experiment, we tested one local algorithm, namely, link prediction algorithm.

\medskip

Indeed, most the the methods for directed graph embeddings are validated via link prediction~\cite{Zhou2021}, where one deletes a subset of edges (typically randomly selected) and then measures how well they can be recovered. A commonly used scoring method is the AUC (area under the ROC curve) obtained from ranking of node pairs from most to least likely to have an edge between them. The ROC curve (receiver operating characteristic curve) is a graph showing the performance of a classification model at all classification thresholds. On the other hand, the AUC provides an aggregate measure of performance across all possible classification thresholds. 
A variation of that approach is known as node recommendation, where one removes some outgoing edges for a subset of nodes and try to recover the most likely missing neighbours. Another score that may potentially be used is to compare precision and recall amongst the top-$k$ candidate node pairs for a specific $k$. This is useful, for example, if one has a limited ``budget'' to investigate if a given pair of nodes are actually linked by an edge or not, so one can only test a small number of pairs.

In our experiments, for each of the four graphs we tested (\textbf{SBM}, \textbf{LFR}, \textbf{noisy-LFR}, and \textbf{EMAIL}), given graph $G$, we randomly selected 5\% of its edges, and removed them to form a graph $G'$. Then we took another random sample of non-adjacent pairs of nodes in $G$, set $E'$. Both classes had the same number of pairs of nodes so that the test set created in such a way was balanced. Our goal was to train a model that uses one of the embeddings of graph $G'$ to detect which pairs of nodes in $E \cup E'$ are adjacent in $G$. We repeated this process independently 5 times (with 5 different seed values) and reported the average AUC scores.

For each of the two algorithms (\textbf{HOPE} and \textbf{Node2Vec}), each combination of their parameters and 16 dimensions we tested, we produced an embedding of graph $G'$. For evaluation purpose, we computed both the global and the local divergence scores for the produced embedding. In order to create a training set for the classifier, we considered all pairs of adjacent nodes in $G'$ (the positive class) as well as a random subset of pairs of non-adjacent nodes (the negative class) that is of the same size as the number of edges in $G'$ (as for the test sets, to keep both classes balanced). Finally, we concatenated embeddings of pairs of nodes representing edges/non-edges into feature vectors to be used for prediction. Such training set was used to train XGBoost model with default hyper-parameter values.  As already mentioned, in order to measure the quality of this simple model, we computed the AUC score that provides a measure of separability, as it tells us how capable the model is of distinguishing between the two classes, edges $E$ and non-edges $E'$, both coming from the original graph $G$. (Note that some pairs of nodes in the test set might overlap with pairs of nodes in the training set. Even more, some positive pair in the test set might be negative in the training one.)

\begin{table}[ht]
\centering
\scalebox{0.7}{
\begin{tabular}{|l|l|l|l|l|l|l|}
\hline
{\textbf{Graph-Embedding}} & \multicolumn{2}{l|}{\textbf{Pearson}} & \multicolumn{2}{l|}{\textbf{Spearman}} & \multicolumn{2}{l|}{\textbf{Kendall-Tau}} \\ \cline{2-7} 
   & \textbf{Global}    & \textbf{Local}   & \textbf{Global}    & \textbf{Local}    & \textbf{Global}      & \textbf{Local}     \\ \hline
SBM10K-HOPE                               & -0.92              & -0.95            & -0.99              & -0.96             & -0.91                & -0.83              \\ \hline 
SBM10K-N2V                                & -0.81              & -0.73            & -0.3               & -0.31             & -0.21                & -0.22              \\ \hline 
LFR10K-HOPE                               & -0.94              & -0.91            & -0.99              & -0.95             & -0.94                & -0.82              \\ \hline 
LFR10K-N2V                                & -0.92              & -0.9             & -0.91              & -0.87             & -0.82                & -0.74              \\ \hline
NLFR10K-HOPE                              & -0.96              & -0.87            & -0.97              & -0.9              & -0.89                & -0.73              \\ \hline 
NLFR10K-N2V                               & -0.91              & -0.95            & -0.96              & -0.9              & -0.86                & -0.74              \\ \hline 
EMAIL-HOPE                                & -0.9               & -0.49            & -0.75              & -0.43             & -0.58                & -0.35              \\ \hline 
EMAIL-N2V                                 & -0.8               & -0.86            & -0.65              & -0.76             & -0.48                & -0.57              \\ \hline
\end{tabular}
}
\caption{Correlation coefficients between the AUC scores in link prediction task and global/local scores (averaged over all parameters).}
\label{tab:corr_auc_scores}
\end{table}

\begin{figure}[h!]
    \centering
    \includegraphics[width=0.4\textwidth]{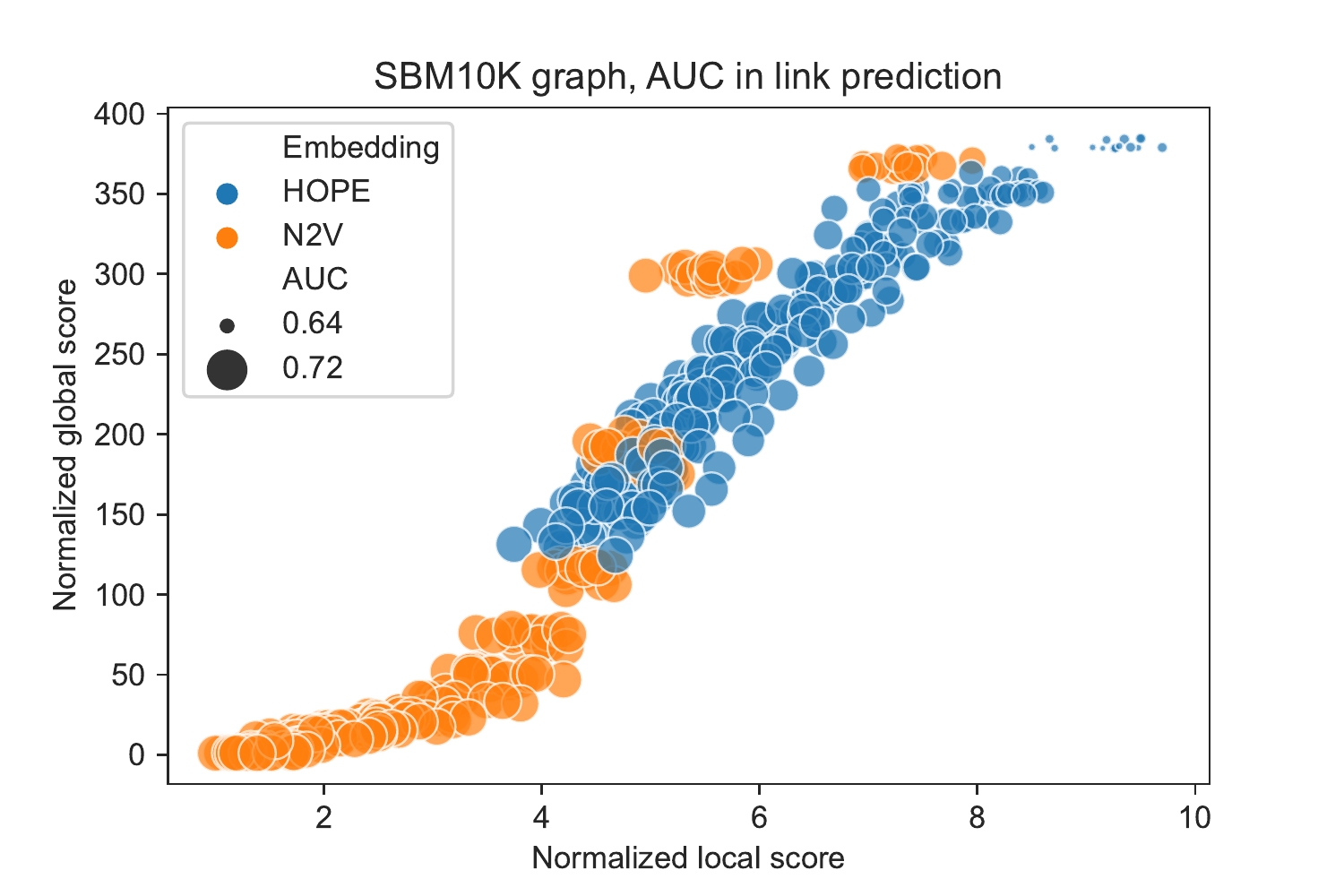}
        \hspace{.1cm}
    \includegraphics[width=0.4\textwidth]{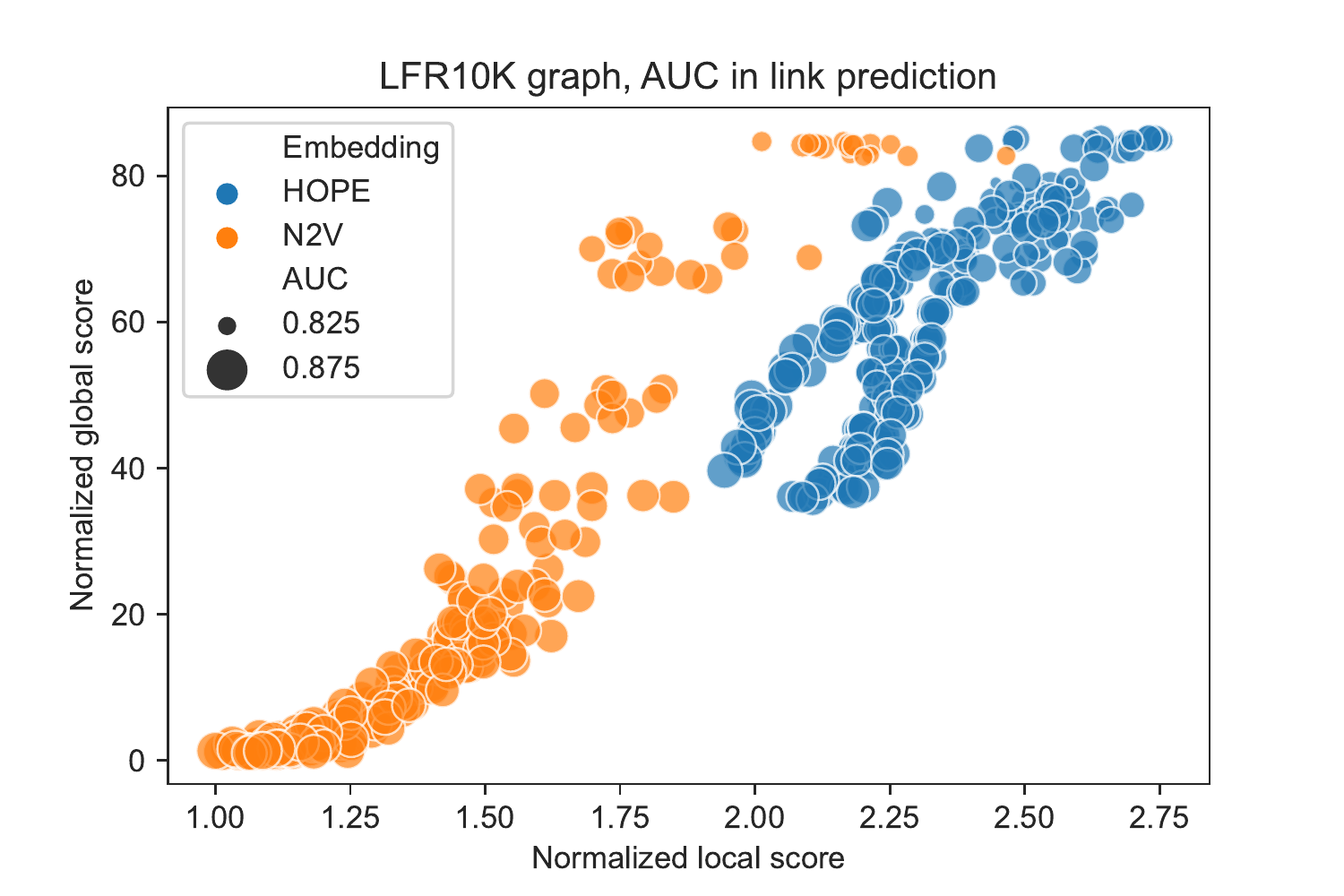}
    \vspace{.1cm}
    \includegraphics[width=0.4\textwidth]{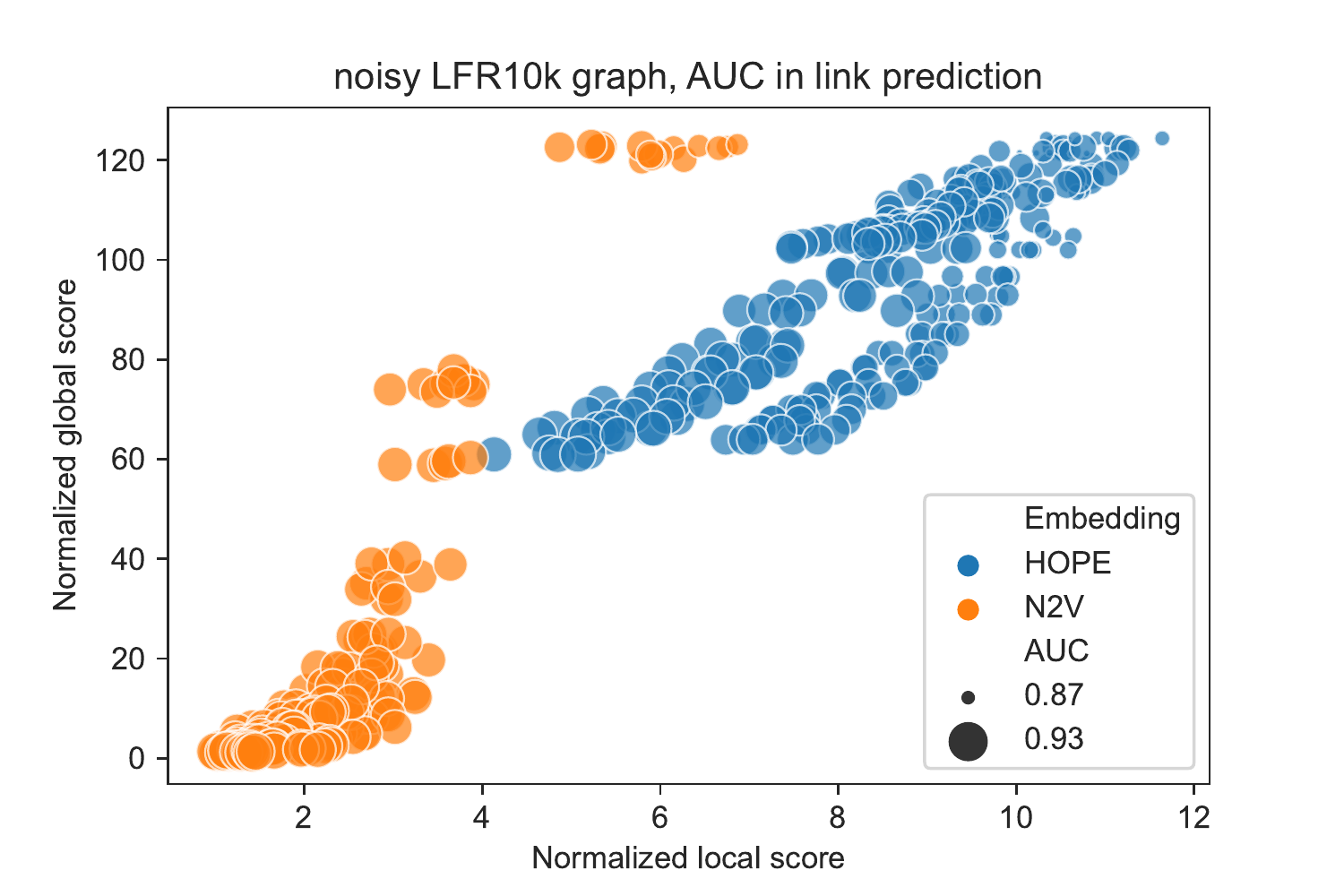}
        \hspace{.1cm}
    \includegraphics[width=0.4\textwidth]{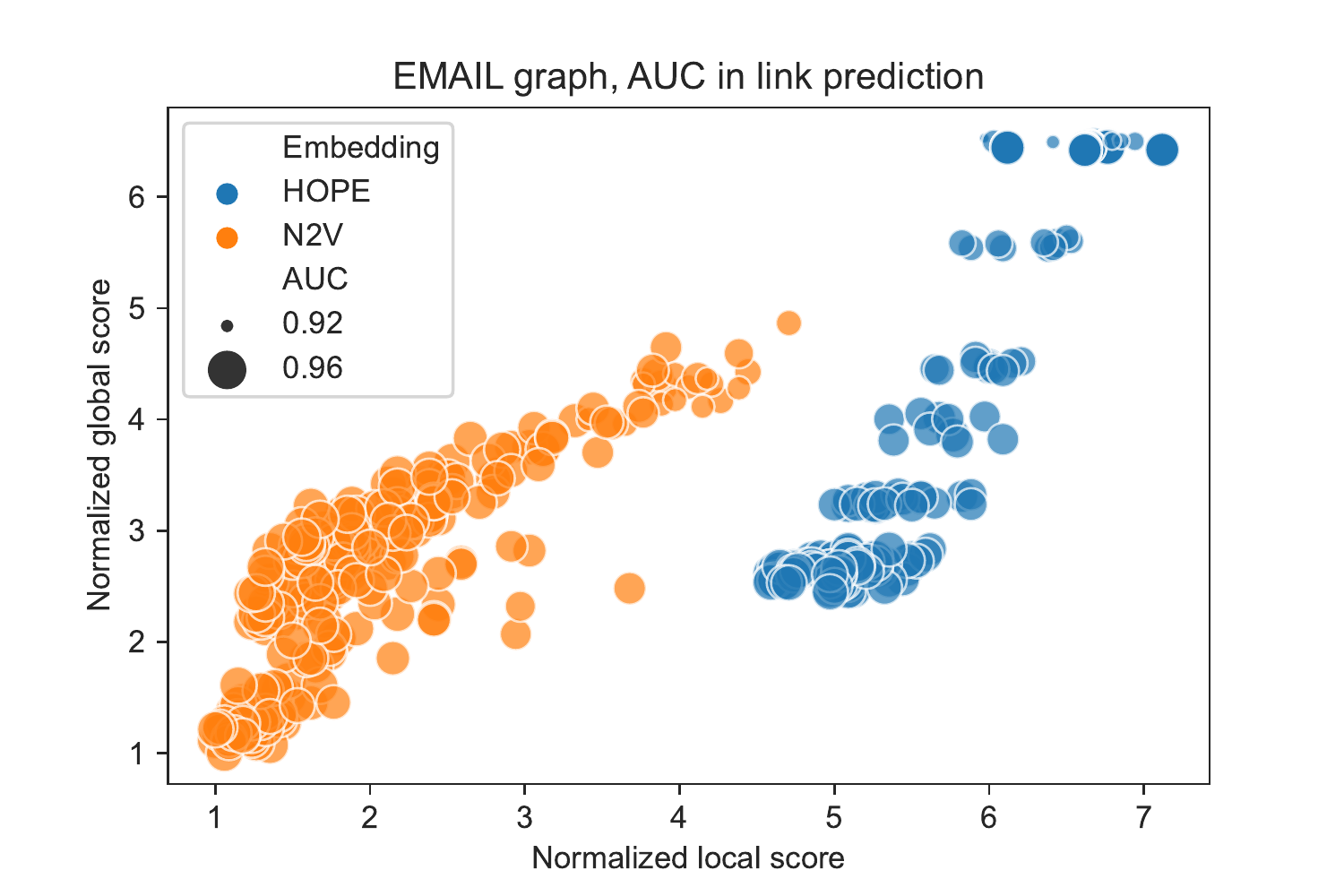}
        \hspace{.1cm}
    \caption{Global and local scores with AUC overlay for \textbf{SBM}, \textbf{LFR}, \textbf{noisy-LFR}, \textbf{EMAIL} graphs and \textbf{HOPE}, \textbf{Node2Vec} embeddings.}
    \label{fig:normalizedlinkpred}
\end{figure}

The results of this experiment are presented on Figure~\ref{fig:normalizedlinkpred}. The embeddings that score well by the framework (left-bottom corner of the plots, near to the auxiliary point $(1,1)$) turn out to perform well in predicting links (large balls representing large values of AUC) and the opposite is true for the embeddings that are identified as poor ones (right-top corner of the plots). More rigorous approach can be found in Table~\ref{tab:corr_auc_scores} where the correlation coefficients between the AUC and the global/local scores are reported. As before, since there are many \textbf{Node2Vec} embeddings that are almost indistinguishable, ranking-based correlation coefficients report weaker correlation. The framework has no chance to predict the exact ranking of the embeddings based on their global/local scores but it is clearly able to identify good embeddings and separate them from bad ones. As usual, more experiments can be found in Appendix~\ref{sec:appendix_experiments}.

\subsection{More Challenging Situations}\label{sec:experiments7}

Based on experiments in Subsection~\ref{sec:correlation_with_global_score}, we see that both global (as expected) and local (far from being obvious) scores correlate well with the quality of both node classification and community detection. Similarly, experiments in Subsection~\ref{sec:correlation_with_local_score} imply that both scores correlate with the quality of link prediction algorithm (this time, we expect this from the local score but not necessarily from the global one). We see such behaviour as the two scores often correlate with one another. Indeed, for a relatively easy graphs to deal with, a good quality embedding should be able to preserve both local and global properties of the network. On the other hand, poor quality embeddings (for example, when the dimension is too small or the parameters are not properly selected) typically fail to produce anything useful, from any of the two perspectives. Our synthetic models and \textbf{EMAIL} graph seem to confirm this. However, it is expected and plausible that large, real-world graphs are more challenging to deal with. For such graphs, it might be difficult if not impossible to find an embedding that scores well from both perspectives. In such situations, one needs to make a decision which embedding to use based on a specific application at hand (using either global or local properties). In order to illustrate such situations, we experimented with an undirected \textbf{ABCD} graph and $32$ dimensional \textbf{Node2Vec} embedding (with parameters set to $p=q=1$). 

\medskip

First, we rewired a specified fraction $p$ of edges within each community. The rationale behind it is that such operation should destroy local properties of the embedding (many edges within communities are now long with respect to the associated embedding and some pairs of close nodes are not adjacent) but global properties should remain unchanged. 

\begin{figure}[h!]
    \centering
    \includegraphics[width=0.45\textwidth]{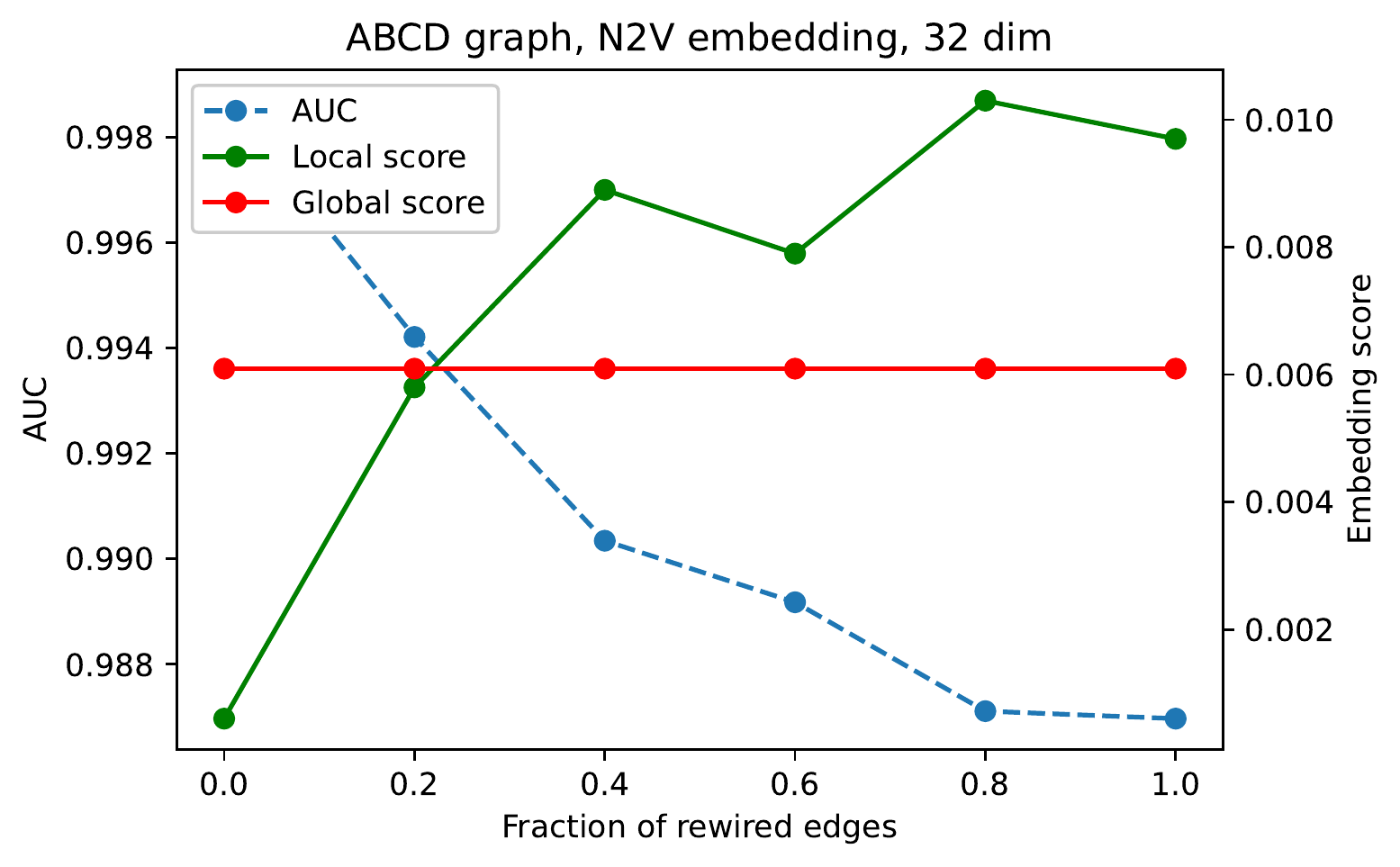}
        \hspace{.1cm}
    \includegraphics[width=0.45\textwidth]{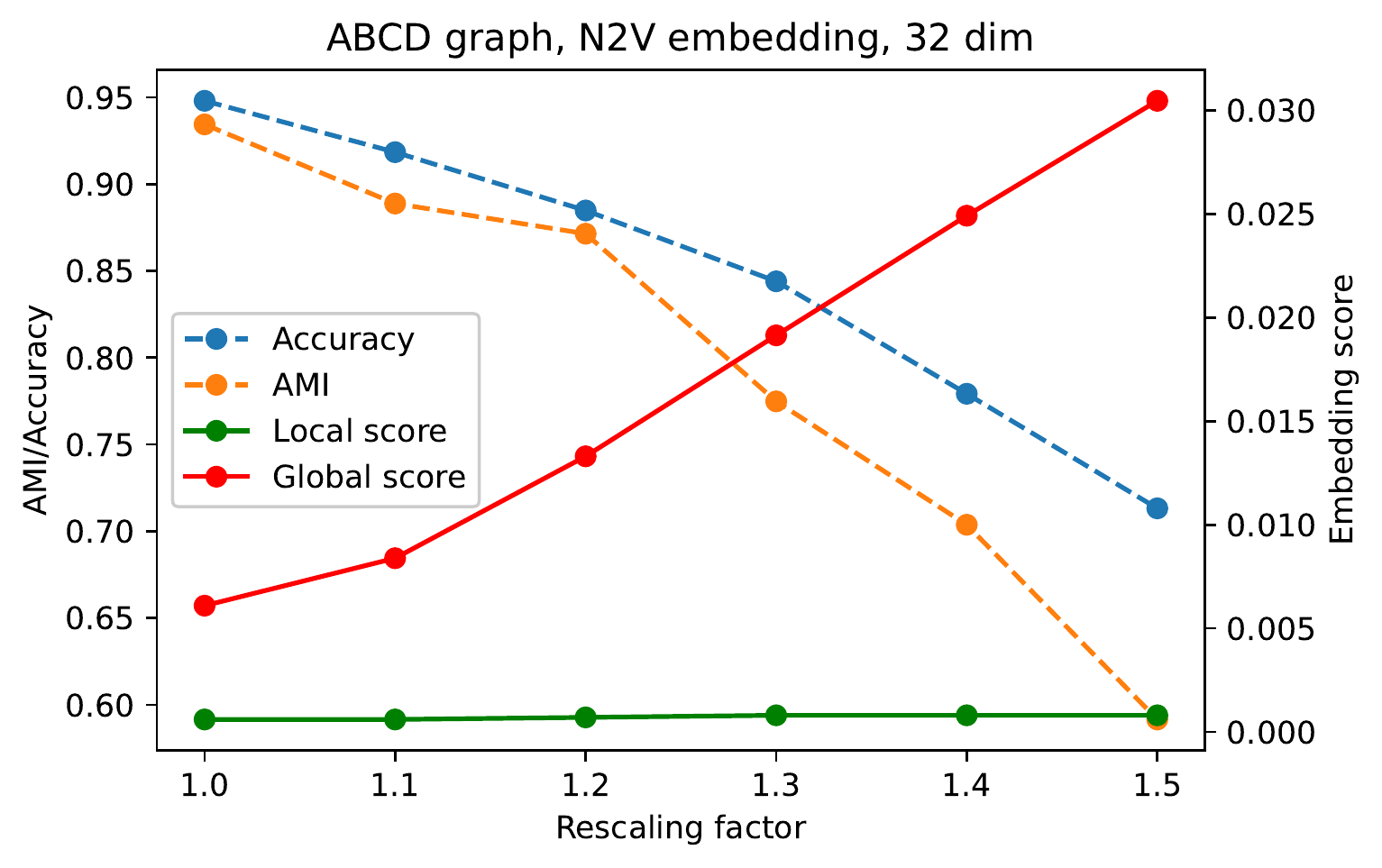}
    \vspace{.1cm}
    \caption{Comparison of local/global scores (and associated quality measures for various tasks) for rewired graphs (left) and rescaled embeddings (right).}
    \label{fig:rewiring}
\end{figure}

We independently generated 5 graphs with the following rewiring fractions: $p=0.2$, $p=0.4$, $p=0.6$, $p=0.8$, and $p=1.0$; note that $p=0$ corresponds to the original \textbf{ABCD} graph. As expected, the global score and the quality of both node classification and community detection algorithms remain the same. More importantly, the local score increases (the quality of the embedding gets worse) and the quality of link prediction algorithm (AUC) gets worse---see Figure~\ref{fig:rewiring} (left). 

\medskip

In order to test the opposite scenario, we kept the original \textbf{ABCD} graph but slightly modified the original embedding. Each community was independently rescaled by a factor of $q>1$. Formally, all points in the embedding that are associated with nodes that belong to a given community increased their distance from the center of mass of this community while keeping the direction from it. The idea behind this approach is that, on average, the distance between nodes from different communities (most pairs of nodes) remain the same, and the distance between nodes within one community is only slightly increased (small fraction of pairs of nodes anyway). As a result, local properties should be preserved almost as well as in the original embedding. On the other hand, inflated communities tend to overlap more in the embedded space than before and, as a result, the global quality measure should decrease. 

We independently generated 5 embeddings with the following re-scaling factors: $q=1.1$, $q=1.2$, $q=1.3$, $q=1.4$, and $q=1.5$; note that $q=1$ corresponds to the original embedding. As expected, the local score and the quality of link prediction algorithm remain the same. More importantly, the global score increases (the quality of the embedding gets worse) and the quality of both node classification (accuracy) and community detection (AMI) get worse---see Figure~\ref{fig:rewiring} (right). 

\section{Conclusions}\label{sec:conclusions}

In this paper, we introduced a framework that assigns two scores (global and local) for embeddings of a given graph. The two scores measure the ability of analyzed embeddings to preserve global and, respectively, local properties of the embedded graph. The code is written in Julia (high-level, high-performance, dynamic programming language) and easily available on-line. The framework is easy to use (knowledge of Julia is not required to use it; default parameters should be suitable for most cases) and flexible (more advanced users may easily play with parameters). 

If the embeddings need to be used for unsupervised tasks, the framework is able to identify the best embedding in an unsupervised way via combined divergence score. Having said that, the user might still want to consider a few embeddings that score well from both perspectives and, if possible, select an embedding generated by a simple/explainable algorithm. In particular, there is usually no need to use very high dimensional embeddings if the improvement is marginal. On the other hand, if the embedding is to be used for supervised tasks, the best course of action might be to consider a number of top-scoring embeddings and select the winner by performing a standard supervised selection for a given application at hand. 

\medskip

Let us also discuss some potential future directions. In~\cite{Arash}, some throughout analysis of the original framework for undirected graphs (and using only the global score) was performed. A~natural next step is to test in more detail the current version of the framework (using the two scores) on both real-world and synthetic networks. More importantly, it would be valuable to show the predictive power and usefulness of the framework for a large data-set and some real-world important application. The industry partners we collaborate with provide a positive feedback so far but having a publishable results of experiments performed on publicly available data-sets will be of high value. 

When analyzing the correlation between the link prediction algorithm and the local score (Subsection~\ref{sec:correlation_with_local_score}), we independently performed a few more experiments investigating the reason why embeddings are able to predict missing links. In particular, we created two additional training sets. The first training set was the same as the original one but with 5 additional columns: in- and out- degrees for both nodes as well as the distance between them. The second training set consisted with only these 5 additional columns. The quality of models obtained by using both such training sets turned out to be comparable to the original one we used for our tests. As expected, the richer set typically gave slightly better quality results but not always. This implies that the embeddings encoded the most important information about the graph via the distances between nodes (or XGBoost was not able to extract more from other sources). This suggests that the Geometric Chung-Lu (GCL) model should be able to accurately model the shape of the embedded networks as they require exactly the 5 columns as the parameters/input. We plan to use the GCL model, along with graph-based communities and other properties such as transitive closure, to build a better quality link prediction algorithm. Another potential and important application is to detect overlapping communities or to detect anomalies. In both cases, the GCL model equipped with a good embedding would be the heart of such an algorithm.

\section{Acknowledgement}

Experiments were conducted using \textbf{SOSCIP}\footnote{\url{https://www.soscip.org/}} Cloud infrastructure.  Launched in 2012, the \textbf{SOSCIP} consortium is a collaboration between Ontario’s research-intensive post-secondary institutions and small- and medium-sized enterprises (SMEs) across the province. Working together with the partners, \textbf{SOSCIP} is driving the uptake of AI and data science solutions and enabling the development of a knowledge-based and innovative economy in Ontario by supporting technical skill development and delivering high-quality outcomes. \textbf{SOSCIP} supports industrial-academic collaborative research projects through partnership-building services and access to leading-edge advanced computing platforms, fuelling innovation across every sector of Ontario’s economy.

For our experiments, we used Compute G4-x8 (8 vCPUs,	32 GB RAM) machines and Ubuntu 18.04 operating system. Computation used for experimentation and calibration of the scripts took approximately 1500 vCPU-hours. For reproducibility purpose, the scripts and results presented in this paper can be found on GitHub repository\footnote{\url{https://github.com/KrainskiL/UnsupervisedFrameworkForComparingGraphEmbeddings}}.

\appendix


\parindent 20pt

\section{Appendix---Geometric Chung-Lu Directed Graph Model is Well-defined} \label{apdx:model}

Let us state the problem in a slightly more general setting. Fix $n=2m$, where $m$ is some natural number. Suppose that for each pair $i, j \in [n]$ we have $a_{ij}=a_{ji} \in \R_+$ if $i\leq m$, $j>m$, $j\neq i+m$, and we have $a_{ij}=0$ otherwise, that is, if $i,j\in [m]$ or $i,j \in [2m] \setminus [m]$ or $j=i+m$. The $a_{ij}$ elements taken together form a matrix $\textbf{A}$, which is symmetric. Finally, for each $i \in [n]$ we have $b_i \in \R_+ \cup \{0\}$, and 
\begin{equation}\label{eq:assumption}
\sum_{i=1}^{m}b_i=\sum_{i=m+1}^{n}b_i>0.
\end{equation}

\medskip

In our application, $m=|V|$ is the number of nodes in a non-empty directed graph $G=(V,E)$, and elements of vector $\mathbf{b} = (b_i)_{i\in[n]}$ correspond to the degree distribution of the graph: for $i\in [m]$, $b_i$ is the out-degree of $v_i$ (that is, $b_i=w_i^{out}$), and for $i\in [2m] \setminus [m]$, $b_i$ is the in-degree of $v_{i-m}$ (that is, $b_i=w_{i-m}^{in}$). The assumption that $\sum_{i=1}^{m}b_i=\sum_{i=m+1}^{n}b_i>0$ is satisfied as the total in-degree is equal to the total out-degree, and the graph is not empty. Positive elements of matrix $\textbf{A}$ satisfy $a_{i,j+m}=a_{j+m,i} = g(d_{i,j}) \in (0,1]$ for $i,j\in[m], i\neq j$, and correspond to the distances between embeddings of the corresponding nodes $v_i$ and $v_j$. The case $i=j$ is excluded as indices $i$ and $i+m$ correspond to the same node in the original directed graph and so $a_{i,i+m}=a_{i+m,i} = g(d_{i,i}) = 0$. Finally, since isolated nodes may be ignored, we may assume that $w_i^{out}+w_i^{in} > 0$, that is,
\begin{equation}\label{eq:non-degenerate}
b_i + b_{i+m} > 0 \qquad \text{ for all } i \in [m].
\end{equation}

Our goal is to investigate if there is a solution, $x_i \in \R_+\cup\{0\}$ for $i \in [n]$, of the following system of equations:
\begin{equation}\label{eq:system}
b_i = x_i \sum_{j=1}^na_{ij}x_j \qquad \text{ for all } i\in[n].
\end{equation}
If there is one, then is this solution unique? The solution to~(\ref{eq:system}) will yield the solution to our problem: for $i \in [m]$, $x_i^{out} = x_i$ and $x_i^{in}=x_{i+m}$.

\medskip

The $m=2$ case is a degenerate case that exhibits a different behaviour but it is easy to investigate. In this case, by assumption~(\ref{eq:assumption}), $b_1 = b_4 = x_1 x_4 a_{12}$ and $b_2 = b_3 = x_2 x_3 a_{21}$. There are infinite number of solutions, each of them being of the form $(x_1, x_2, x_3, x_4) = (s, t, b_2/(a_{12}t), b_1/(a_{12}s))$ for some $t, s \in \R_+$. Having said that, in our application, all of these solutions yield the same random graph with the following distribution: $p_{12} = x_1 x_4 a_{12} = b_1 = w^{out}_1$ and $p_{21} = x_2 x_3 a_{21} = b_2 = w^{out}_2$.

\medskip

Suppose now that $m \ge 3$. We will show that the desired solution of~(\ref{eq:system}) exists if
\begin{equation}\label{eq:condition1}
\sum_{i=m+1}^nb_i > b_j + b_{j+m} \quad \text{for} \quad j\in[m],
\end{equation}
and
\begin{equation}\label{eq:condition2}
\sum_{i=1}^mb_i > b_j + b_{j-m} \quad \text{for} \quad j\in[2m] \setminus [m]
\end{equation}
(also recall that by assumption~(\ref{eq:assumption}), $\sum_{i=1}^{m}b_i=\sum_{i=m+1}^{n}b_i>0$, and by assumption~(\ref{eq:non-degenerate}), $b_i+b_{i+m}>0$ for all $i \in [m]$).
In other words, the condition is that the in-degree of any node $v_i$ is smaller than the sum of out-degrees of nodes other than $v_i$, and the out-degree of $v_i$ is smaller than the sum of in-degrees of nodes other than $v_i$. This is a very mild condition that holds in our application. Indeed, properties~(\ref{eq:condition1})--(\ref{eq:condition2}) with non-strict inequalities are satisfied for \emph{all} directed graphs $G$, and strict inequalities are satisfied \emph{unless} $G$ has an independent set of size $n-1$, that is, $G$ is a star with one node being part of \emph{every} edge. 

These degenerate cases can be ignored in further analysis, as such configurations of $\mathbf{b}$ allow to reconstruct the graph deterministically. The implementation of the framework identifies such cases and returns the corresponding 0/1 values of $p_{ij}$. For degenerate cases, the corresponding system of equations~(\ref{eq:system}) might or might not have a solution depending on the parameters. For example when $m=3$ and $\mathbf{b}=(2,0,0,0,1,1)$ the system has infinitely many solutions of the form $(t,0,0,s,1/(a_{51}t), 1/(a_{61}t)))$ for any $t,s>0$, all of them yielding the same deterministic graph: $p_{12}=p_{13}=1$, $p_{21}=p_{23}=p_{31}=p_{32}=0$. However, for $m=3$, $\mathbf{b}=(2,1,1,2,1,1)$ and non zero elements $a_{ij}$ equal to $1$ the system has no solutions. We do not try to classify cases when the system has the solution in the proof as they lead to deterministic graphs and we handle them in the implementation of the framework separately anyway.

\medskip

Let us make the following observations that will be useful later on. Provided that properties~(\ref{eq:condition1})--(\ref{eq:condition2}) are satisfied, the following properties hold:
\begin{itemize}
    \item $x_i=0$ if and only if $b_i=0$. Indeed, if $x_i = 0$, then trivially $b_i =0$. Suppose then that $b_i=0$ and by symmetry we may assume that $i\in[m]$. By properties~(\ref{eq:condition1})--(\ref{eq:condition2}), $b_j > 0$ for at least one value of $j \in [2m]\setminus[m]$ and $j\neq i+m$. It follows that $x_j > 0$, $a_{ij} > 0$, and so $b_i \ge x_i a_{ij} x_j$. As a result, $x_i$ has to be equal to zero in order for $b_i$ to be zero. 
    \item we may assume that $x_1=1$. Indeed, one can reorder the nodes so that $b_1>0$. Then, one can multiply $x_i$ for all $i\in[m]$ by any positive constant $\alpha \in \R_+$ and divide $x_i$ for all $i \in [2m] \setminus [m]$ by $\alpha$ and the solution will not change.
\end{itemize}

The last observation means that we need to introduce the constraint $x_1=1$ if we ever hope to prove the uniqueness of the solution. If we do not do that, then there will be an infinite number of solutions but all of them will yield the same edge distribution for the random graph as the particular solution we are searching for.

\medskip

We will start with proving the uniqueness. After that, we will show that~(\ref{eq:condition1})--(\ref{eq:condition2}) are sufficient conditions.

\subsection{Uniqueness}

Let us assume that $m \ge 3$. For a contradiction, suppose that we have two different solutions: $\mathbf{x} = (x_i)_{i\in[n]}$ ($x_i \in \R_+$, $i \in [n]$) with $x_1=1$ and $\mathbf{y} = (y_i)_{i \in [n]}$ ($y_i \in \R_+$, $i \in [n]$) with $y_1=1$. It follows that for all $i \in [n]$ we have
$$
b_i = f_i(\mathbf{x}) = f_i(\mathbf{y}), \quad \text{ where } f_i(\mathbf{x}) = x_i \sum_{j=1}^n a_{ij} x_j.
$$
Let us analyze what happens at point $\mathbf{z} = t\mathbf{x} + (1-t)\mathbf{y}$ for some $t \in [0,1]$ (that is, $z_i = tx_i+(1-t)y_i$, $i \in [n]$). For each $i \in [n]$ we get
\begin{eqnarray*}
f_i(\mathbf{z}) &=& (tx_i+(1-t)y_i) \sum_{j=1}^n a_{ij} (tx_j+(1-t)y_j) \\
&=& \sum_{j=1}^n a_{ij} \left( t^2 x_i x_j+ t(1-t)(x_i y_j + x_j y_i) + (1-t)^2 y_i y_j \right) \\
&=& f_i(\textbf{x}) t^2 + \frac {t(1-t)}{x_i y_i} (f_i(\textbf{y}) x_i^2 + f_i(\textbf{x}) y_i^2) + f_i(\textbf{y}) (1-t)^2 \\
&=& b_i \left( t^2 + \frac {t(1-t)}{x_i y_i} (x_i^2 + y_i^2) + (1-t)^2 \right) \\
&=& b_i \left( 1 - 2t(1-t) + \frac {t(1-t)}{x_i y_i} (x_i^2 + y_i^2) \right) \\
&=& b_i \left( 1 + \frac {t(1-t)}{x_i y_i} (x_i^2 - 2x_iy_i + y_i^2) \right) \\
&=& b_i \left( 1+t(1-t)\frac{(x_i-y_i)^2}{x_iy_i} \right) =: g_i(t).
\end{eqnarray*}
Note that $g_i'(1/2)=0$ for all $i$ (as either $x_i-y_i$ vanishes and so $g_i(t)$ is a constant function or it does not vanish but then $g_i(t)$ is a parabola with a maximum at $t=1/2$). For convenience, let $\mathbf{v} = (\mathbf{x} + \mathbf{y})/2$ and $\mathbf{s} = (\mathbf{x} - \mathbf{y})/2$ (that is, $v_i=(x_i+y_i)/2$ and $s_i=(x_i-y_i)/2$ for all $i \in [n]$). It follows that 
$$
\frac{d g_i}{dh} \Big( \mathbf{v}+h\mathbf{s} ~|~ h=0 \Big)=0.
$$
On the other hand,
$$
g_i(\mathbf{v}+h\mathbf{s}) = \sum_{j=1}^na_{ij}(v_i+hs_i)(v_j+hs_j)
$$
and so
$$
\frac{d g_i}{dh}(\mathbf{v}+h\mathbf{s}) = \sum_{j=1}^na_{ij}(s_i(v_j+hs_j)+s_j(v_i+hs_i)).
$$
Combining the two observations, we get that
\begin{equation}\label{eq:condition_for_s}
0=\frac{d g_i}{dh} \Big( \mathbf{v}+h\mathbf{s} ~|~ h=0 \Big) = \sum_{j=1}^na_{ij}(s_iv_j+s_jv_i).
\end{equation}

Now, for $i \ge 1$, let $u_i = s_i / v_i$, provided that $v_i \neq 0$.
Recall that in particular $s_1=(x_1-y_1)/2=0$ and $v_1=(x_1+y_1)/2=1$, as we assumed that $x_1=y_1=1$, and so $u_1=0$.
Additionally, if $v_i=0$ (that is, the corresponding node has in-degree 0 or out-degree 0), then we may take $u_i=0$, as it will cancel out anyway. Denote the set of indices $i$ when $v_i=0$ by $Z$. Substituting it to~(\ref{eq:condition_for_s}) we get:
$$
\sum_{j=1}^na_{ij}v_iv_j(u_i+u_j) = 0.
$$

Notice that we can rescale all $u_i$'s by the same multiplicative factor so that $u_{\ell}=1$ for some $\ell \in [n]$ and for all other indices $|u_i|\leq1$ (potentially with negative rescaling factor). For index $\ell$ we have:
$$
\sum_{j=1}^n a_{\ell j} v_{\ell}v_j(u_{\ell}+u_j) = \sum_{j=1}^n a_{\ell j} v_{\ell}v_j(1+u_j) = 0.
$$
But this possible only if $\ell\in[m]$, as otherwise the left hand side of the above equation is at least its first term, namely, $a_{\ell 1}v_{\ell} v_1(1+u_1) =a_{\ell 1}v_{\ell} > 0$. If $\ell \in[m]$, then we get
$$
\sum_{j=m+1}^n a_{\ell j}v_{\ell} v_j (u_{\ell}+u_j) = \sum_{j=m+1}^n a_{\ell j}v_{\ell} v_j (1+u_j) = 0
$$
as $a_{\ell j}=0$ for $j\in[m]$. But this means that $u_j=-1$ for $j\in[n]\setminus([m]\cup Z)$.

Let us concentrate on any index $\ell'\in [n]\setminus([m]\cup Z)$ (note that this set is non-empty).
For this index, we have the following condition:
$$
\sum_{j=1}^m a_{\ell' j}v_{\ell'}v_j(u_{\ell'}+u_j) = \sum_{j=1}^m a_{\ell' j}v_{\ell'}v_j(-1+u_j) = 0.
$$
However, since $|u_j|\leq 1$ for all $j \in[m]$, all entries of the sum are non-negative and so they would all have to be equal to $0$ for the sum to be $0$. This is not possible as $u_1=0$ and so $a_{\ell'1}v_{\ell'}v_1(-1+u_1) =-a_{\ell'1}v_{\ell'}<0$. The desired contradiction shows that the solution is unique.





\subsection{Sufficiency}

We will continue assuming that $m \geq 3$. For a contradiction, suppose that there exists a vector $\mathbf{b} = (b_i)_{i \in [n]}$, that satisfies~(\ref{eq:condition1})--(\ref{eq:condition2}), and $\sum_{i=1}^{m}b_i=\sum_{i=m+1}^{n}b_i>0$ (assumption~(\ref{eq:assumption})) but for which there is no solution to the system~(\ref{eq:system}). Without loss of generality, since one can reorder nodes and relabel in- and out-degrees if needed, we may assume that $b_1$ is a largest value in vector $\mathbf{b}$. We will call such vectors \emph{infeasible}. On the other hand, vectors that yield a solution $\mathbf{x} = (x_i)_{i \in [n]}$, with $x_i\geq0$ for all $i$, will be called \emph{feasible}. As proved earlier, if the solution exists, then it must be unique (remember that we assume that $x_1=1$). We will introduce more vectors $\textbf{b}$ (both feasible and infeasible) below but we assume that matrix $\textbf{A}$ is fixed.

Let us now construct another vector $\mathbf{b}' = (b'_i)_{i \in [n]}$ for which there exists a solution to~(\ref{eq:system}) (that is, $\mathbf{b}'$ is feasible) but also $b'_1=b_1$ is a largest element in $\mathbf{b}'$. Indeed, it can be done easily by, for example, taking $x'_1=1$, $x'_i = s$ for $i\in[m]\setminus\{1\}$ ($s$ is a fixed but sufficiently small positive constant for the inequalities below to hold), and $x'_i = b_1 / (\sum_{j \in [n]} a_{1j}) = b_1 / (\sum_{j \in [n] \setminus [m]} a_{1j})$ for $i\in[n]\setminus[m]$. Vector $\mathbf{b}'$ is now defined by the system~(\ref{eq:system}). We immediately get that
$$
b'_1 = x'_1 \sum_{j \in [n]} a_{1j} x'_j = \sum_{j \in [n] \setminus [m]} a_{1j} x'_j = b_1.
$$
Also, $s$ can be made arbitrarily small so that $b'_i<b_1$ for $i\in[m]\setminus\{1\}$.
Finally, for $i \in[n]\setminus[m]$ we have
$$
b'_i = x'_i \sum_{j \in [m]} a_{ij} x'_j = x'_i a_{i1} + x'_i \sum_{j=2}^m a_{ij} s = b_1 \frac{a_{i1}} {\sum_{j \in [n]} a_{1j}} + x'_i \sum_{j=2}^m a_{ij} s.
$$
Since $m \ge 3$, the first term is smaller than $b_1$. Hence, since $s$ can be made arbitrarily small, we can ensure that $b'_i < b_1$. The desired properties hold.

We will consider points along the line segment between $\mathbf{b}'$ and $\mathbf{b}$, namely,
$$
\mathbf{b}(t) = (b_i(t))_{i \in [n]} = (1-t)\mathbf{b}'+t\mathbf{b}, \qquad \text{ for } t\in[0,1].
$$
Since $\mathbf{b}'$ is feasible and we already proved that~(\ref{eq:condition1})--(\ref{eq:condition2}) are necessary conditions, we know that $\mathbf{b}'$ satisfies~(\ref{eq:condition1})--(\ref{eq:condition2}). But, as a result, not only $\mathbf{b}$ and $\mathbf{b}'$ satisfy these properties but also $\mathbf{b}(t)$ satisfies them for any $t \in [0,1]$. In particular, it follows that there exists a universal constant $\eps > 0$ such that for any $t \in [0,1]$ we have 
\begin{equation}\label{eq:separation}
(1-\eps)\sum_{i=m+2}^nb_i(t) > b_1(t).
\end{equation}

Fix $t \in [0,1)$ and suppose that $\mathbf{b}(t)$ is feasible. Let $\mathbf{x}(t) = (x_i(t))_{i\in [n]}$ be the (unique) solution for $\mathbf{b}(t)$. From the analysis performed in the proof of uniqueness of the solution it follows that our transformation is a local diffeomorphism, that is, the differential of the transformation is bijective for the admissible values of $x_i$ and $b_i$. (Note that this also covers the case $t=0$. This case is on the boundary of the considered range of $t$ but it is an interior point of the domain of the mapping.) In the following considerations, we assume that point $x_1$ and $b_1$ are removed from the analysis (as they are fixed) and also that the indices from the set $Z$ (that is, as defined above, the set of indices $i$ for which $v_i=0$) are excluded as they are fixed. As a result we may move to a manifold of a dimension $n'<n$ by dropping the dimensions that are fixed. In the considered manifold, any open set in $\R^{n'}$ containing (part of) $\mathbf{x}(t)$ is mapped to an open set in $\R^{n'}$ containing (part of) $\mathbf{b}(t)$. In particular, there exists $\delta > 0$ such that $\mathbf{b}(s)$ is feasible for any $t - \delta \le s \le t + \delta$. Combining this observation with the fact that $\mathbf{b}'=\mathbf{b}(0)$ is feasible, $\mathbf{b}=\mathbf{b}(1)$ is \emph{not} feasible we get that there exists $T \in (0,1]$ such that $\mathbf{b}(T)$ is not feasible but $\mathbf{b}(t)$ is feasible for any $t \in [0,T)$. Indeed, if no such $T$ exists (that is, there is no minimal infeasible $t \in (0,1]$), then there would exist a decreasing sequence of infeasible values of $t$ that converges to a feasible $t$. This is not possible as in some neighbourhood of a feasible point $t$, points are also feasible.

Consider any sequence $(t_i)_{i \in \N}$ of real numbers $t_i \in [0,T)$ such that $t_i\to T$ as $i \to \infty$; for example, $t_i = T(1-1/i)$. All limits from now on will be for $i \to \infty$. Recall that $\mathbf{b}(t_i)$ is feasible and so $\mathbf{x}(t_i)$ is well-defined.

Before we move forward, let us show that there exists a sufficiently large but universal constant $\Delta$ such that for all $t \in [0,T)$ and all $i$ (except possibly $i=m+1$), we have $x_i(t) \le \Delta$. Indeed, by our assumption on the solution, $x_1(t)=1\le \Delta$. By the equation~(\ref{eq:system}) for $b_1(t) = b_1$, we have for $i\in[n]\setminus[m+1]$
$$
x_i(t) = \frac {1}{a_{1i}} \cdot x_1(t) a_{1i} x_i(t) < \frac {1}{a_{1i}} \cdot x_1(t) \sum_{j=1}^n a_{1j} x_j(t) = \frac {b_1(t)}{a_{1i}} = \frac {b_1}{a_{1i}} \le \Delta.
$$
But this immediately means that $x_i(t)$ are also bounded for $i\in[m]$ by considering any equation for $b_i(t) \le b_1(t) =b_1$ where $i\in[n]\setminus([m]\cup Z)$. This implies that only $x_{m+1}(t)$ can potentially be unbounded.

If $x_{m+1}(t)$ is bounded for all $t \in [0,T)$, then by the Bolzano-Weierstrass theorem the sequence $t_i$ has a subsequence $(\mathbf{x}(t_{s_i}))_{i \in [n]}$ such that $\Vert \mathbf{x}(t_{s_i}) \Vert \to c$ for some $c \in \R$. However, if this is the case then, by continuity of our transformation, the limiting value $\mathbf{b}(T)$ would be feasible, giving us the desired contradiction. It remains to consider the case when $x_{m+1}(t_{s_i}) \to \infty$ for some subsequence $s_i$.
However, this implies that $x_{j}(t_{s_i})\to 0$ for all $j\in[m]\setminus\{1\}$. This means that in the limit we have $x_i(T)=b_i(T)/a_{i1}$ for $i\in[n]\setminus[m+1]$. Substituting it into the first equation we get 
$$
b_1(T) = x_1(T)\sum_{i=m+2}^na_{1i}x_i(T) = 1\cdot\sum_{i=m+2}^na_{1i}\cdot b_i(T)/a_{i1} = \sum_{i=m+2}^nb_i(T).
$$
This contradicts (\ref{eq:separation}), which concludes the proof.

\bigskip

As a final note, let us observe that the proof implies that if $b_1$ gets close to $\sum_{i=m+2}^nb_i$ (from below) and $b_{m+1}>0$, then indeed $x_{m+1}$ will grow to be a large number. This consideration has a numerical impact as it might affect the convergence of numeric algorithms finding $x_i$ due to floating point computation precision issues. The proof also shows that the case when the conditions~(\ref{eq:condition1})--(\ref{eq:condition2}) are not satisfied (that is, we would have an equality instead of inequality) will have a solution if $b_{m+1}=0$ and otherwise will not have a solution (this corresponds to the two examples we have given earlier).

\subsection{Model with Loops}

In order to accommodate loops that are present in the graph, we relax the assumption that $a_{i,i+m}=0$ for $i\in[m]$ and now assume that $a_{i,i+m} \ge 0$ for $i\in[m]$. However, using the notation from the previous section we will additionally assume that $a_{1, m+1}>0$, that is, for the largest element of $\mathbf{b}$, we assume that the corresponding node has a loop. This auxiliary assumption is satisfied in our application as, in fact, all landmarks have loops.

The $m=2$ case continues to be a degenerate case that has to be delt with independently. Consider the following set of equations:
$$
\left[\begin{matrix}
b_1 \\ b_2 \\ b_3 \\ b_4
\end{matrix}\right] =
\left[\begin{matrix}
a_{13} & a_{14} & 0 & 0 \\
0 & 0 & a_{23} & a_{24} \\
a_{13} & 0 & a_{23} & 0 \\
0 & a_{14} & 0 & a_{24}
\end{matrix}\right]
\left[\begin{matrix}
x_1x_3 \\ x_1x_4 \\ x_2x_3 \\ x_2x_4
\end{matrix}\right].
$$
As before, we assume that $x_1=1$ and, since $b_1+b_2=b_3+b_4$ the system reduces to: 
$$
\left[\begin{matrix}
b_1 \\ b_2 \\ b_3
\end{matrix}\right] =
\left[\begin{matrix}
a_{13} & a_{14} & 0 & 0 \\
0 & 0 & a_{23} & a_{24} \\
a_{13} & 0 & a_{23} & 0
\end{matrix}\right]
\left[\begin{matrix}
x_3 \\ x_4 \\ x_2x_3 \\ x_2x_4
\end{matrix}\right] =
\left[\begin{matrix}
a_{14} & 0 & a_{13} & 0 \\
0 & a_{24} & 0 & a_{23} \\
0 & 0 & a_{13} & a_{23} 
\end{matrix}\right]
\left[\begin{matrix}
x_4  \\ x_2x_4 \\ x_3 \\ x_2x_3
\end{matrix}\right]
.
$$
Equivalently, since $b_1$ is the largest element of $\mathbf{b}$, for some non-negative $p_i$ and positive $q=a_{23}$: 
$$
\left[\begin{matrix}
p_1 \\ p_2 \\ p_3
\end{matrix}\right] =
\left[\begin{matrix}
1 & 0 & 0 & -q \\
0 & 1 & 0 & q \\
0 & 0 & 1 & q
\end{matrix}\right]
\left[\begin{matrix}
x_4  \\ x_2x_4 \\ x_3 \\ x_2x_3
\end{matrix}\right].
$$
If $x_2=0$, we get a unique positive solution for $x_3$ and $x_4$ and it exists only if $p_2=0$ (which happens for $b_2=0$).
On the other hand, if $x_2>0$ (recall that, by assumption, $x_i \ge 0$), then in the above system of equations one can reduce $x_3$ and $x_4$ leaving only $x_2$ as a variable in the quadratic equation:
$$
Q(x_2)=q(p_1+p_3)x_2^2+(-p_2q+p_3q+p_1)x_2-p_2 = 0.
$$
Since $x_2 > 0$, we get that $p_2 =x_2(x_4+qx_3) > 0$. (Note that if $x_3=x_4=0$, then $b_3=b_4=0$ and, as a consequence, $b_1=b_2=0$ which gives as a contradiction as $b_1>0$.) As the term $(p_1+p_3)q$ is positive and $-p_2$ is negative we get that there exists exactly one positive solution $x_2$ of this equation. Indeed, since $Q(0) = -p_2 <0$ and the parabola $Q(x_2)$ has the coefficient $(p_1+p_3)q>0$ associated with the quadratic term, there is exactly one positive solution $x_2 > 0$. Now, assuming that $x_2>0$, from the third equation we see that $x_3>0$ and from the first equation we get that $x_4>0$. 

In summary, for $m=2$, subject to the constraint $x_1=1$, the solution of the system always exists and is unique.
If $m\geq3$ we also show that the solution exists always. The part of the proof of uniqueness remains unchanged. The part for sufficiency also remains unchanged until we reach the case where we consider $x_{m+1}(t_{s_i}) \to\infty$. However as $a_{1,m+1}>0$ this is not possible since $x_1=1$ and $b_1$ is fixed. So we are left with the cases that $\Vert \mathbf{x}(t_{s_i}) \Vert \to c$ for some $c \in \R$ which means that $\mathbf{b}(t)$ always converges to a feasible solution as $t\to1$ (even if in conditions~(\ref{eq:condition1})--(\ref{eq:condition2}) we have an equality).

\section{Appendix---Scalable Implementation with Landmarks}\label{apdx:landmarks}

Recall that in \textbf{Step~1} of the algorithm, we obtain a partition $\textbf{C}$ of the set of nodes $V$ into $\ell$ communities: $C_1, \ldots, C_\ell$. The partition is then carefully refined by repeatedly splitting some parts of it with the goal to reach precisely $n'=4 \sqrt{n}$ parts; $n'$ might be adjusted by more experienced user, if needed. (The number of communities is typically relatively small. However, if $\ell \ge 4 \sqrt{n}$, then of course there is no need to do the refinement. However, in such rare cases each part in the initial partition is forced to be split into $s$ parts anyway. We fixed $s=4$ as a default value.) The heuristic algorithm is quite involved as it needs to find a good compromise between the quality of the approximation and the speed. The reader is directed to~\cite{Embedding_Complex_Networks_Scalable} for more details on how the refinement is obtained. 

Once the partition is refined, each part $C_i$ is replaced by its landmark $u_i$. The procedure depends on whether we deal with undirected or directed graphs. Let us start with undirected graphs. The position of landmark $u_i$ in the embedded space $\R^k$ is assigned as follows:
\begin{equation}\label{eq:position_landmark}
\emb(u_i) = \frac {\sum_{j \in C_i} w_j \ \emb(v_j)}{\sum_{j \in C_i} w_j}.
\end{equation}
In order to measure a variation within a cluster, we also compute the weighted sum of squared errors:
\begin{equation}\label{eq:error_landmark}
e_i = \sum_{j \in C_i} w_j \ \dist \big( \emb(u_i), \emb(v_j) \big)^2.
\end{equation}
The expected degree of landmark $u_i$ (that we denote as $w'_i$ in order to distinguish it from $w_i$, the expected degree of node $w_i$) is the sum of the expected degrees of the associated nodes in the original model, that is, $w'_i := \sum_{j \in C_i} w_j$.

The approximated algorithm uses the auxiliary \emph{Geometric Chung-Lu} (GCL) model on the set of landmarks $V = \{ u_1, \ldots, u_{n'} \}$ in which each pair of landmarks $u_i, u_j$, independently of other pairs, forms an edge with probability $p'_{i,j}$, where
\begin{equation*}
p'_{i,j} = x_i' x'_j g(d_{i,j}) 
\end{equation*}
for some carefully tuned weights $x'_i \in \R_+$. Additionally, for $i\in[n']$, the probability of creating a self loop around landmark $u_i$ is equal to 
$$
p'_{i,i} = (x'_i)^2 g(d_{i,i}), \qquad \text{ where } \qquad d_{i,i} = \sqrt{ \frac{e_i}{\sum_{j \in C_i} w_j} }.
$$
Note that the ``distance'' $d_{i,i}$ from landmark $u_i$ to itself is an approximation of the unobserved weighted average distance $d_{a,b}$ over all pairs of nodes $a$ and $b$ associated with $u_i$.
The weights are selected such that the expected degree of landmark $u_i$ is $w'_i$; that is, for all $i \in [n']$
$$
w'_i = \sum_{j \in [n']} p'_{i,j} =  x'_i \sum_{j \in [n']} x'_j g(d_{i,j}).
$$
The relationship between the weights in the auxiliary model and the original one is expected to be as follows: for any node $v_k \in C_i$ associated with landmark $u_i$ we have
$$
x_k \approx x'_i \ \frac {w_k}{\sum_{j \in C_i} w_j}.
$$

The adjustment for directed graph is straightforward. We use the same algorithm and positions for landmarks, that is, we still use~(\ref{eq:position_landmark}) and~(\ref{eq:error_landmark}) but with $w_j$ being the total degree of landmark $u_j$, namely, $w_j = w_j^{in} + w_j^{out}$. The probability for an ordered pair of landmarks $u_i, u_j$ to form a directed edge in the auxiliary \emph{Geometric Chung-Lu Directed Graph Model} is equal to 
$$
p'_{i,j} = {x'}_i^{out} {x'}_j^{in} g(d_{i,j}) 
$$
and
$$
p'_{i,i} = {x'}_i^{out} {x'}_i^{in} g(d_{i,i}), \qquad \text{ where } \qquad d_{i,i} = \sqrt{ \frac{e_i}{\sum_{j \in C_i} w_j} }.
$$
As before, any node $v_k \in C_i$ associated with landmark $u_i$ inherits a fraction of its weights, that is, we expect that the original weights are well approximated by the following:
$$
x_k^{out} \approx {x'}_i^{out} \ \frac {w_k^{out}}{\sum_{j \in C_i} w_j^{out}} \qquad  \text{and} \qquad x_k^{in} \approx {x'}_i^{in} \ \frac {w_k^{in}}{\sum_{j \in C_i} w_j^{in}}.
$$


\section{Additional Experiments}\label{sec:appendix_experiments}

\begin{figure}[ht]
     \centering
     \includegraphics[width=0.35\textwidth]{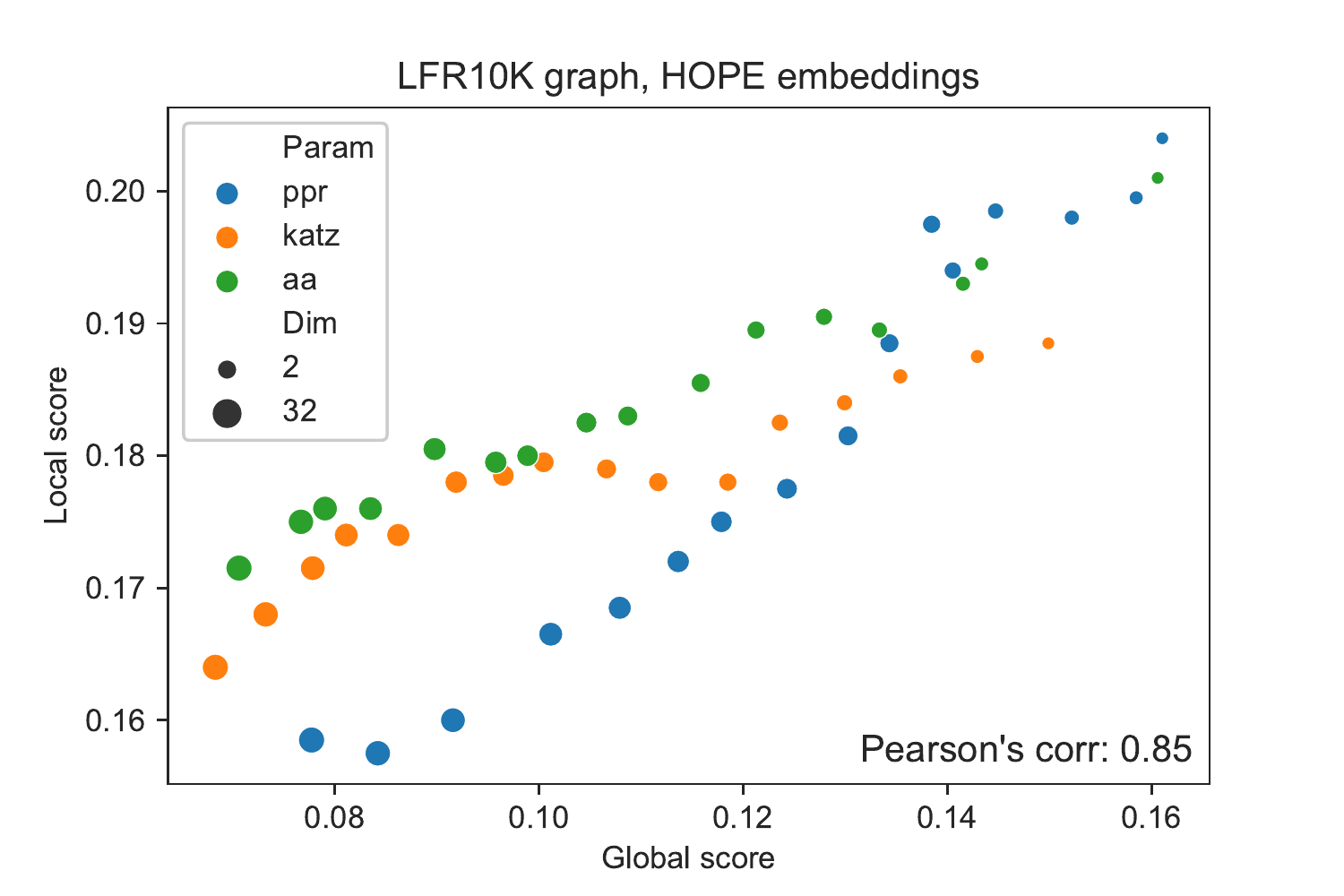}
         \hspace{.1cm}
     \includegraphics[width=0.35\textwidth]{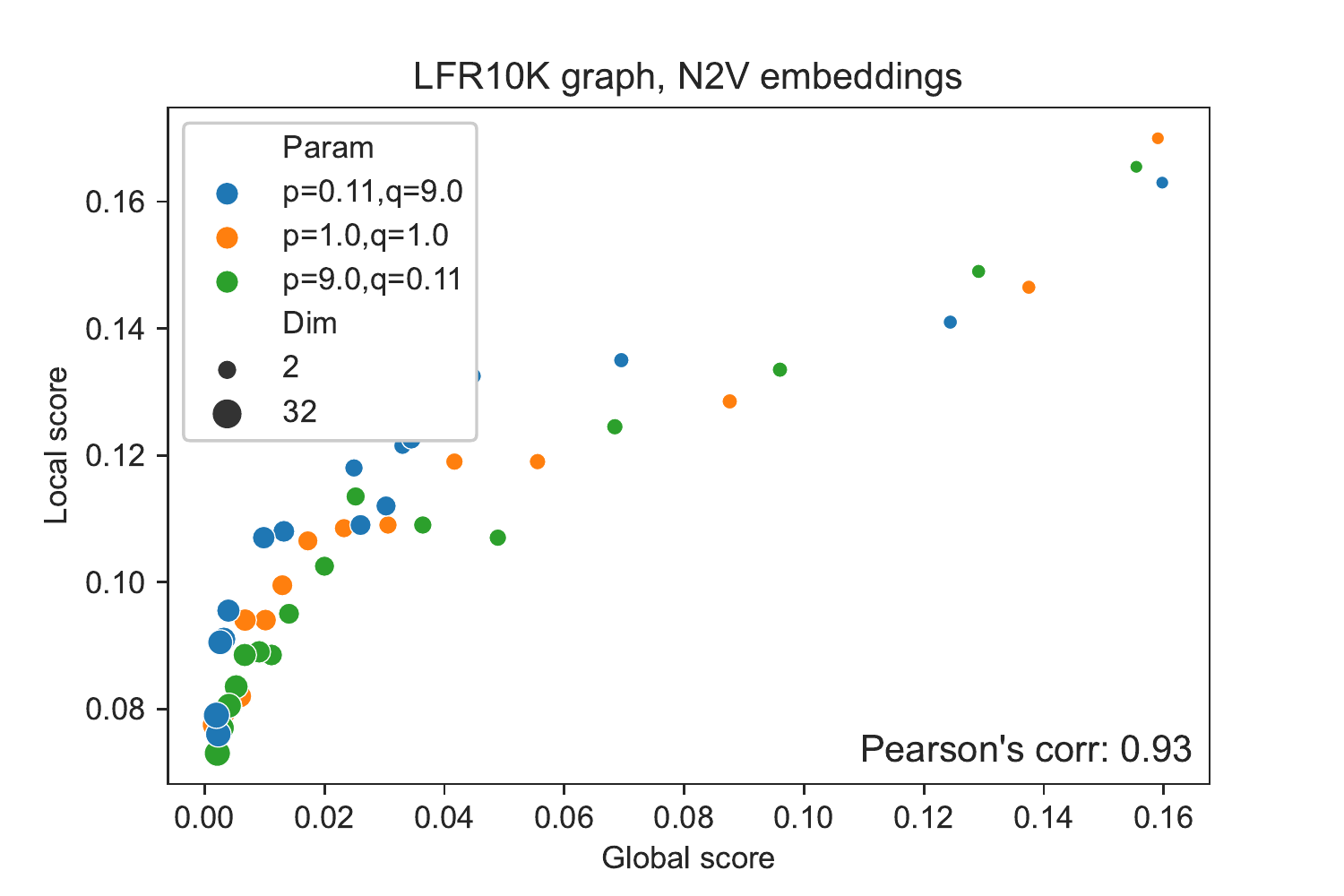}
     \vspace{.1cm}
     \includegraphics[width=0.35\textwidth]{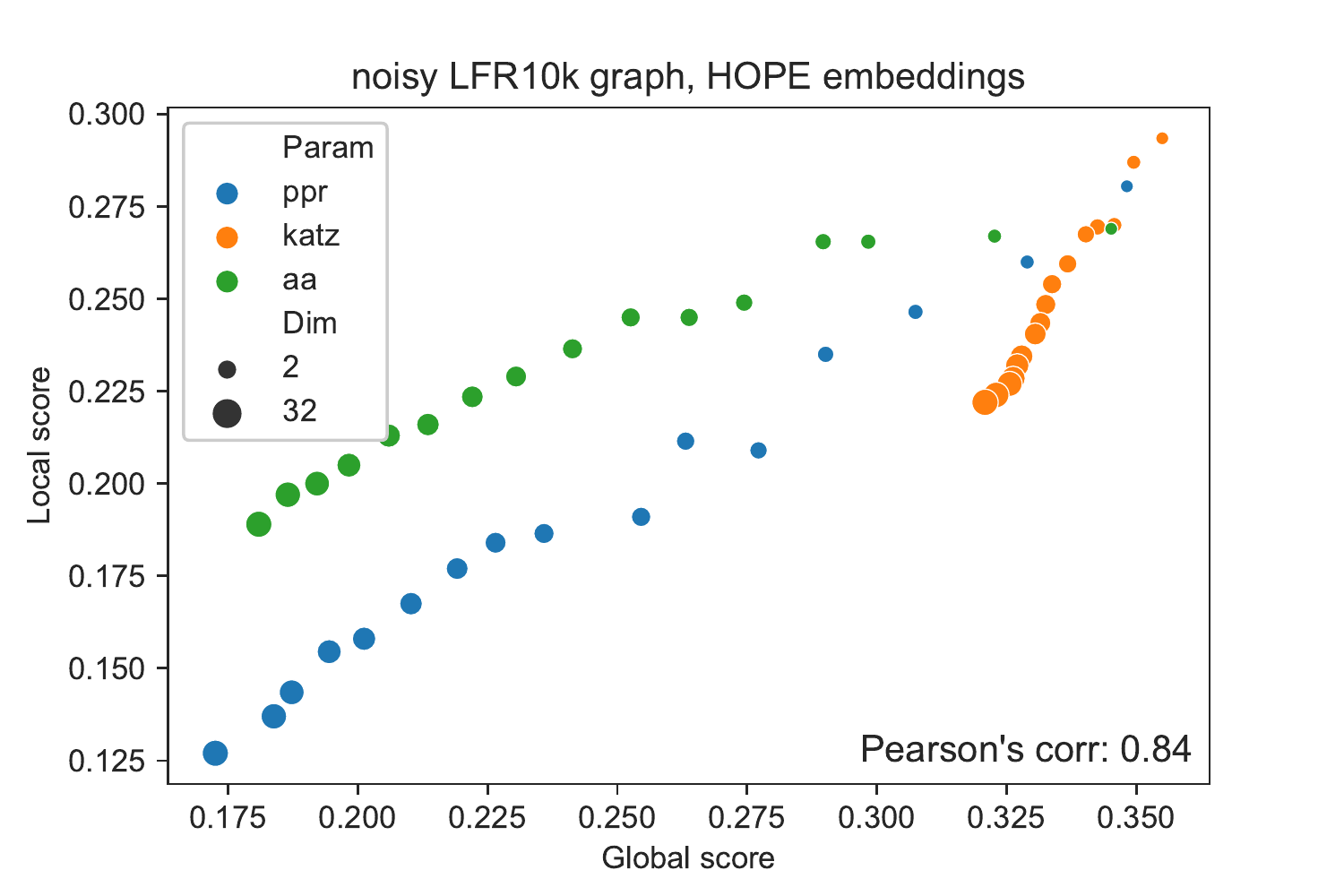}
         \hspace{.1cm}
     \includegraphics[width=0.35\textwidth]{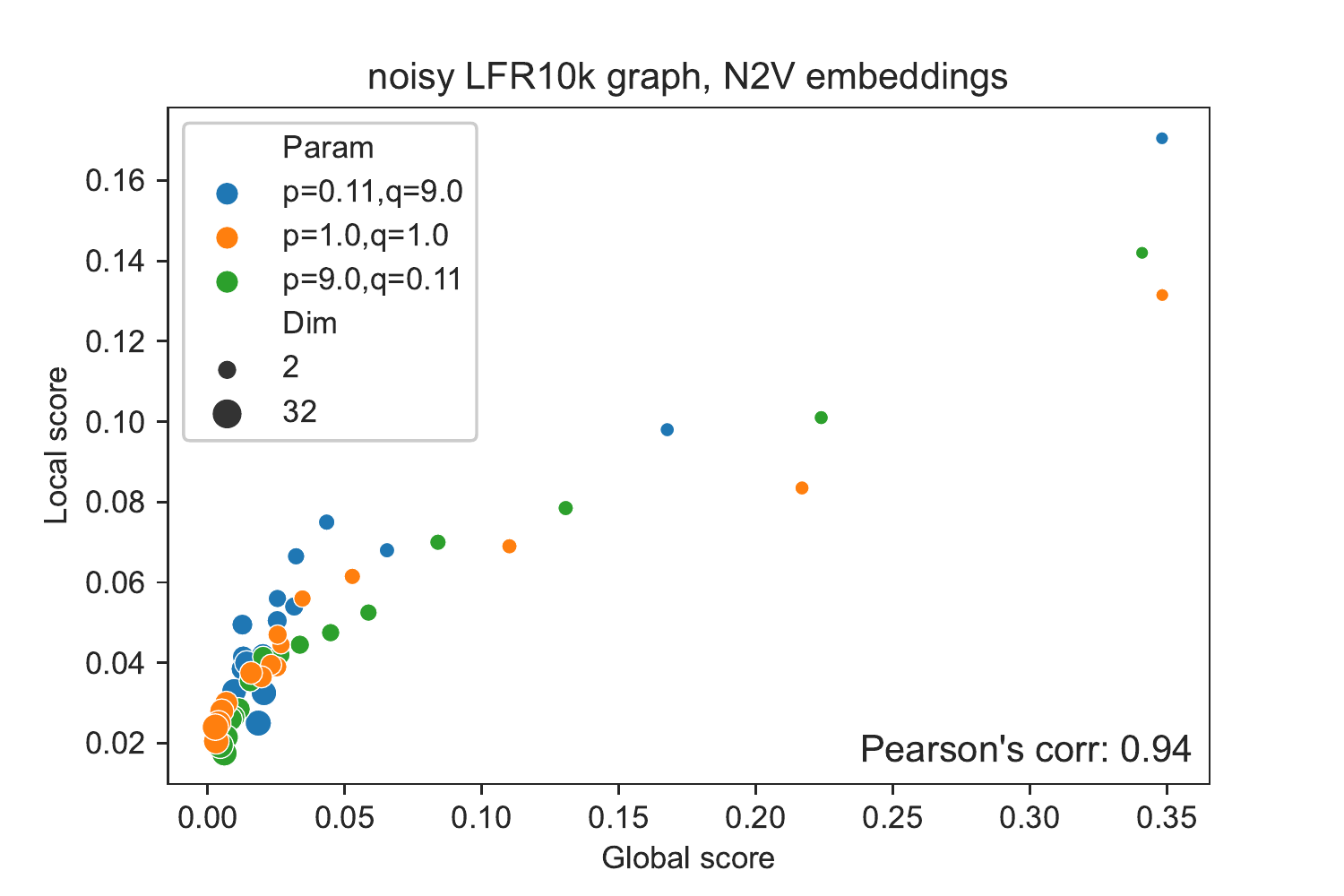}
     \vspace{.1cm}
     \includegraphics[width=0.35\textwidth]{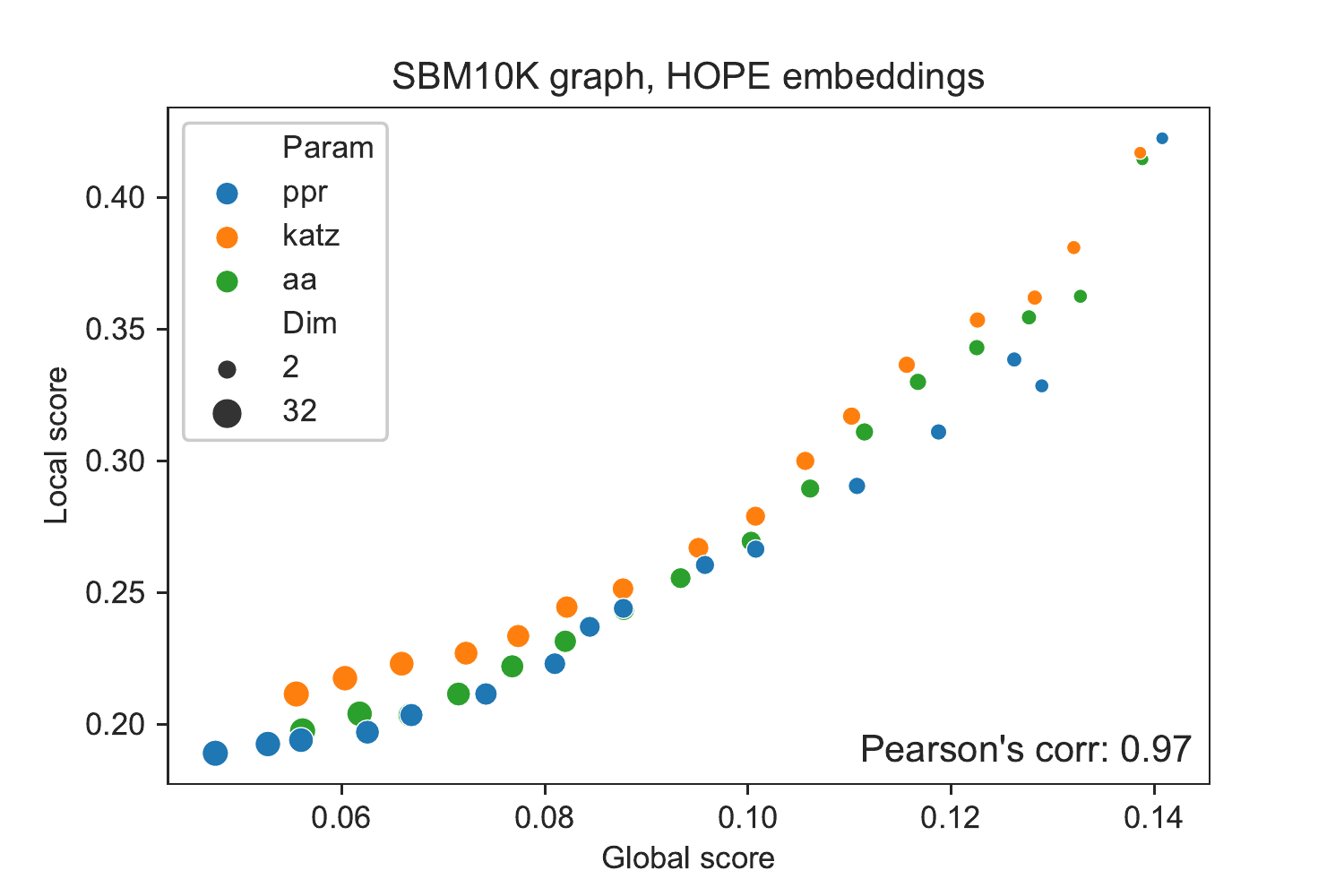}
         \hspace{.1cm}
     \includegraphics[width=0.35\textwidth]{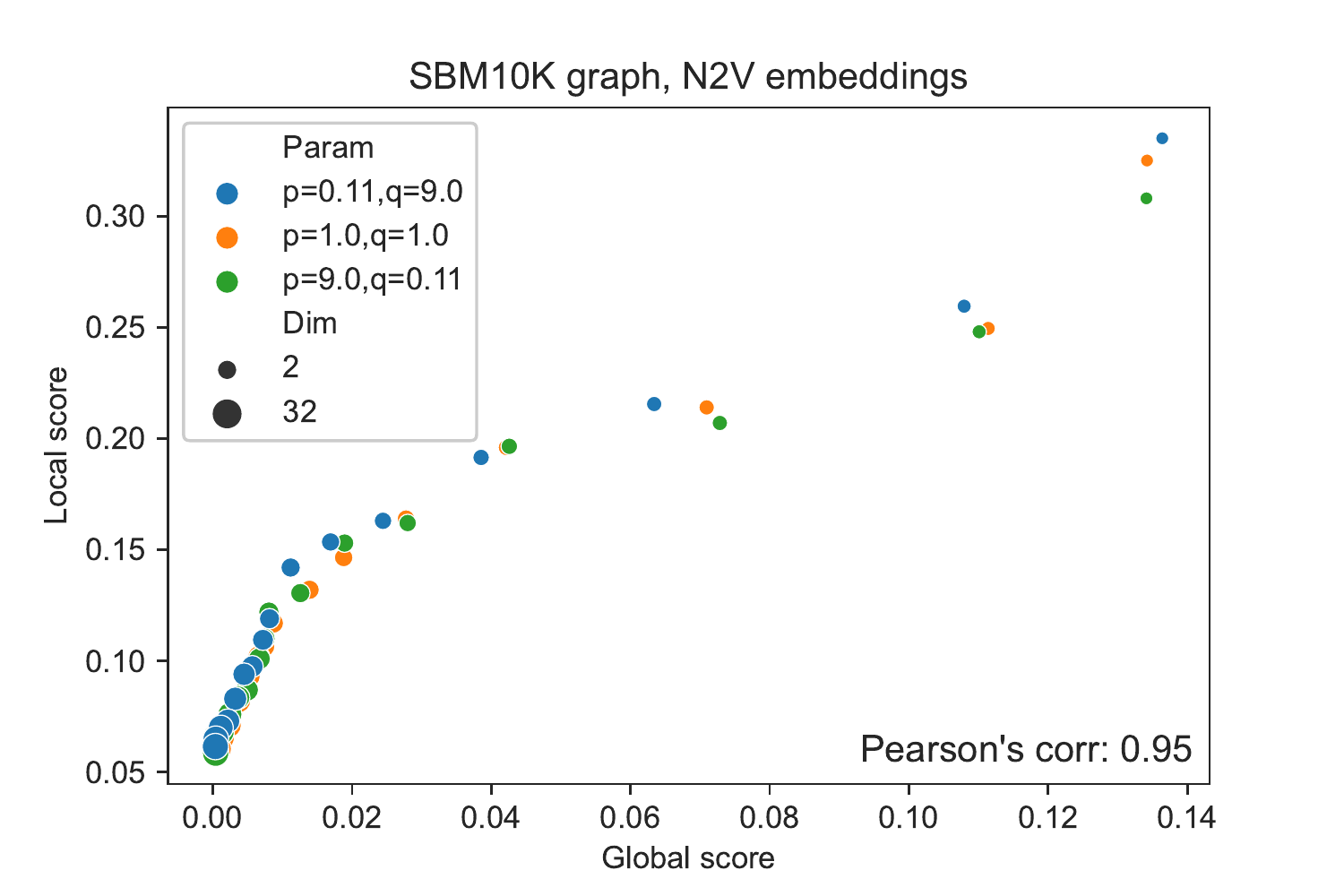}
     \vspace{.1cm}
     \includegraphics[width=0.35\textwidth]{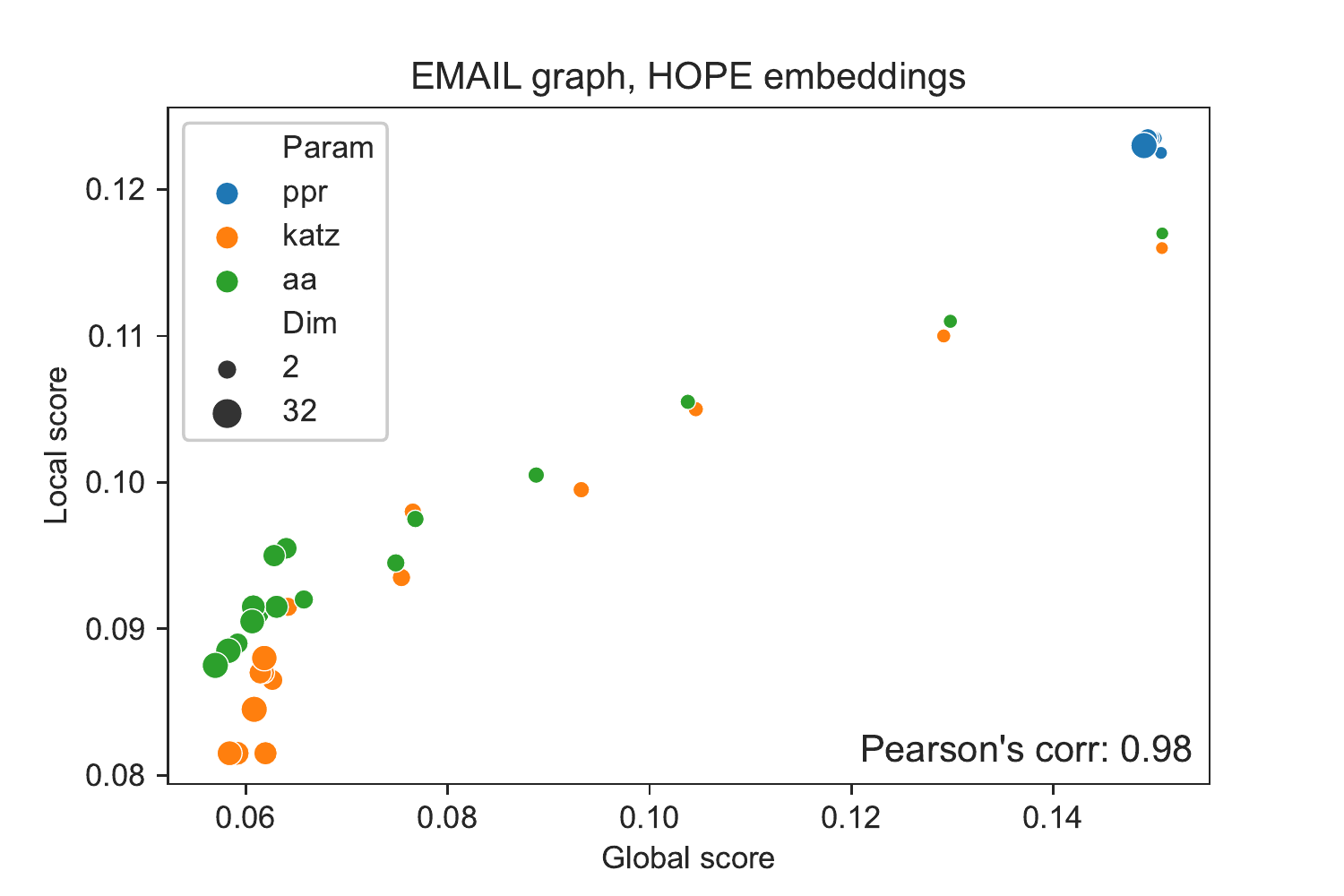}
         \hspace{.1cm}
     \includegraphics[width=0.35\textwidth]{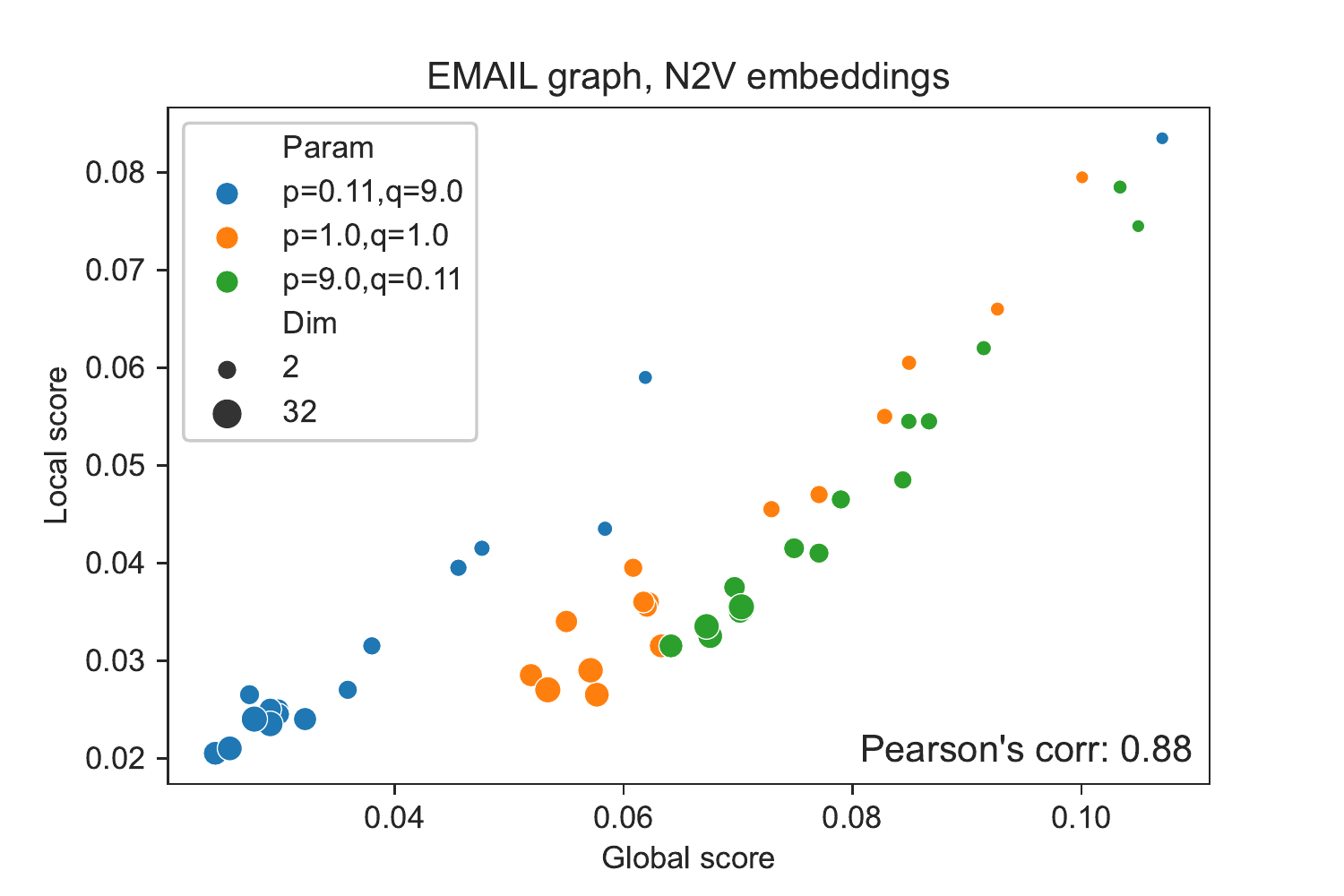}
     \caption{Global and local scores for \textbf{SBM}, \textbf{LFR}, \textbf{noisy-LFR}, \textbf{EMAIL} graphs with \textbf{HOPE} (left) and \textbf{Node2Vec} (right) embeddings.}
     \label{fig:10kexact_2measseed}
 \end{figure}

\begin{figure}[ht]
     \centering
     \includegraphics[width=0.45\textwidth]{Plots/exp_div_landmarks_exact_10k_nosplit_seed/lfr10k_hope_div.pdf}
         \hspace{.1cm}
     \includegraphics[width=0.45\textwidth]{Plots/exp_div_landmarks_exact_10k_nosplit_seed/lfr10k_n2v_div.pdf}
     \vspace{.1cm}
     \includegraphics[width=0.45\textwidth]{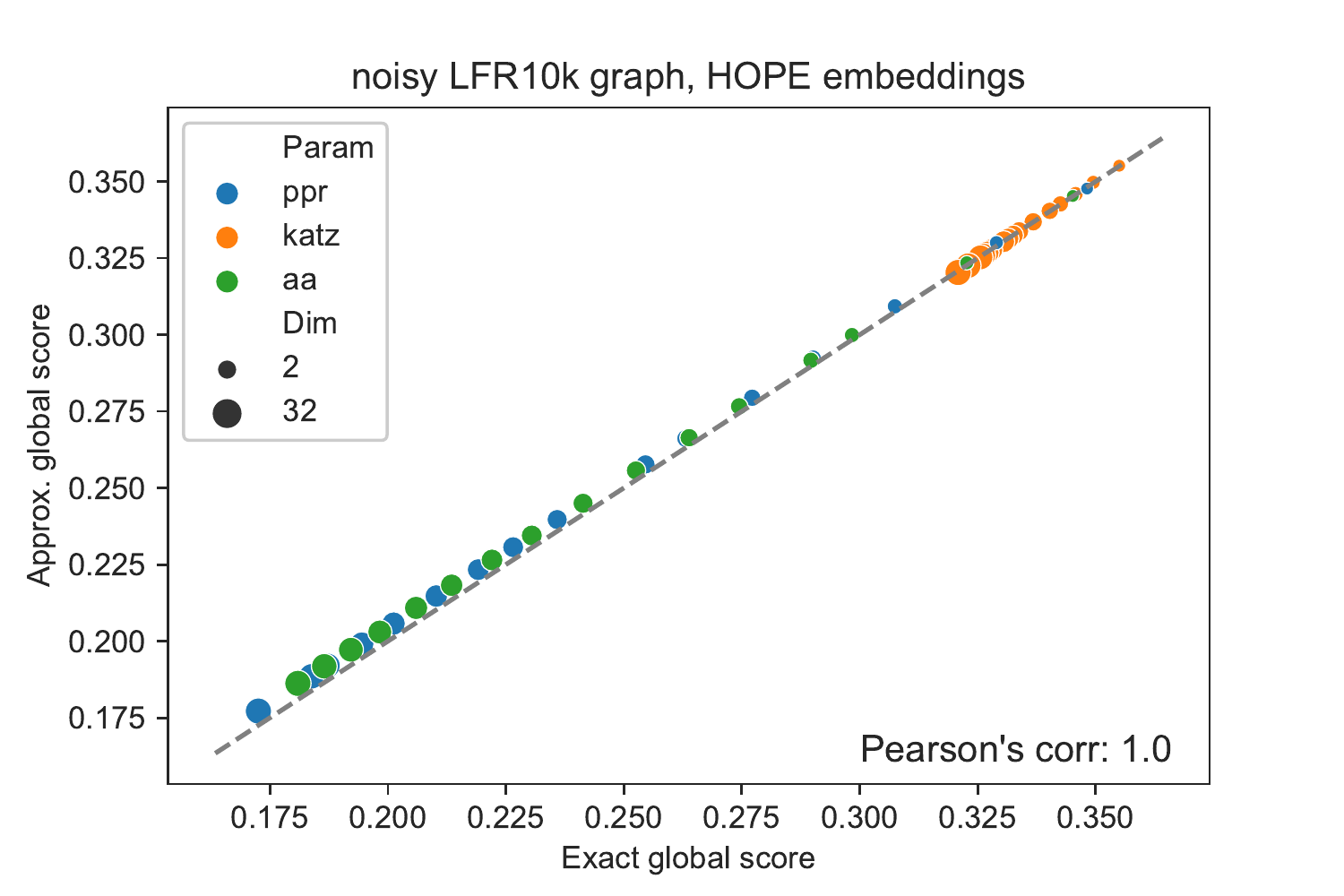}
         \hspace{.1cm}
     \includegraphics[width=0.45\textwidth]{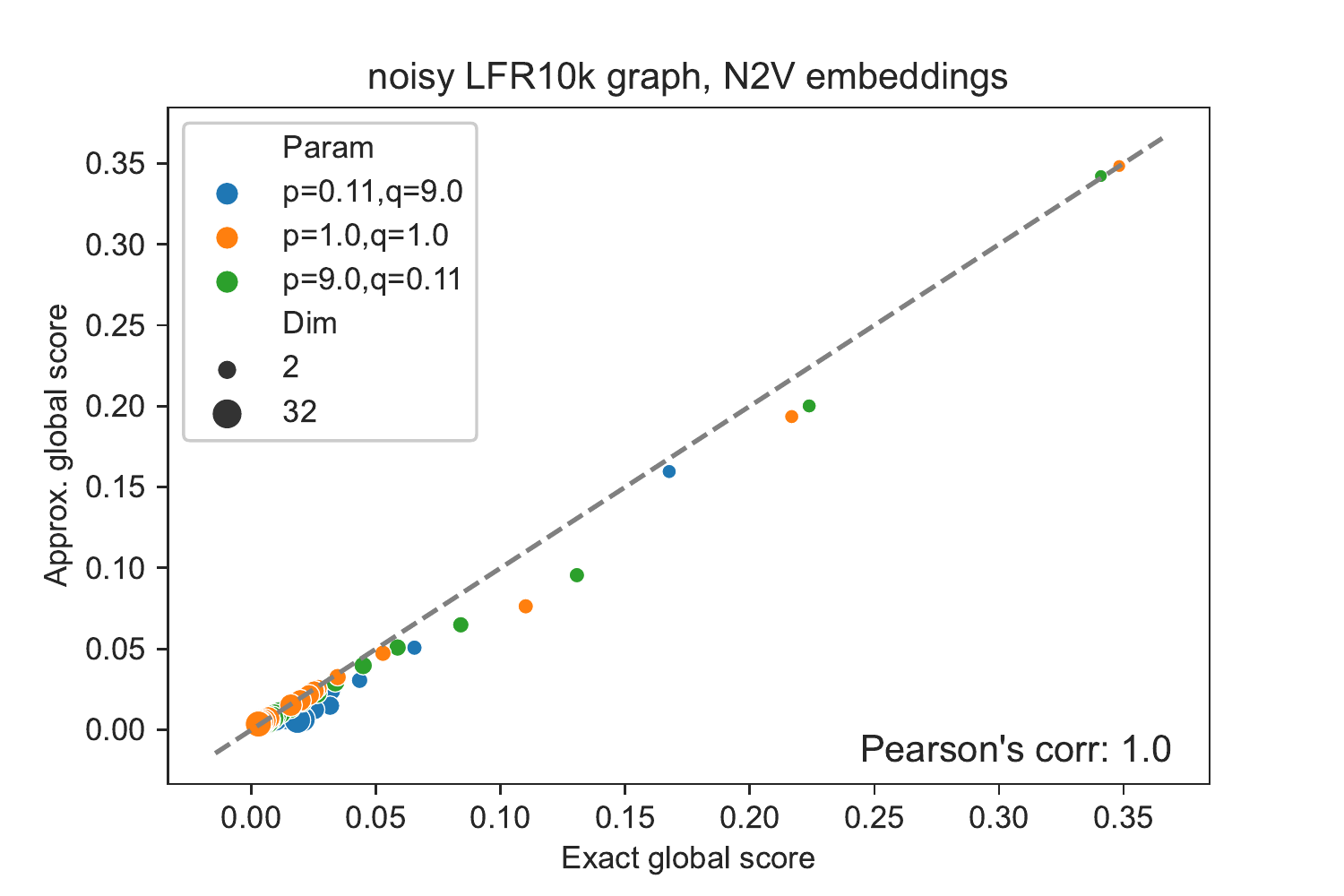}
     \vspace{.1cm}
     \includegraphics[width=0.45\textwidth]{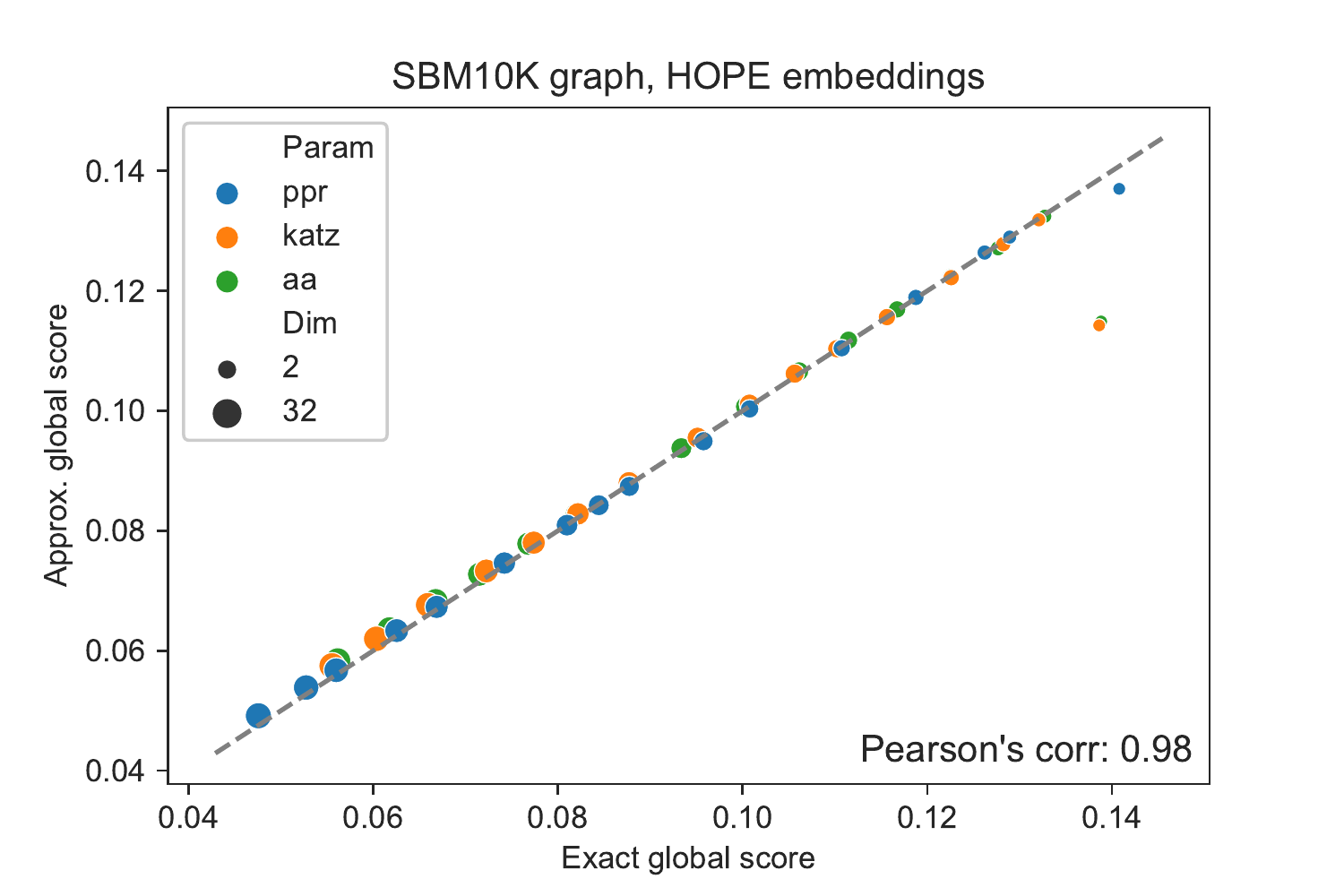}
         \hspace{.1cm}
     \includegraphics[width=0.45\textwidth]{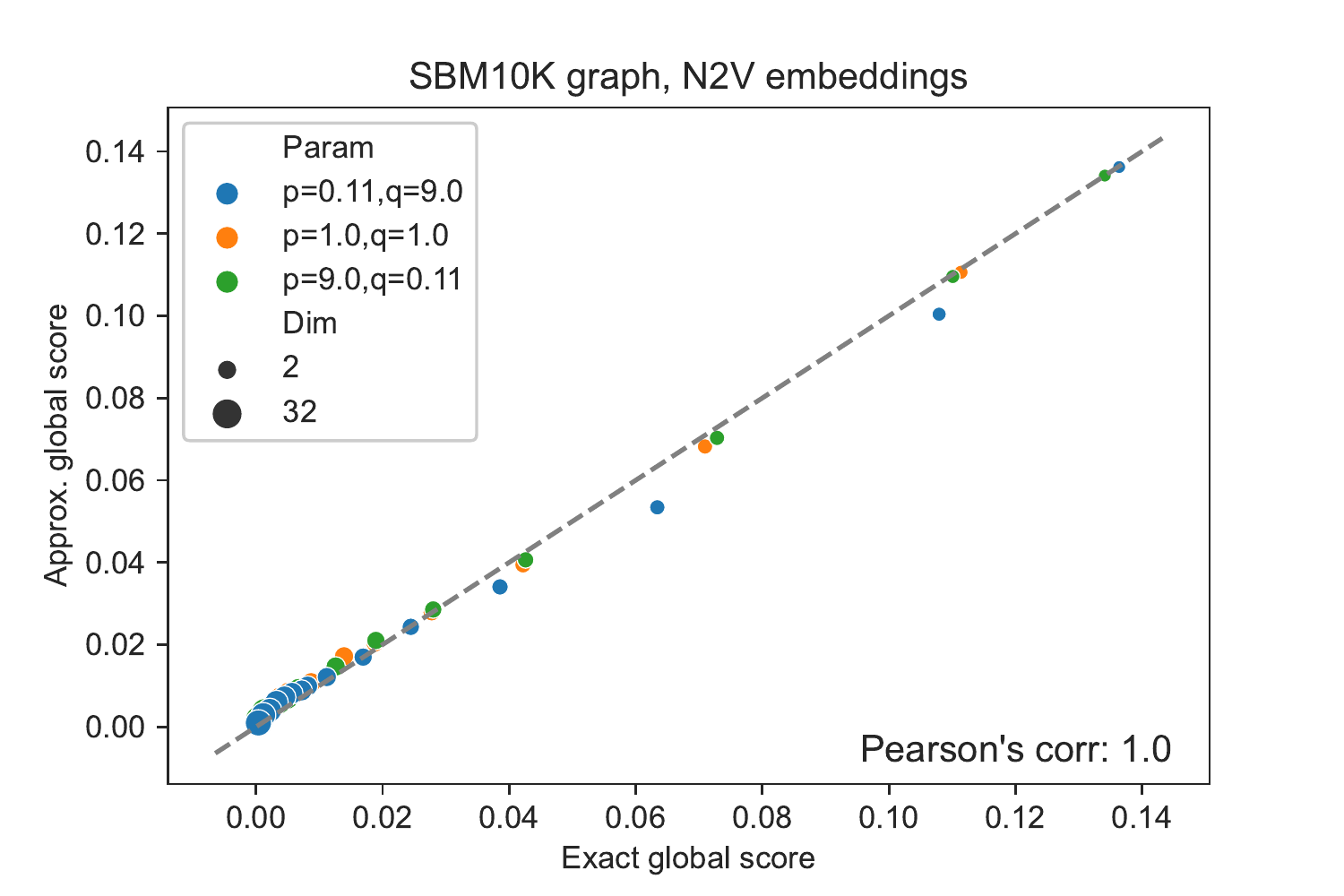}
     \caption{Approximated vs.\ exact global scores for \textbf{SBM}, \textbf{LFR}, \textbf{noisy-LFR} graphs and \textbf{HOPE} (left), \textbf{Node2Vec} (right) embeddings.}
     \label{fig:10kapprox_exact_divseed}
 \end{figure}

\begin{figure}[ht]
     \centering
     \includegraphics[width=0.45\textwidth]{Plots/exp_auc_landmarks_exact_10k_nosplit_seed/lfr10k_hope_auc.pdf}
         \hspace{.1cm}
     \includegraphics[width=0.45\textwidth]{Plots/exp_auc_landmarks_exact_10k_nosplit_seed/lfr10k_n2v_auc.pdf}
     \vspace{.1cm}
     \includegraphics[width=0.45\textwidth]{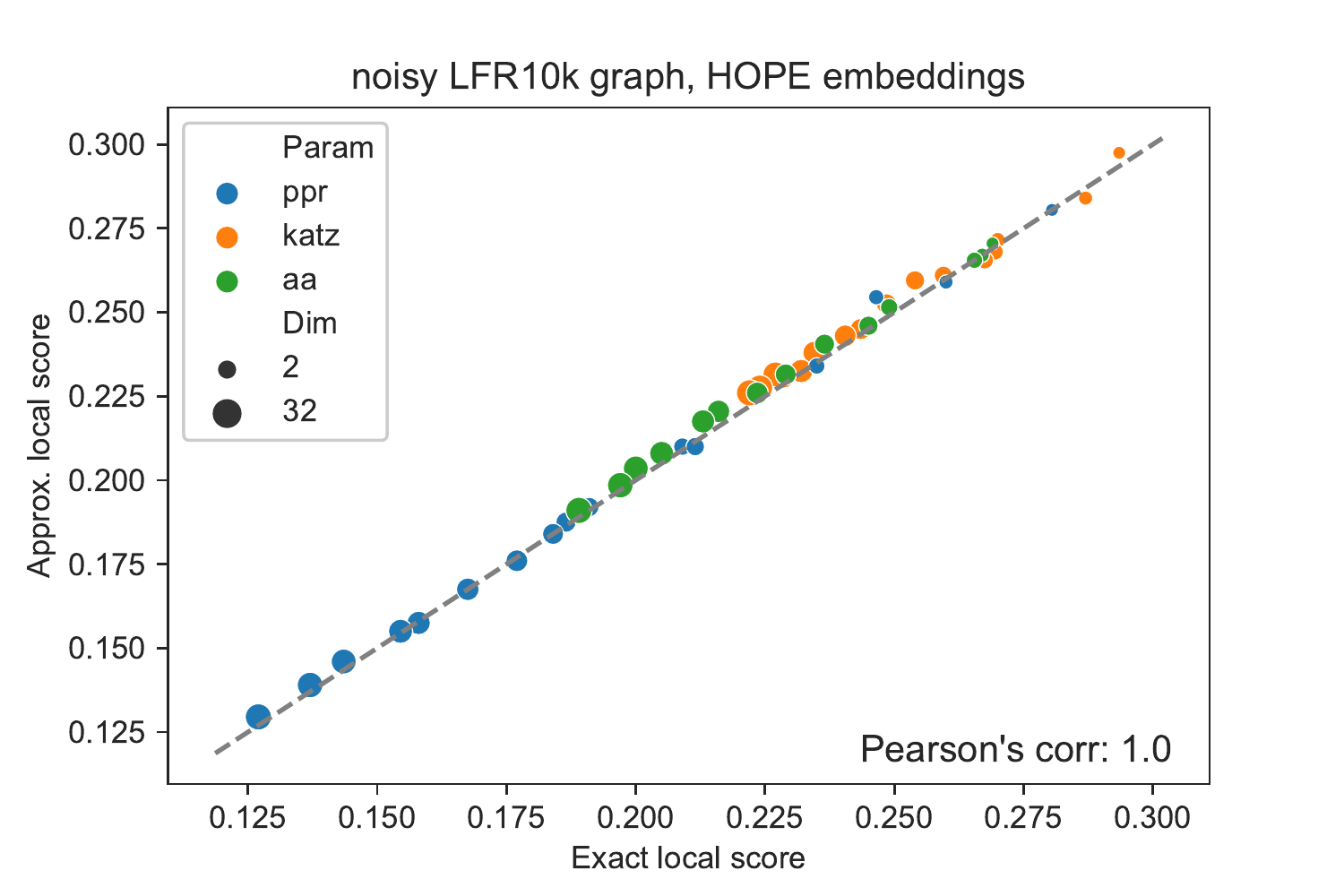}
         \hspace{.1cm}
     \includegraphics[width=0.45\textwidth]{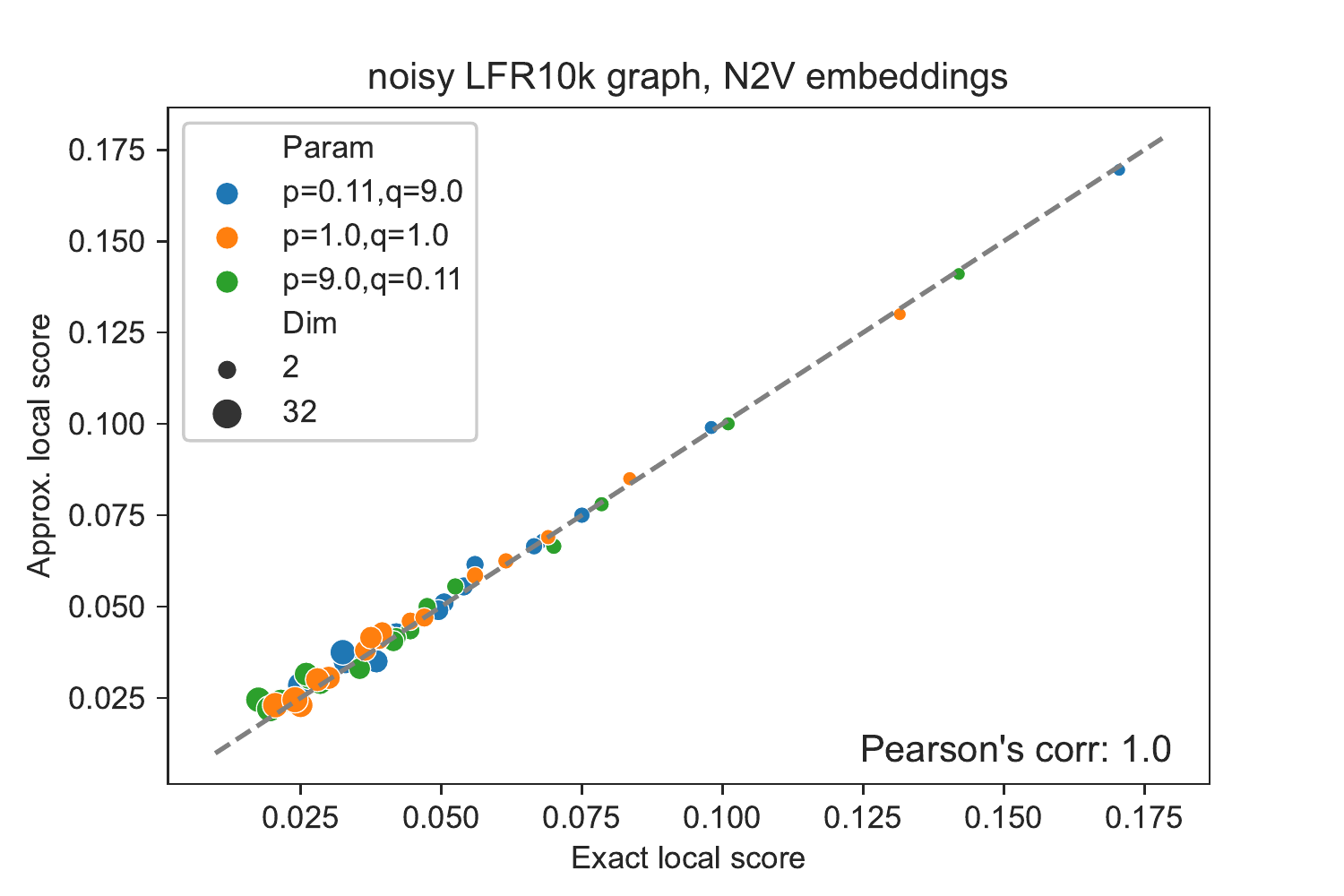}
     \vspace{.1cm}
     \includegraphics[width=0.45\textwidth]{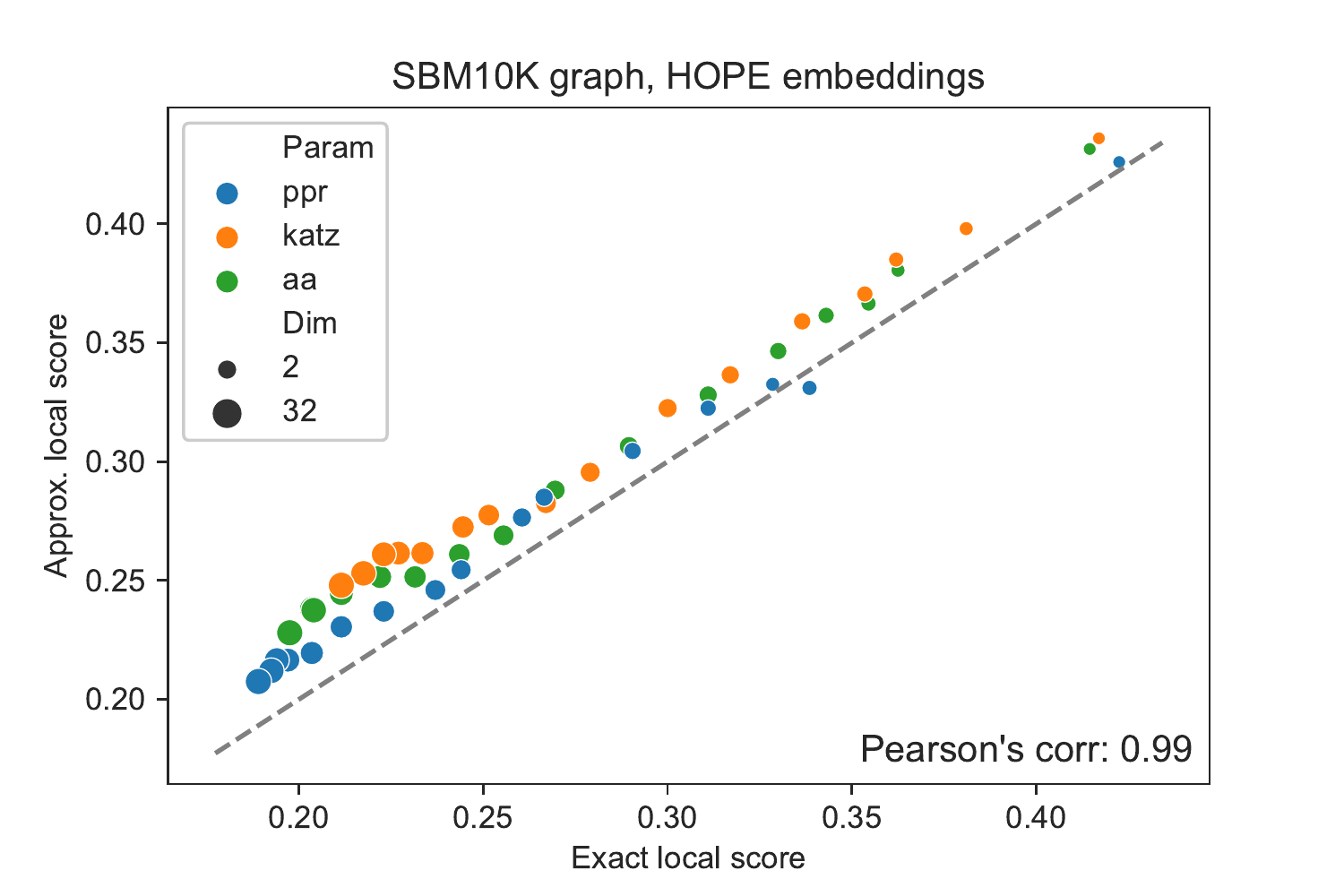}
         \hspace{.1cm}
     \includegraphics[width=0.45\textwidth]{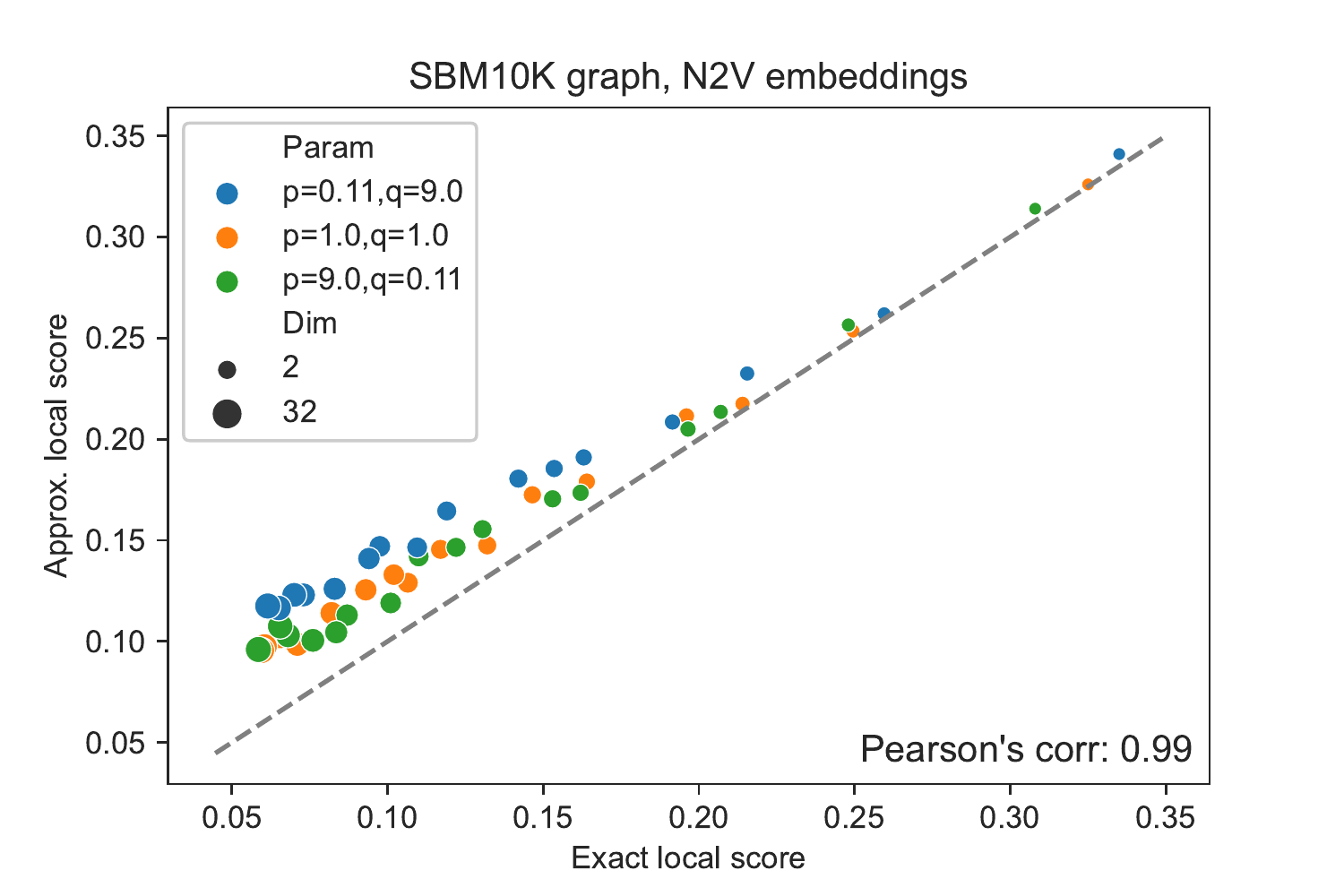}
     \caption{Approximated vs.\ exact local scores for \textbf{SBM}, \textbf{LFR}, \textbf{noisy-LFR} graphs and \textbf{HOPE} (left), \textbf{Node2Vec} (right) embeddings.}
     \label{fig:10kapprox_exact_aucseed}
 \end{figure}

\begin{figure}[ht]
    \centering
    \includegraphics[width=0.35\textwidth]{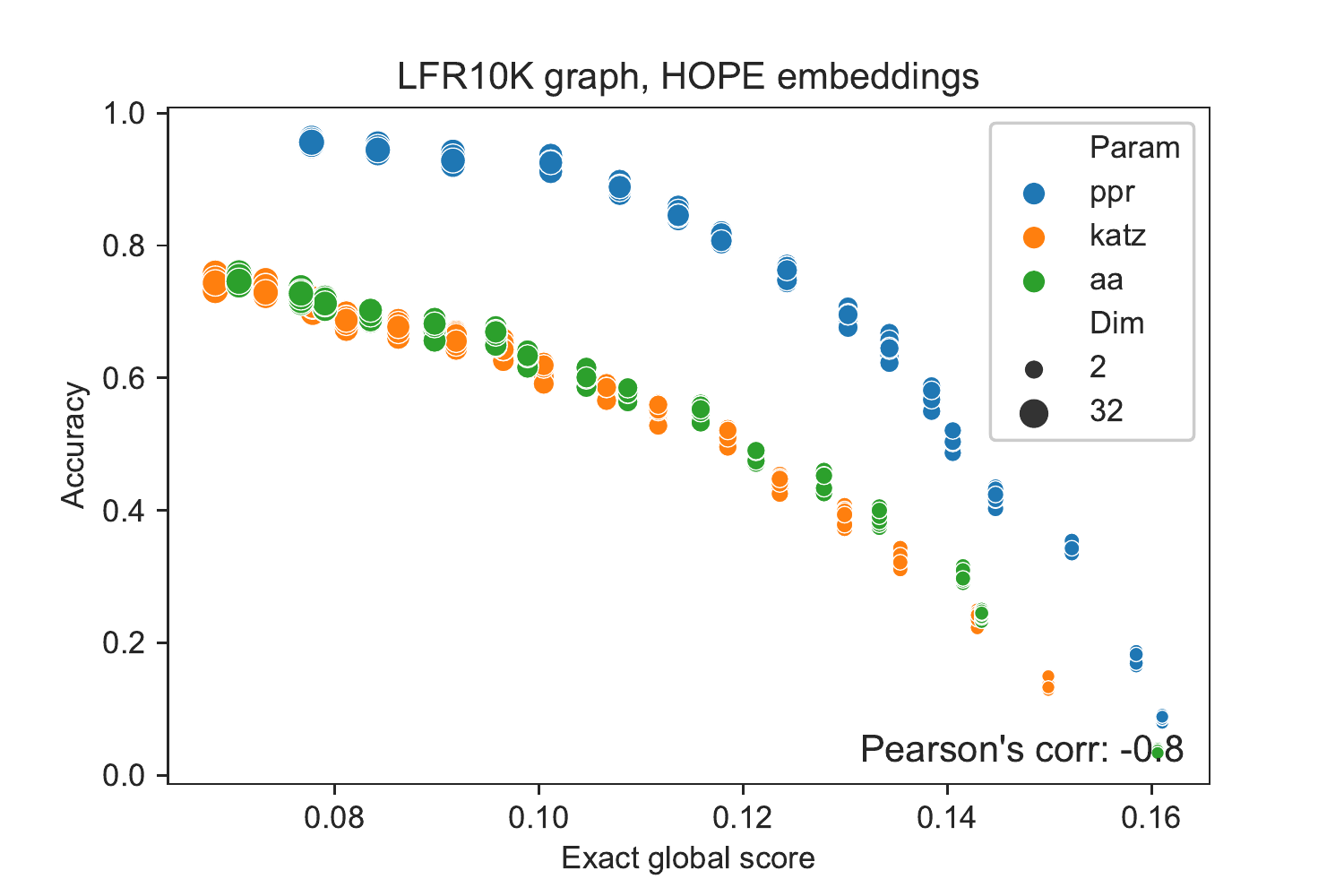}
        \hspace{.1cm}
    \includegraphics[width=0.35\textwidth]{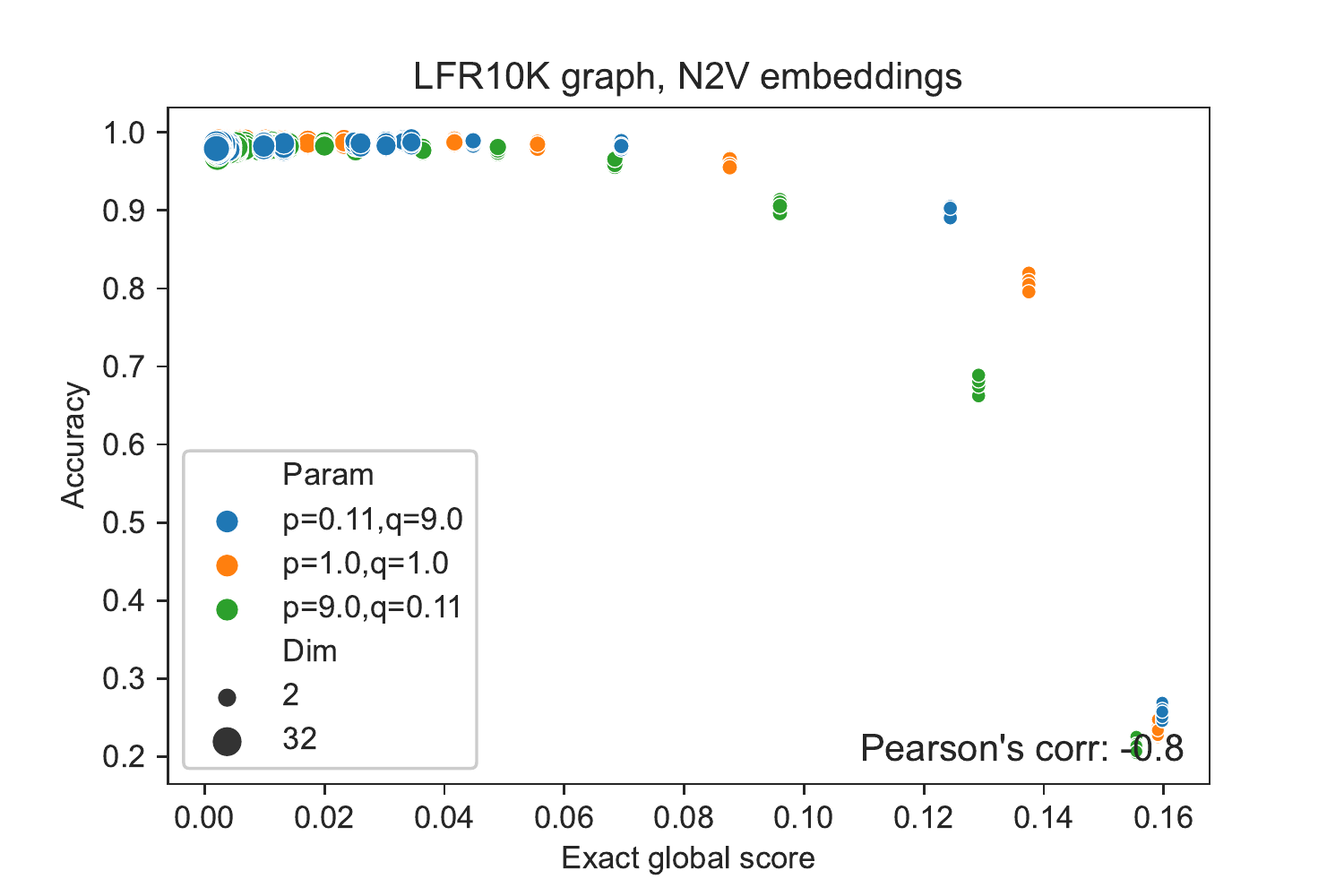}
    \vspace{.1cm}
    \includegraphics[width=0.35\textwidth]{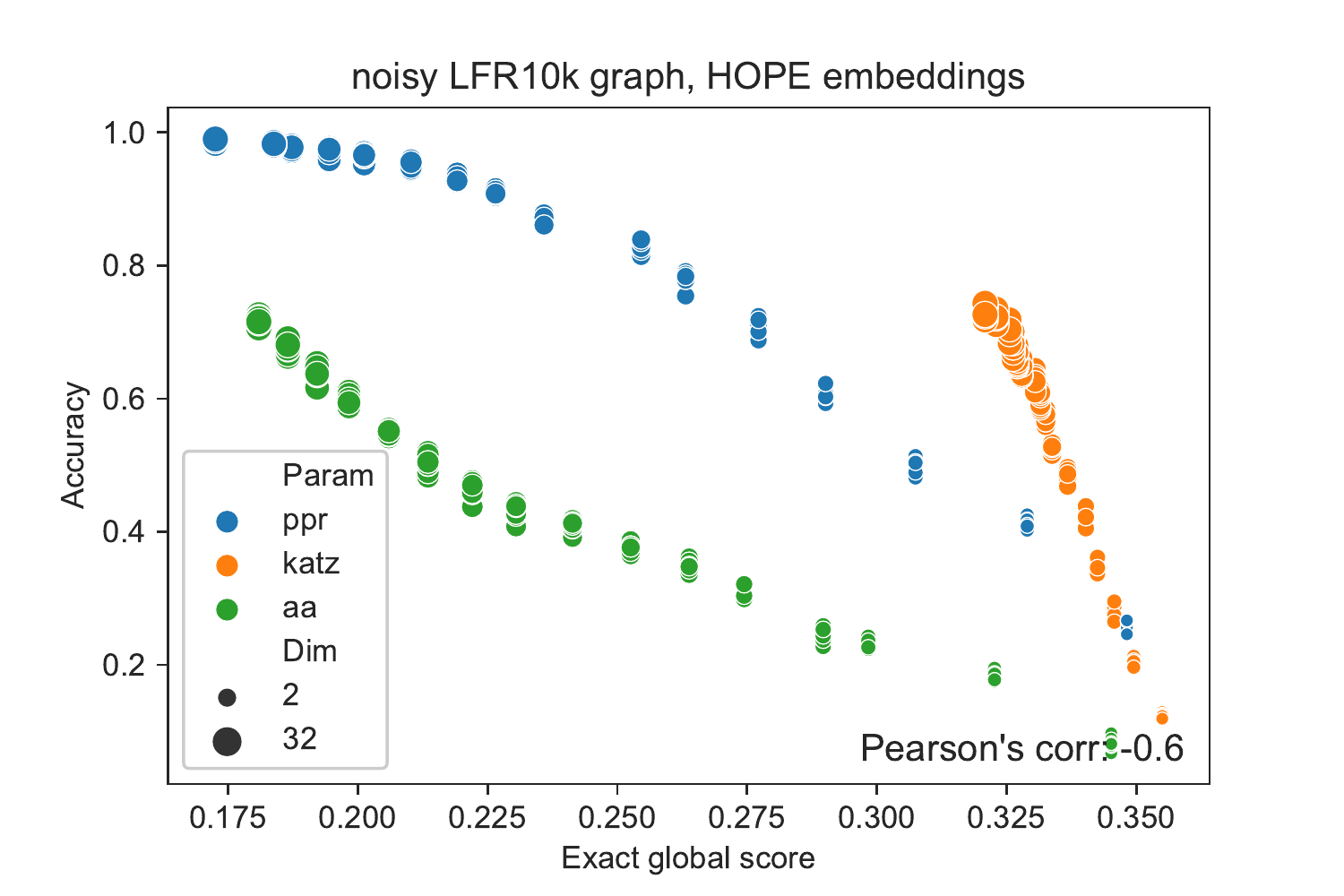}
        \hspace{.1cm}
    \includegraphics[width=0.35\textwidth]{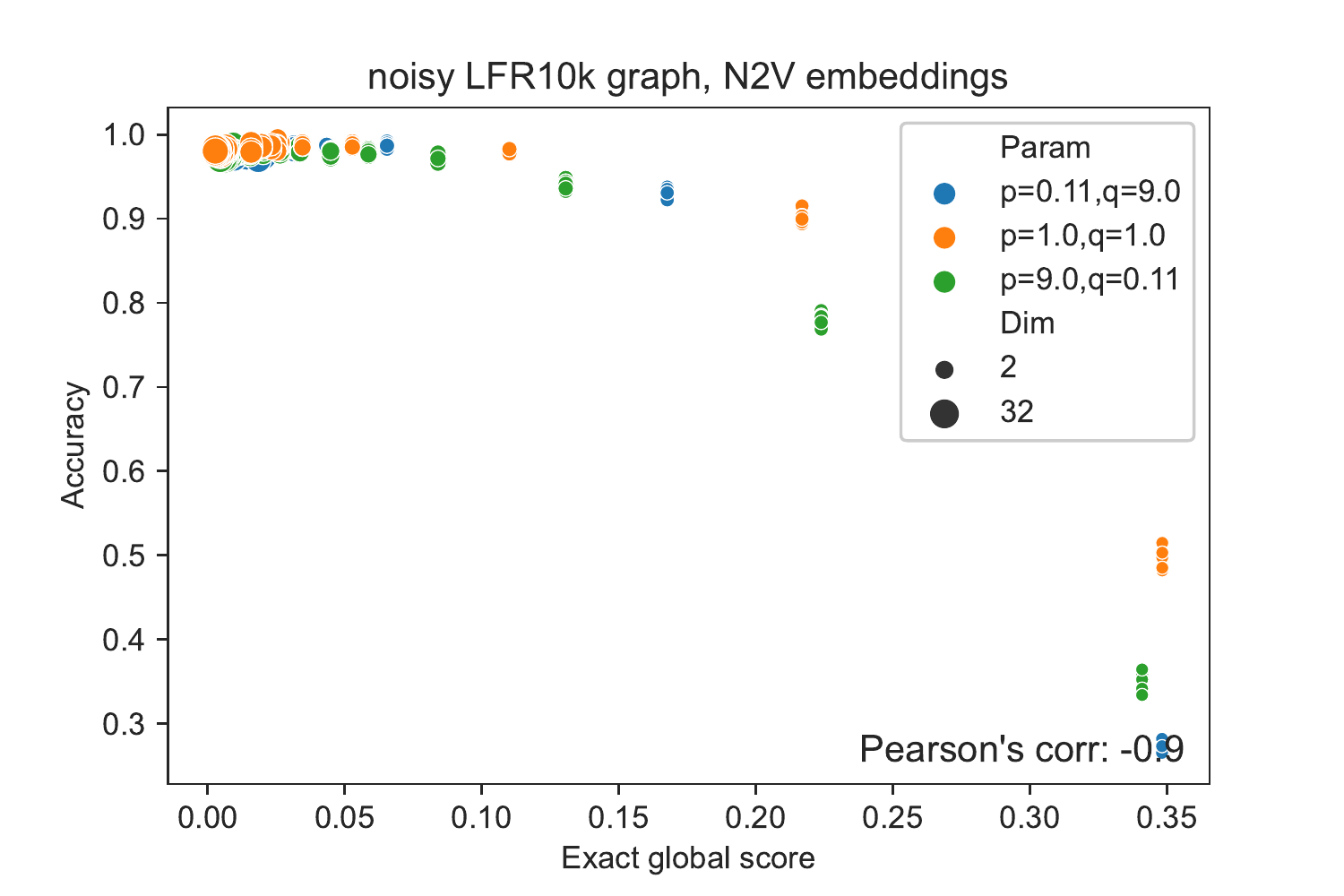}
    \vspace{.1cm}
    \includegraphics[width=0.35\textwidth]{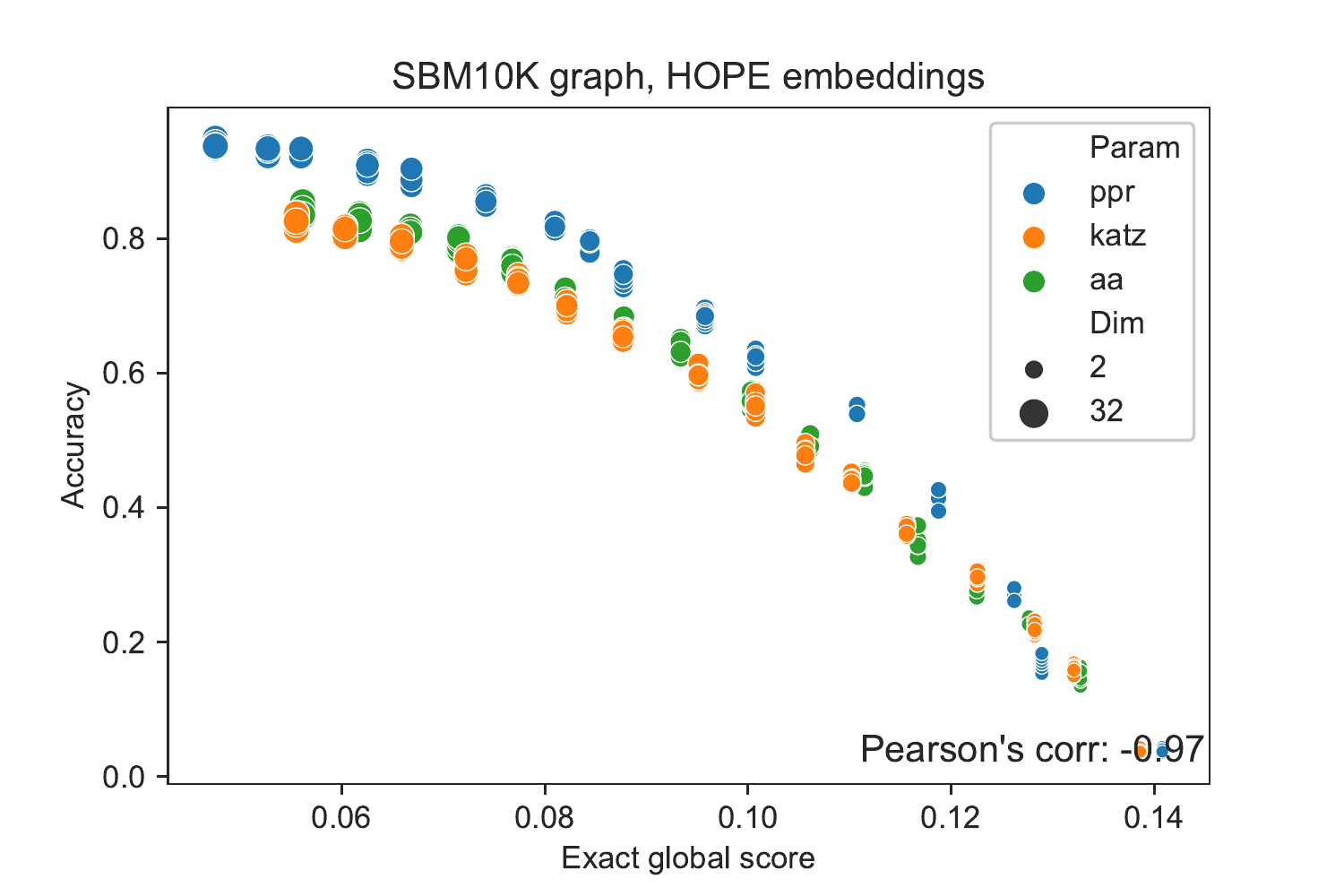}
        \hspace{.1cm}
    \includegraphics[width=0.35\textwidth]{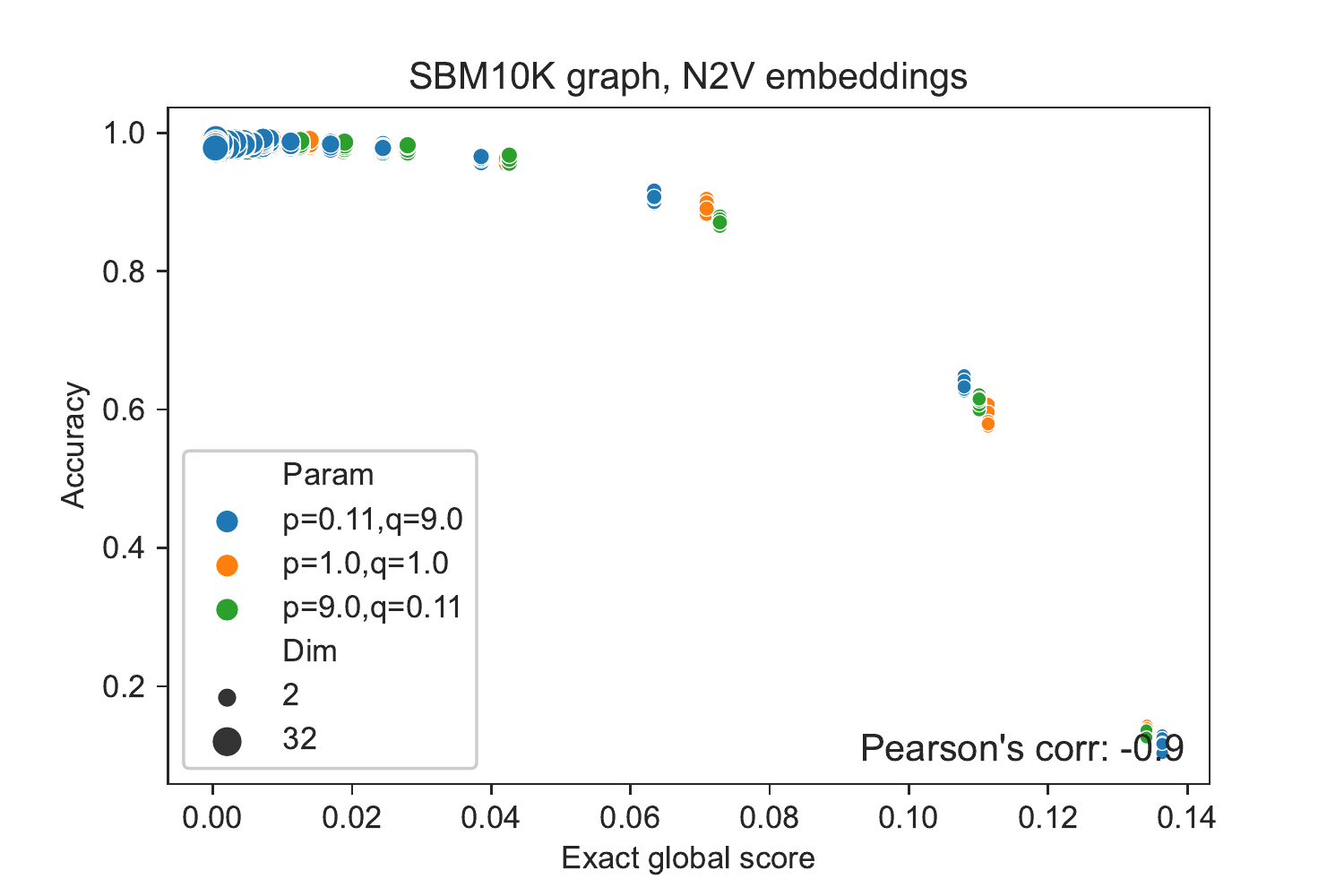}
        \hspace{.1cm}
    \includegraphics[width=0.35\textwidth]{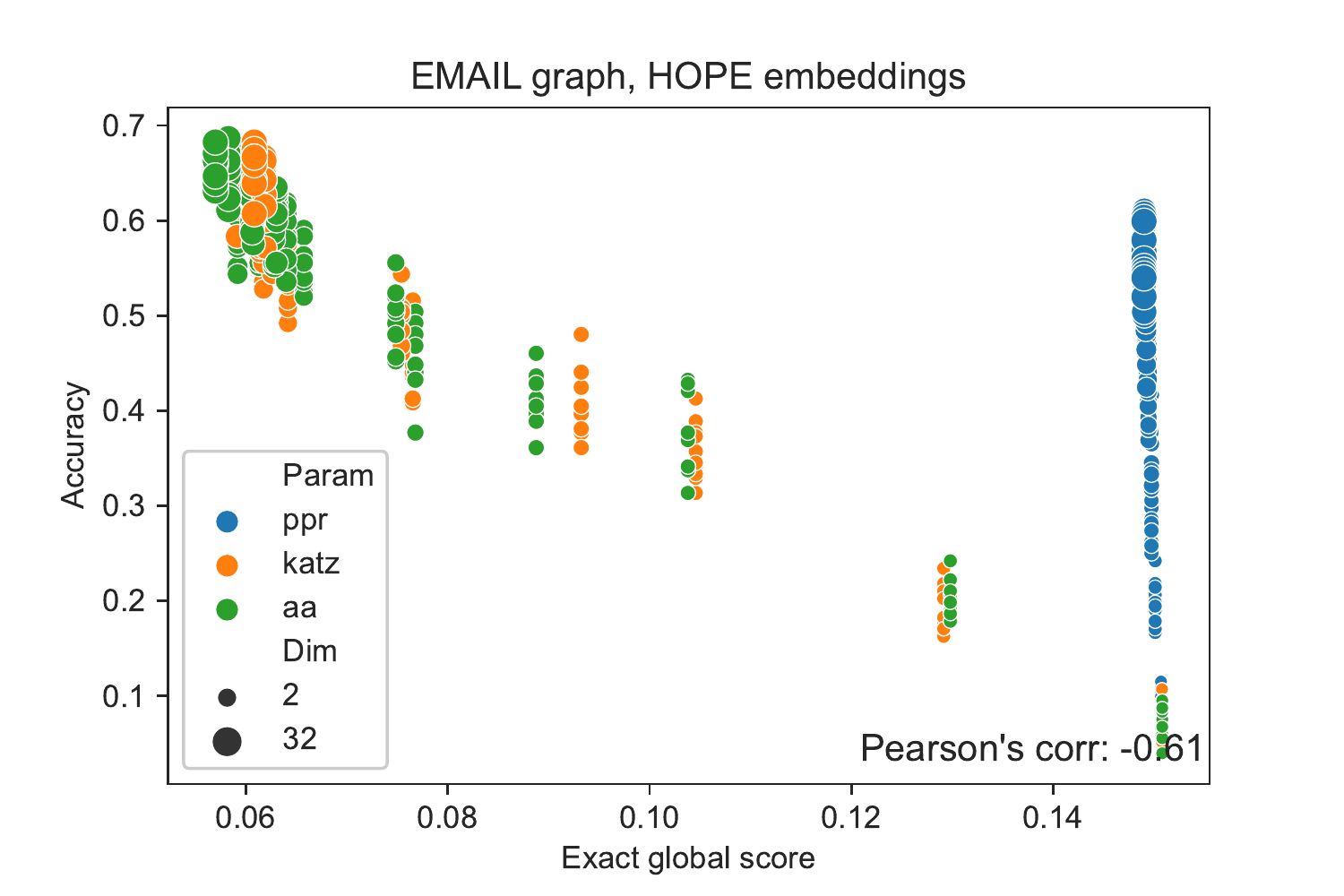}
        \hspace{.1cm}
    \includegraphics[width=0.35\textwidth]{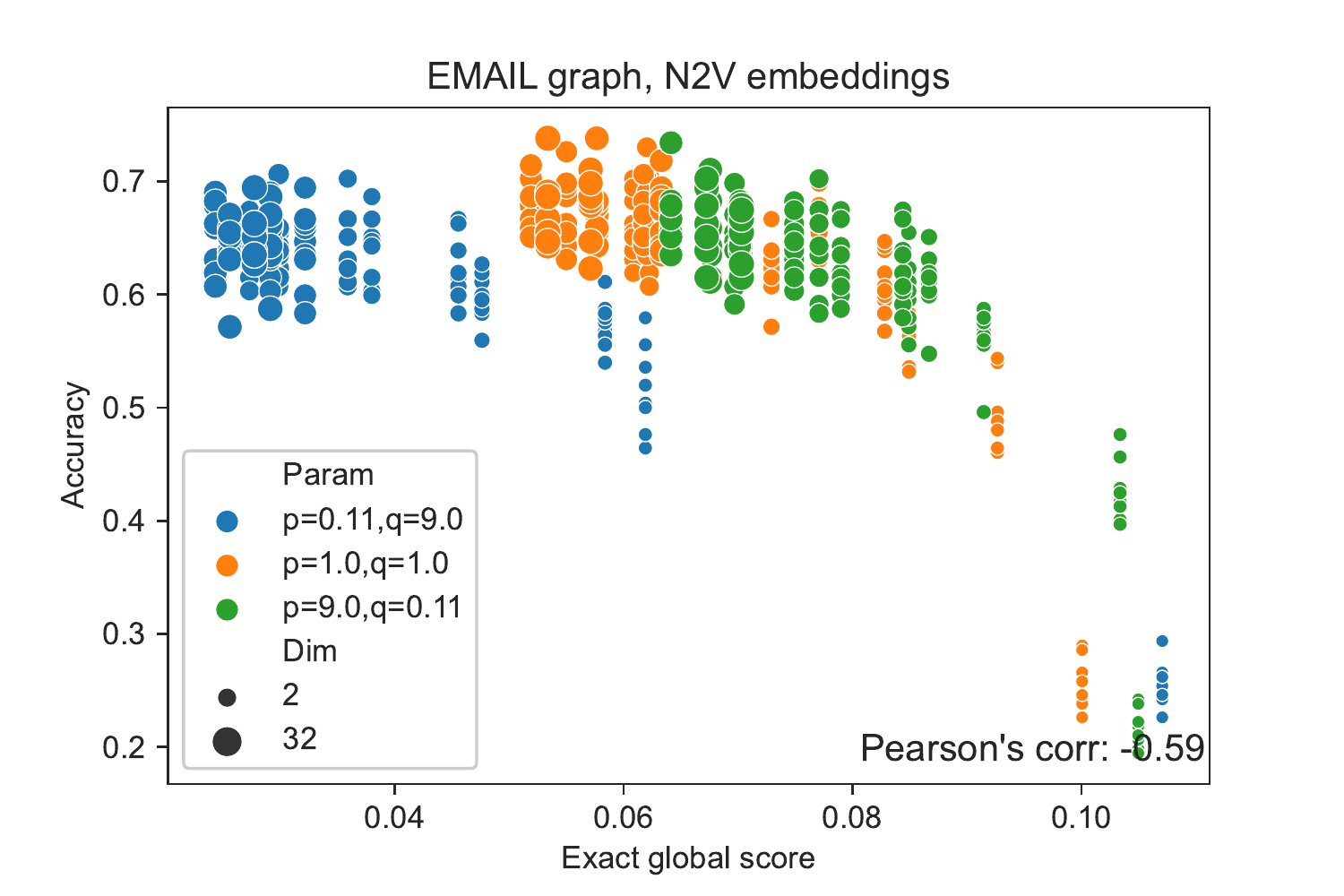}
    \caption{Node classification: global score vs.\ accuracy for \textbf{SBM}, \textbf{LFR}, \textbf{noisy-LFR}, \textbf{EMAIL} graphs and \textbf{HOPE} (left), \textbf{Node2Vec} (right) embeddings.}
    \label{fig:10kxgboost}
\end{figure}

\begin{figure}[ht]
    \centering
    \includegraphics[width=0.35\textwidth]{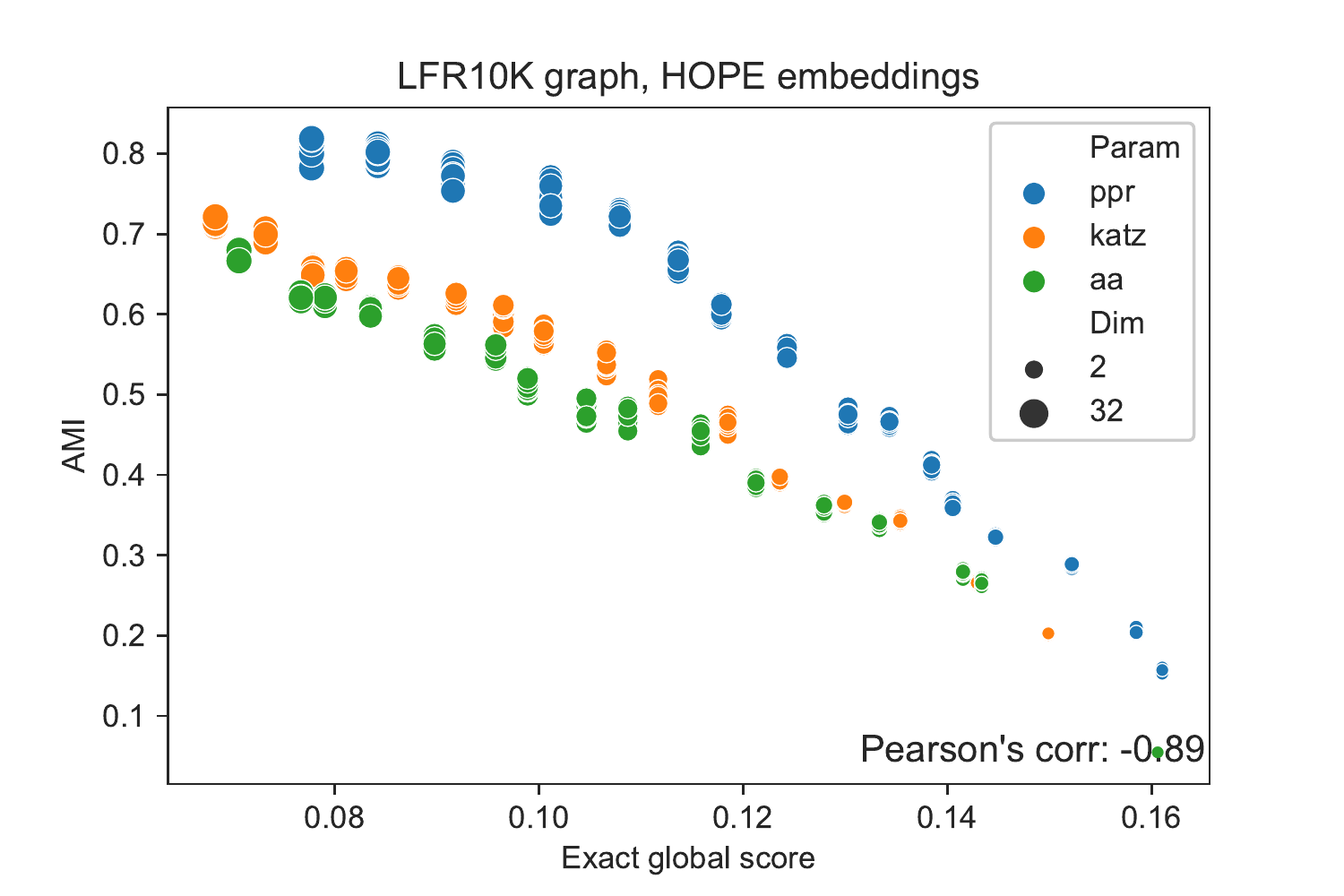}
        \hspace{.1cm}
    \includegraphics[width=0.35\textwidth]{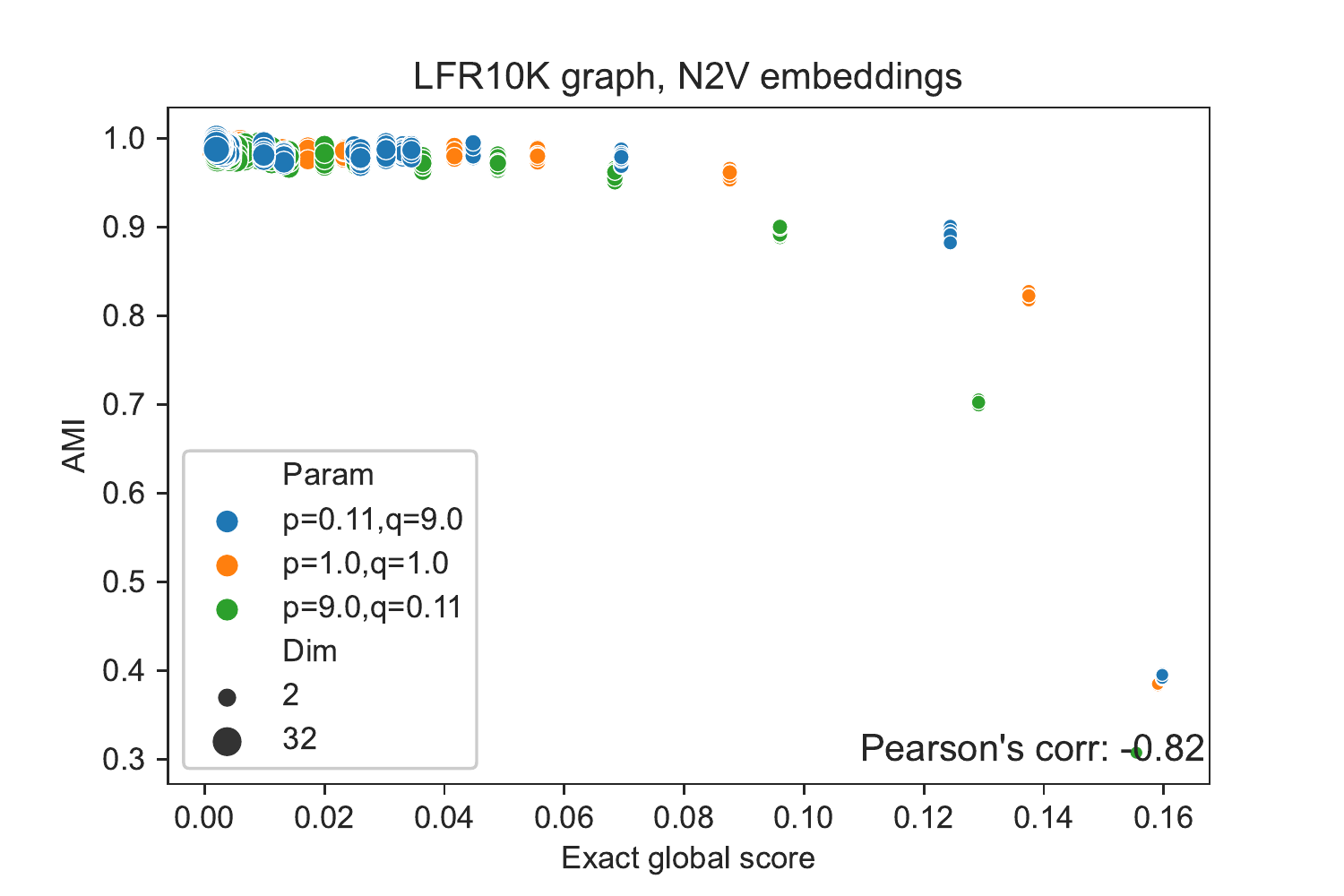}
    \vspace{.1cm}
    \includegraphics[width=0.35\textwidth]{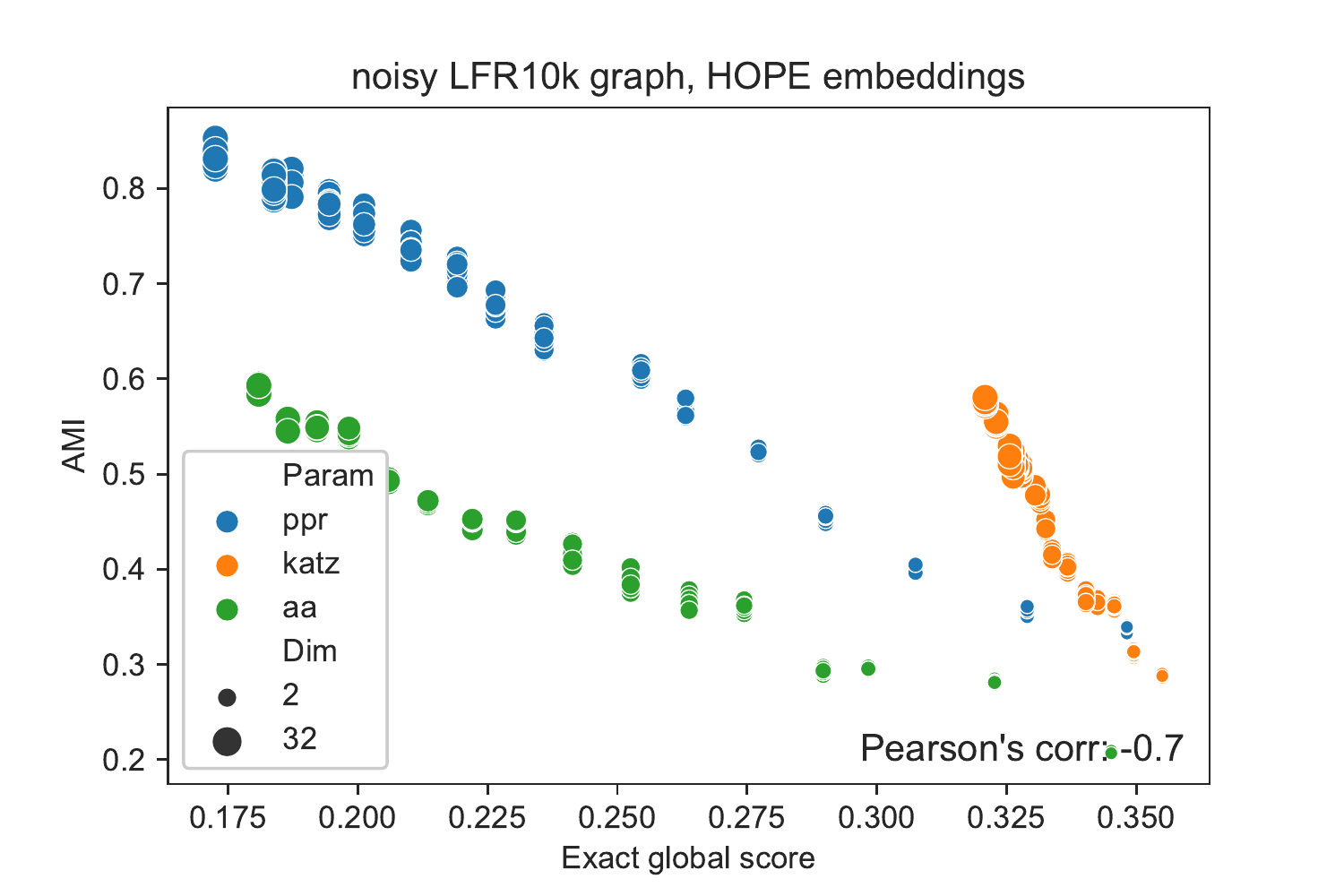}
        \hspace{.1cm}
    \includegraphics[width=0.35\textwidth]{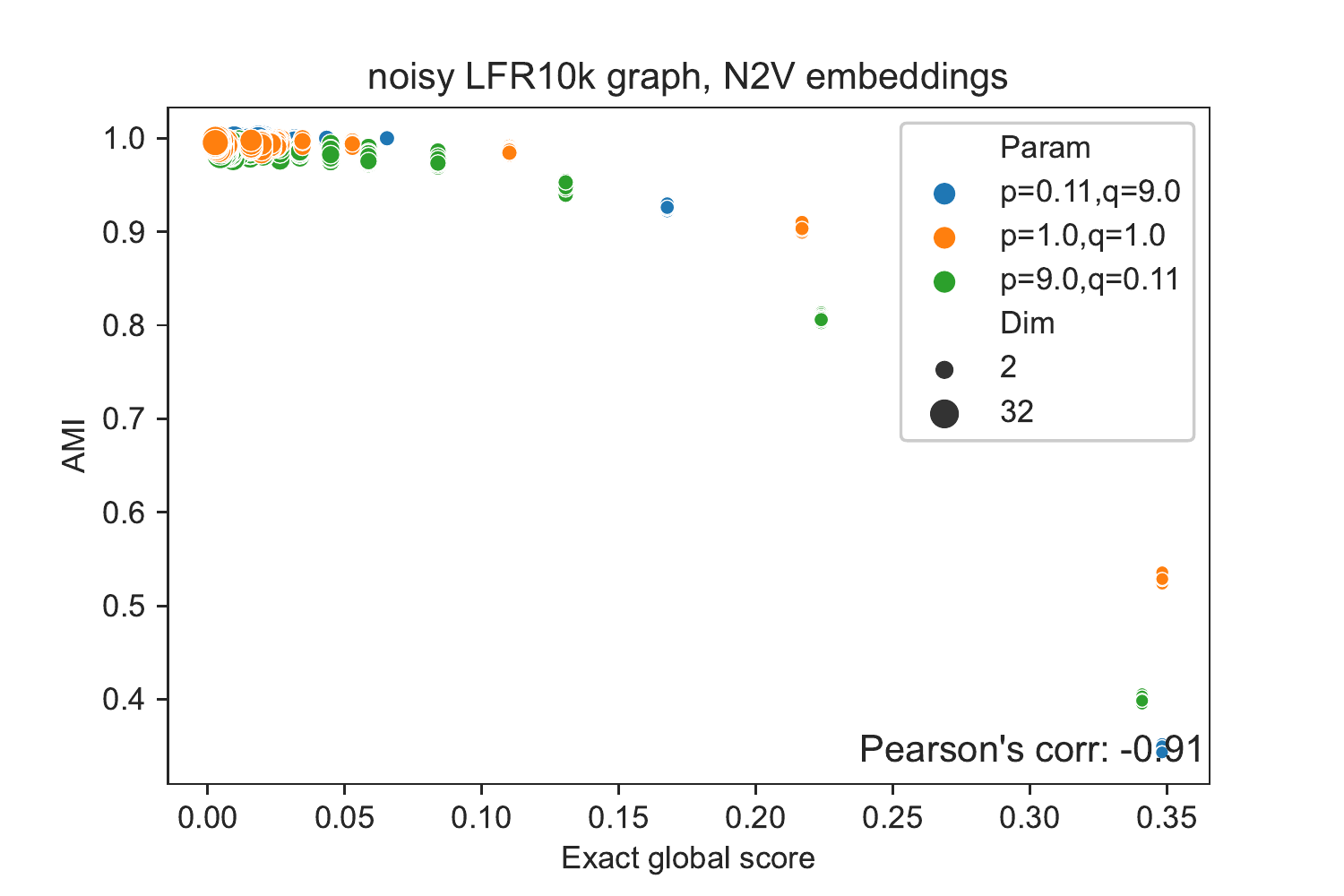}
    \vspace{.1cm}
    \includegraphics[width=0.35\textwidth]{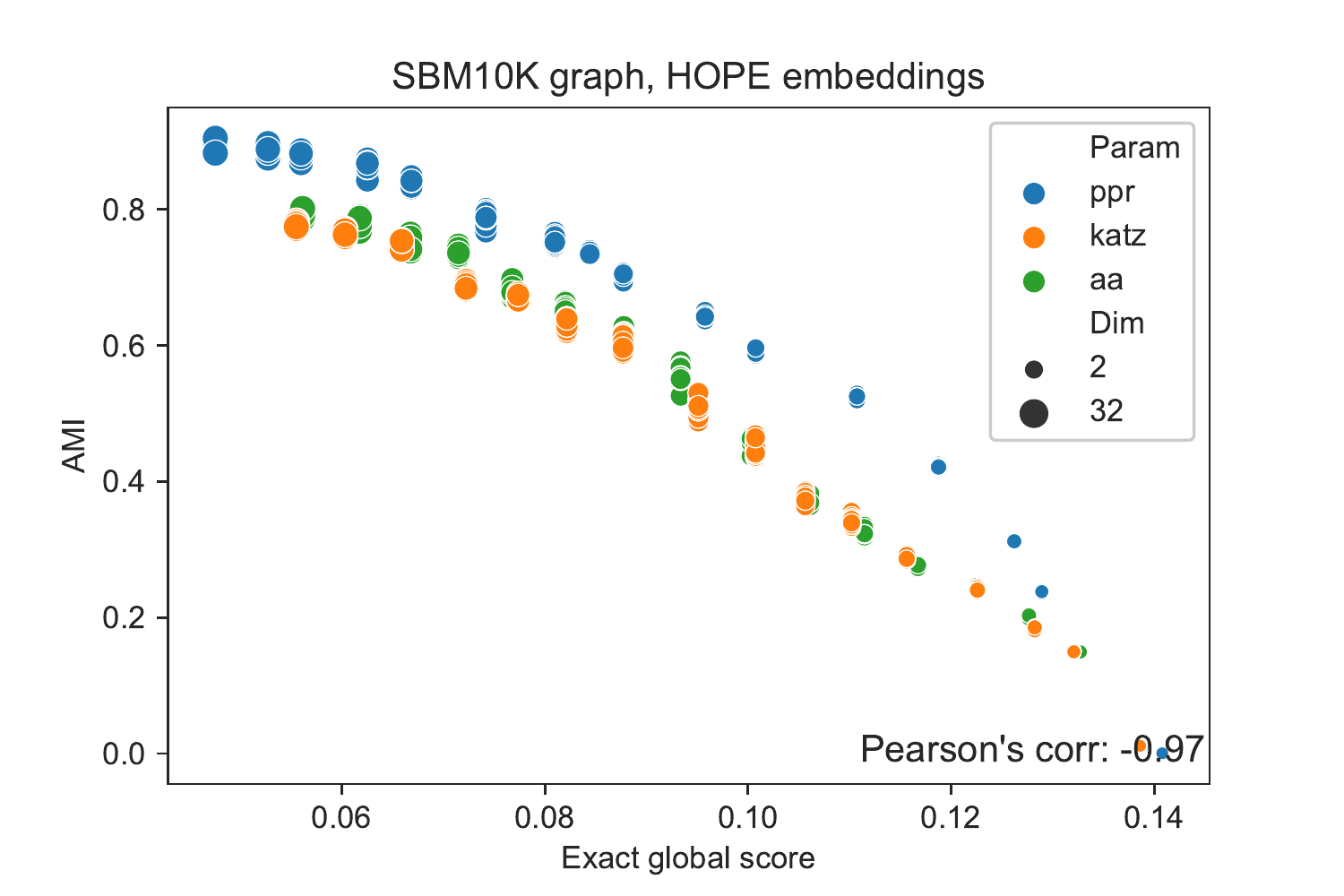}
        \hspace{.1cm}
    \includegraphics[width=0.35\textwidth]{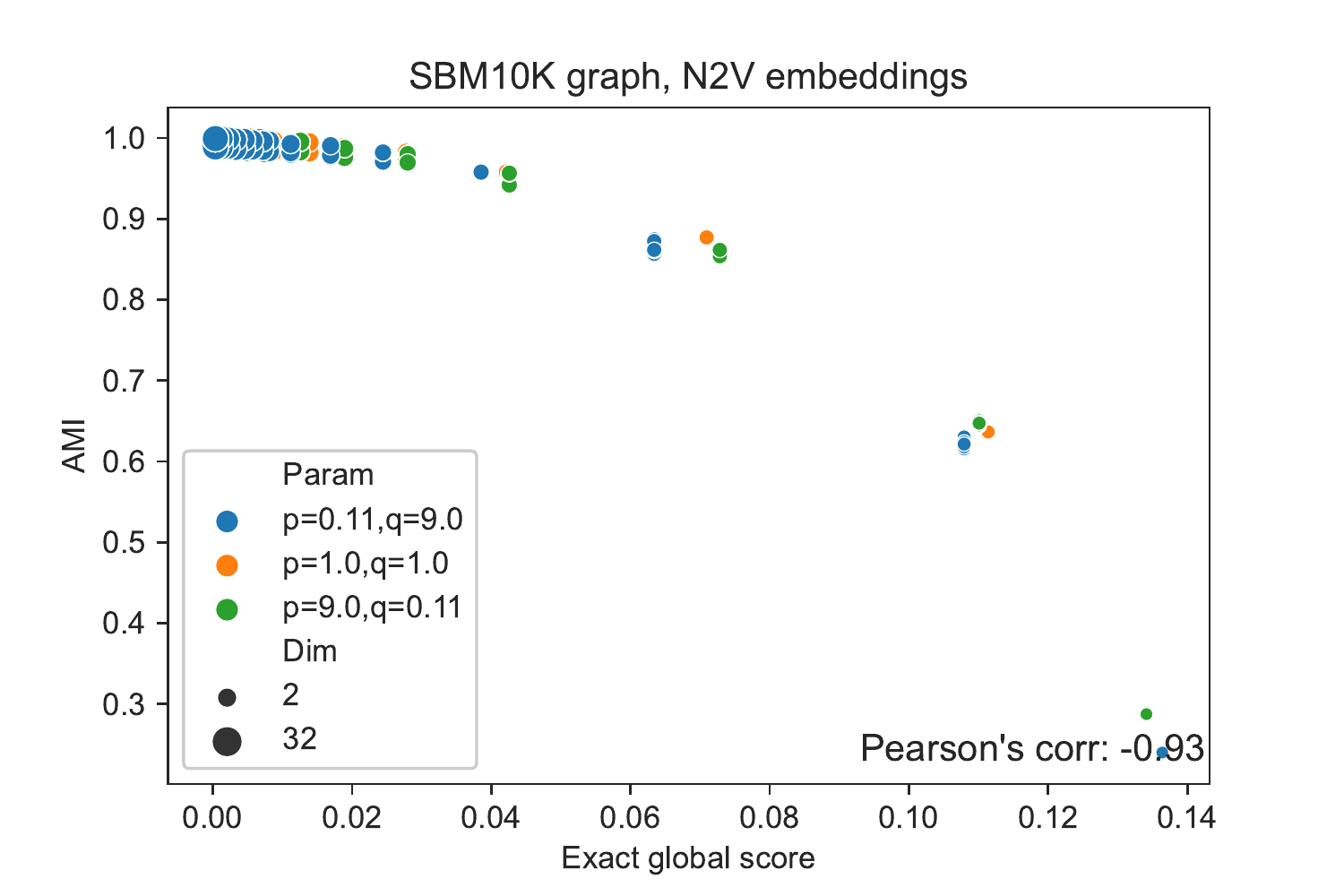}
    \vspace{.1cm}
    \includegraphics[width=0.35\textwidth]{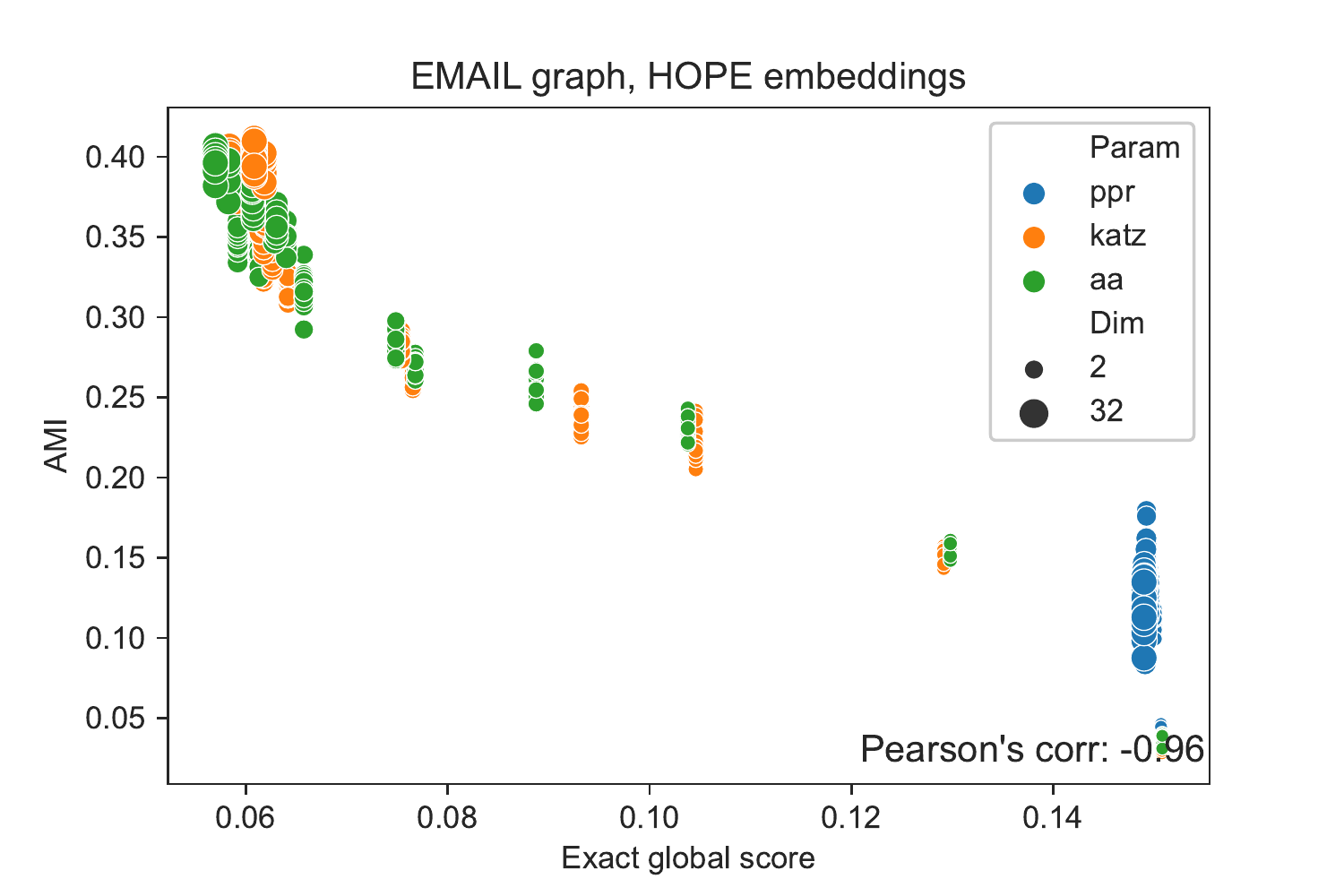}
        \hspace{.1cm}
    \includegraphics[width=0.35\textwidth]{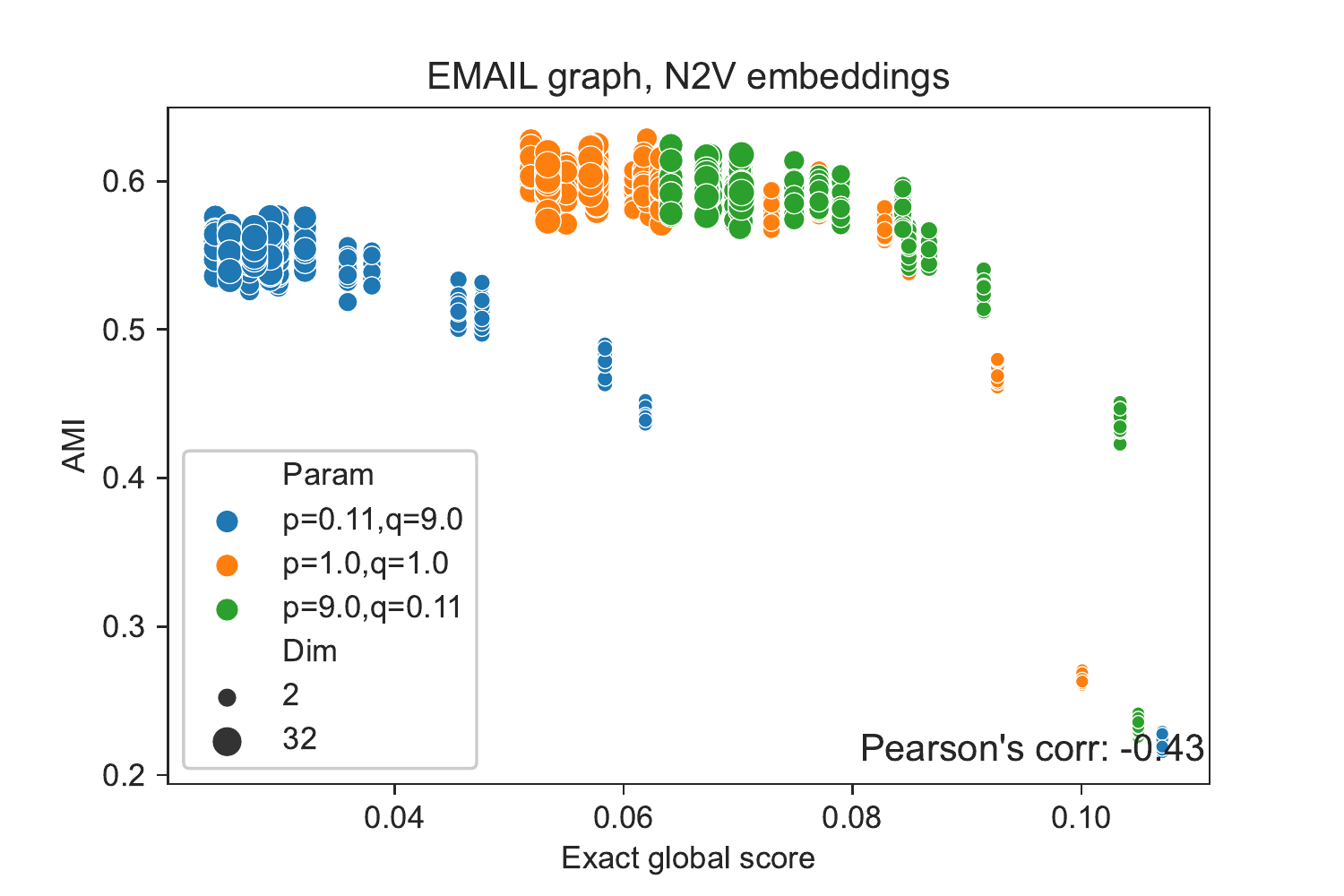}
    \caption{Community detection: global score vs.\ AMI for \textbf{SBM}, \textbf{LFR}, \textbf{noisy-LFR}, \textbf{EMAIL} graphs and \textbf{HOPE} (left), \textbf{Node2Vec} (right) embeddings.}
    \label{fig:10kkmeans}
\end{figure}

\begin{figure}[ht]
    \centering
    \includegraphics[width=0.35\textwidth]{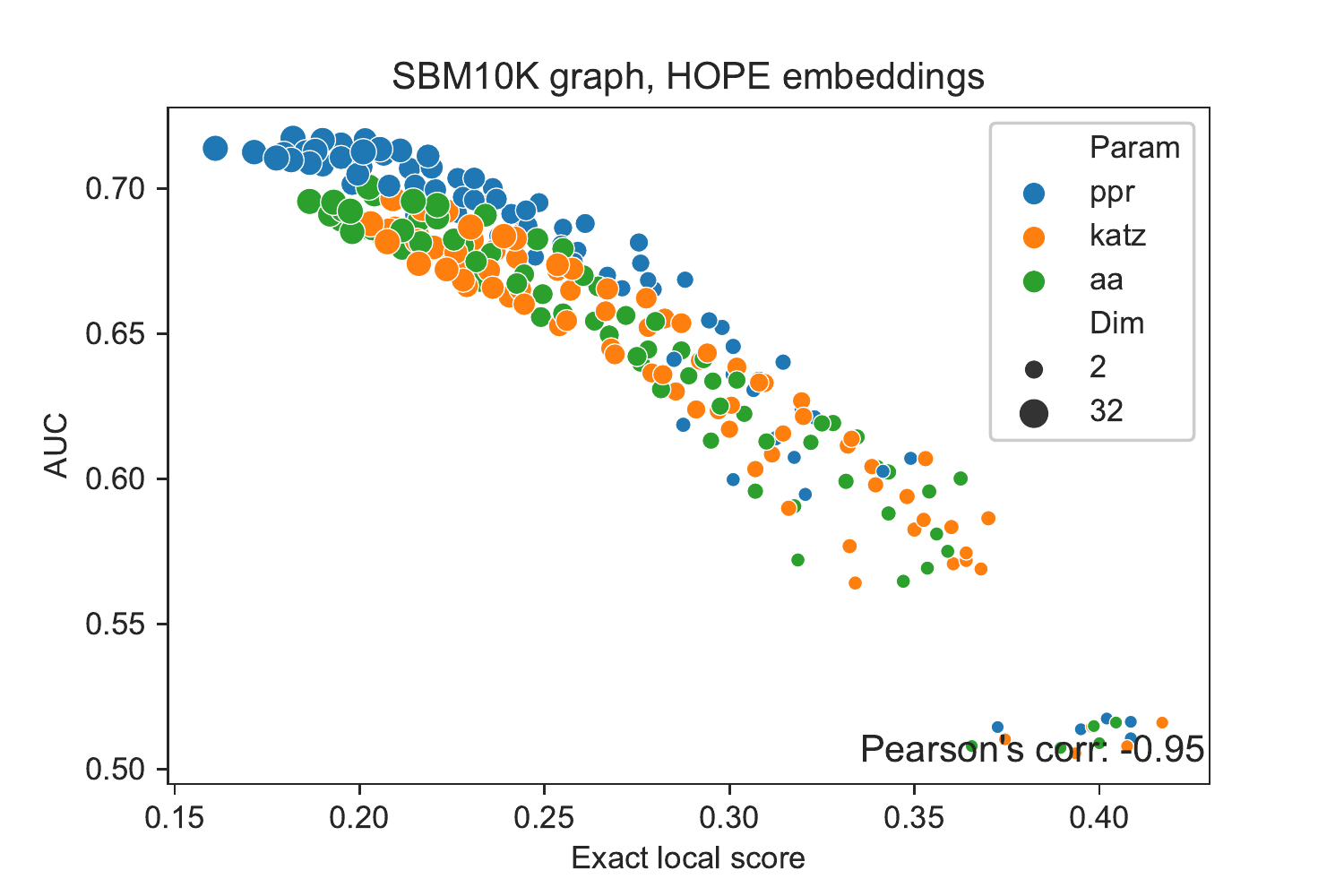}
        \hspace{.1cm}
    \includegraphics[width=0.35\textwidth]{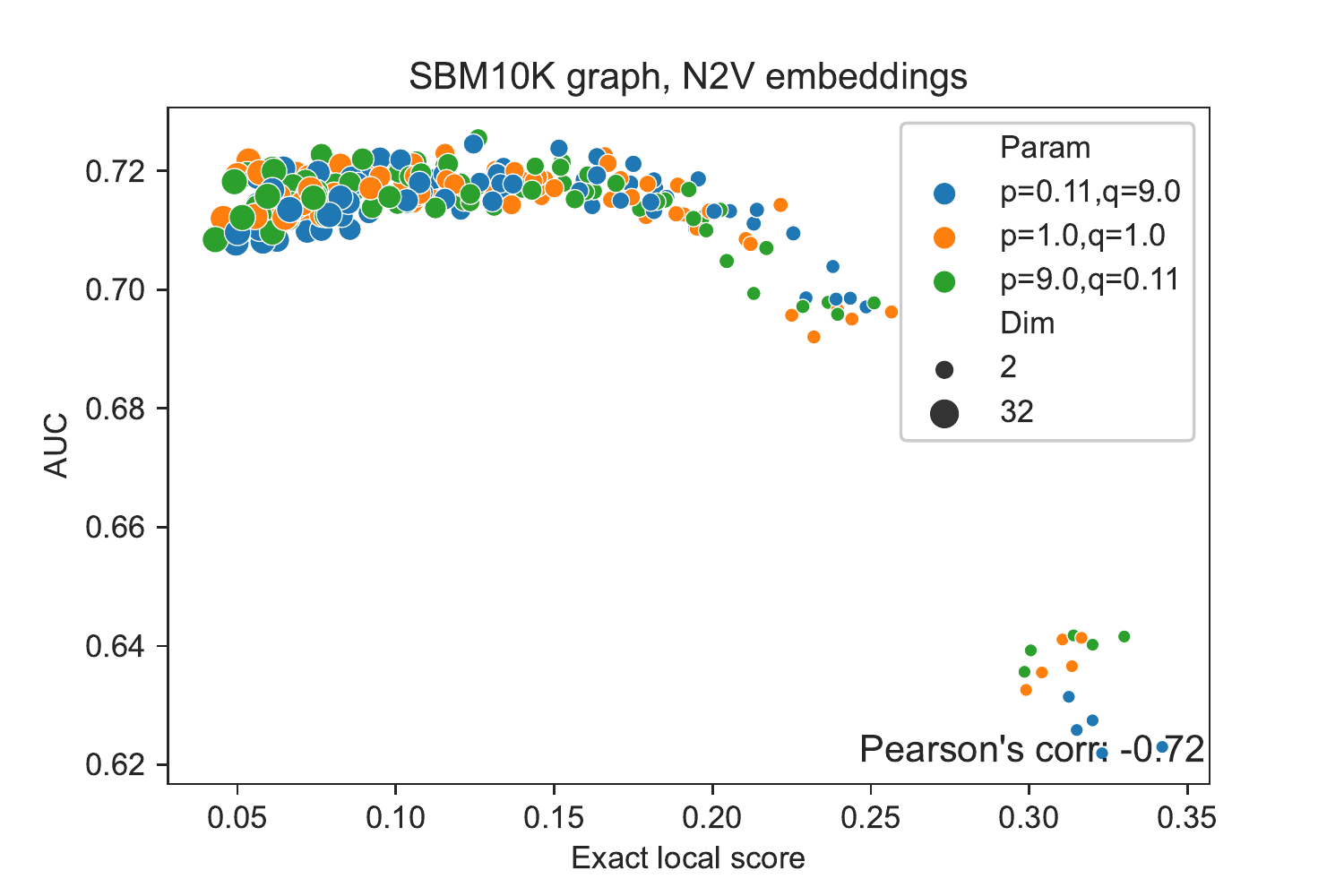}
    \vspace{.1cm}
    \includegraphics[width=0.35\textwidth]{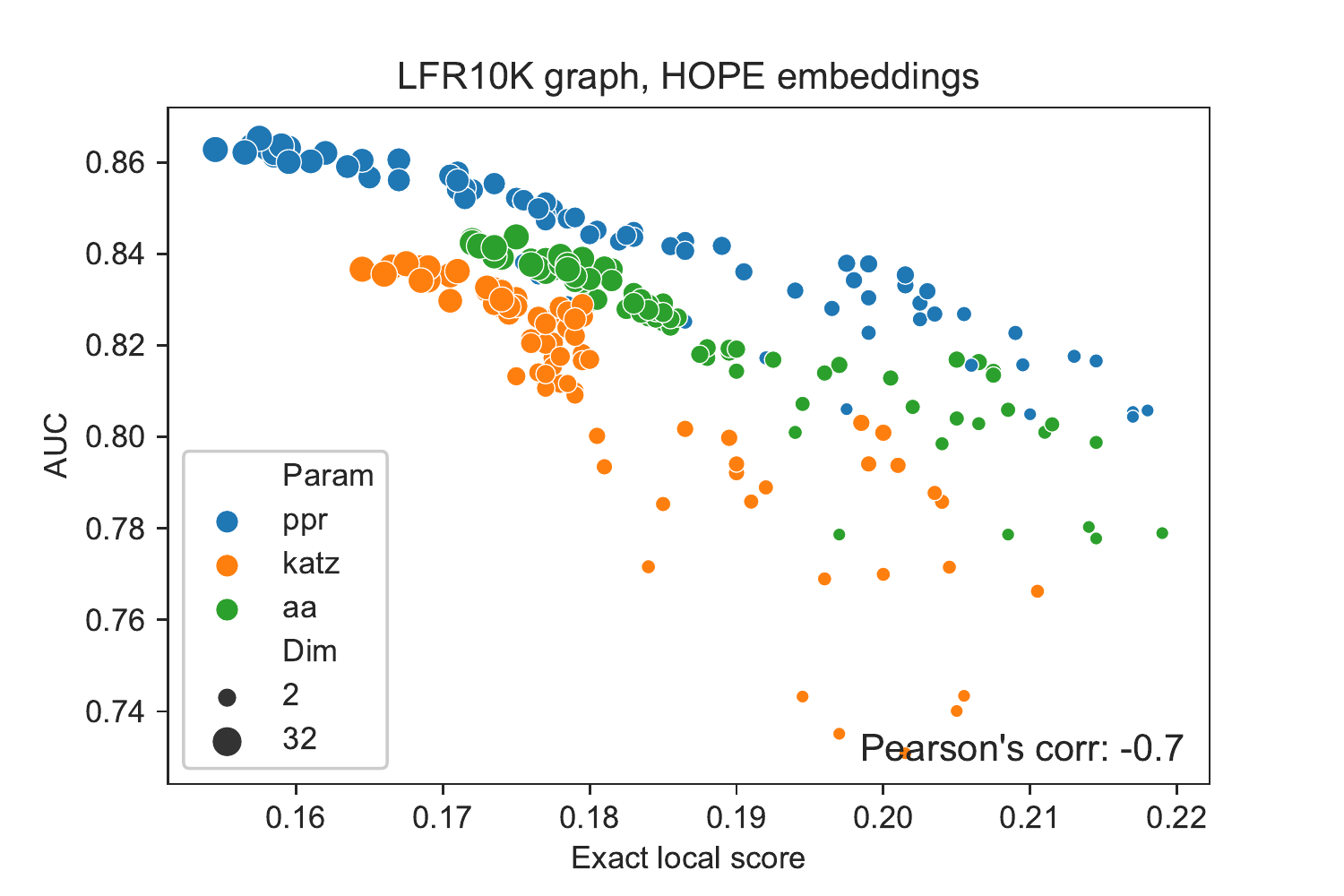}
        \hspace{.1cm}
    \includegraphics[width=0.35\textwidth]{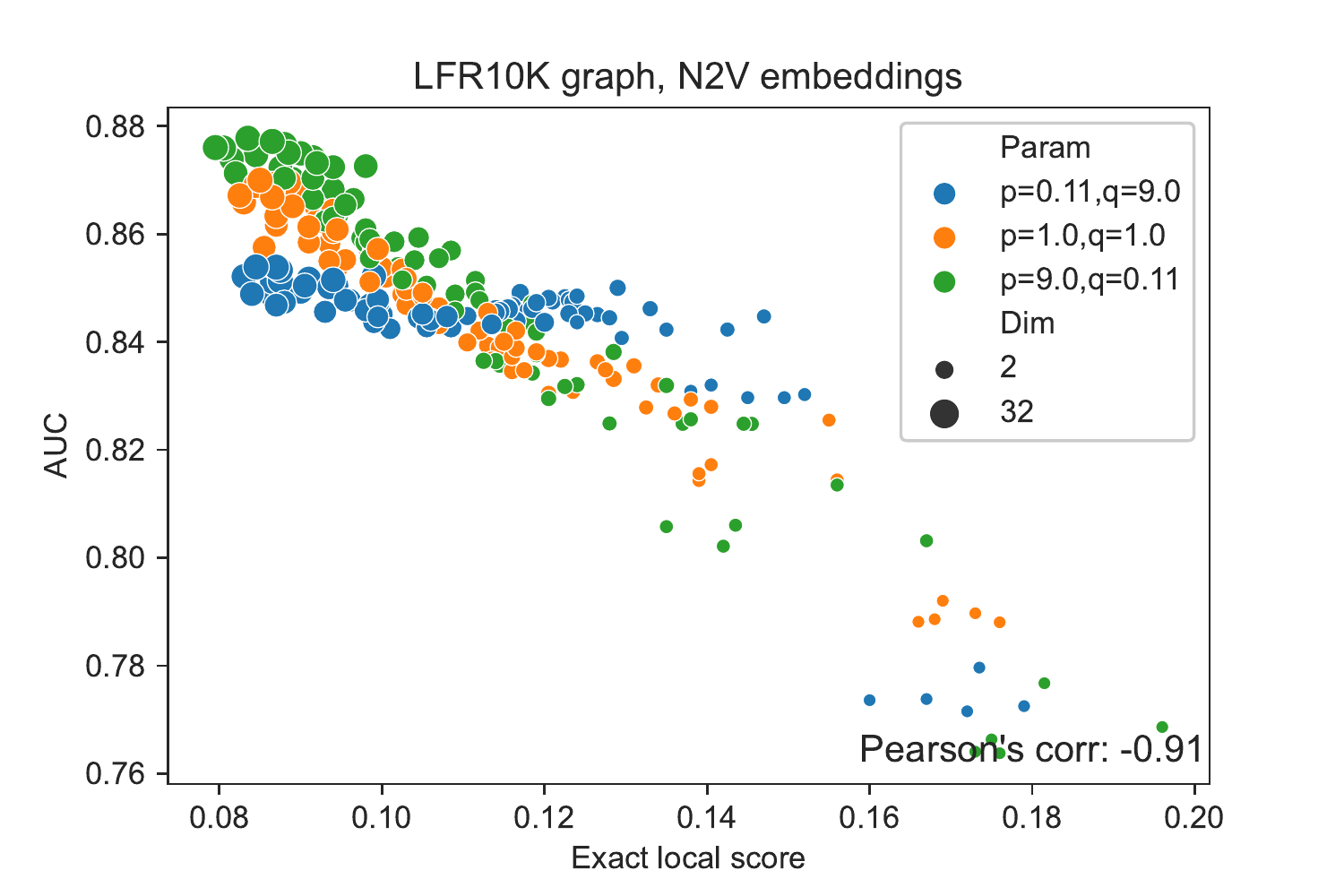}
    \vspace{.1cm}
    \includegraphics[width=0.35\textwidth]{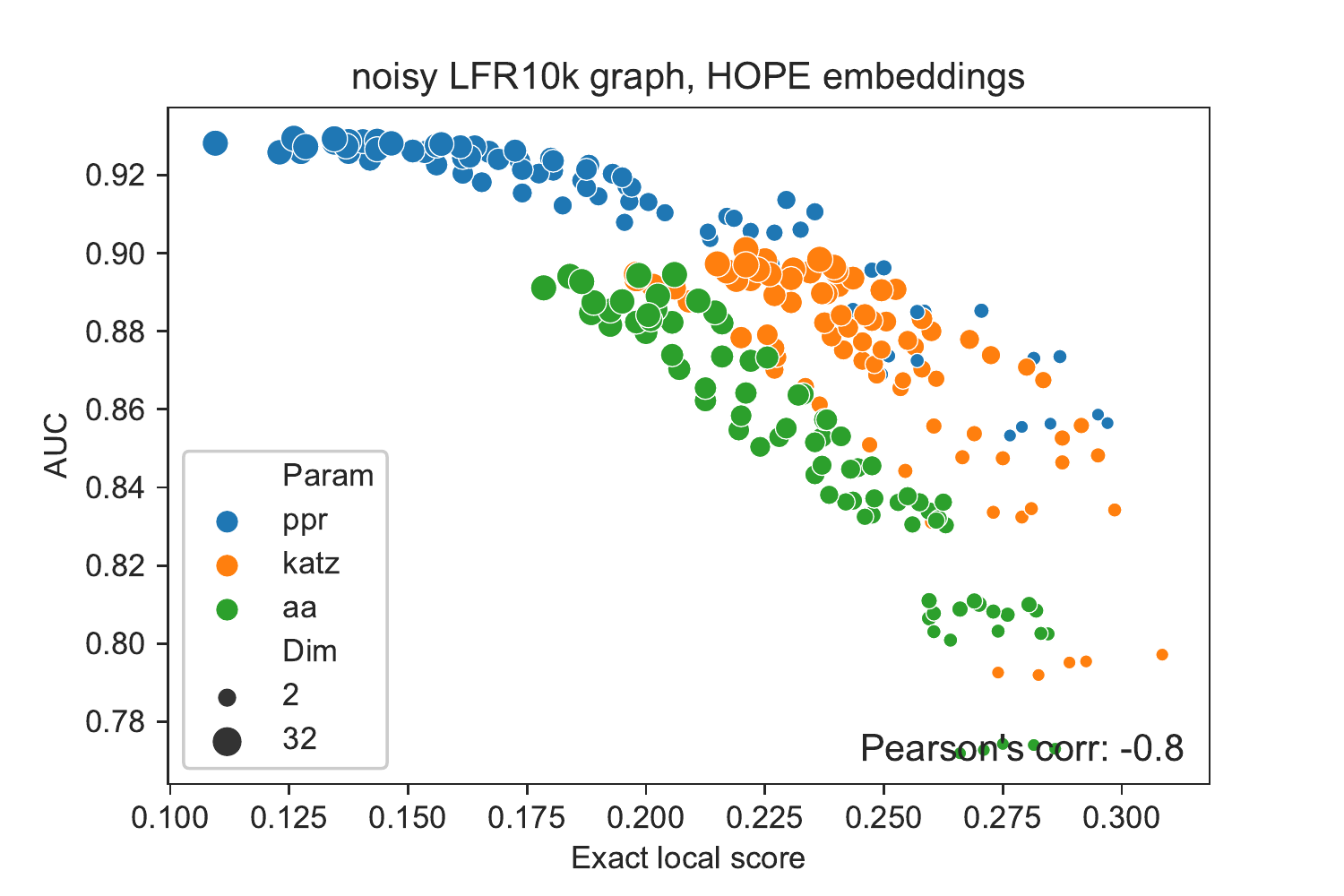}
        \hspace{.1cm}
    \includegraphics[width=0.35\textwidth]{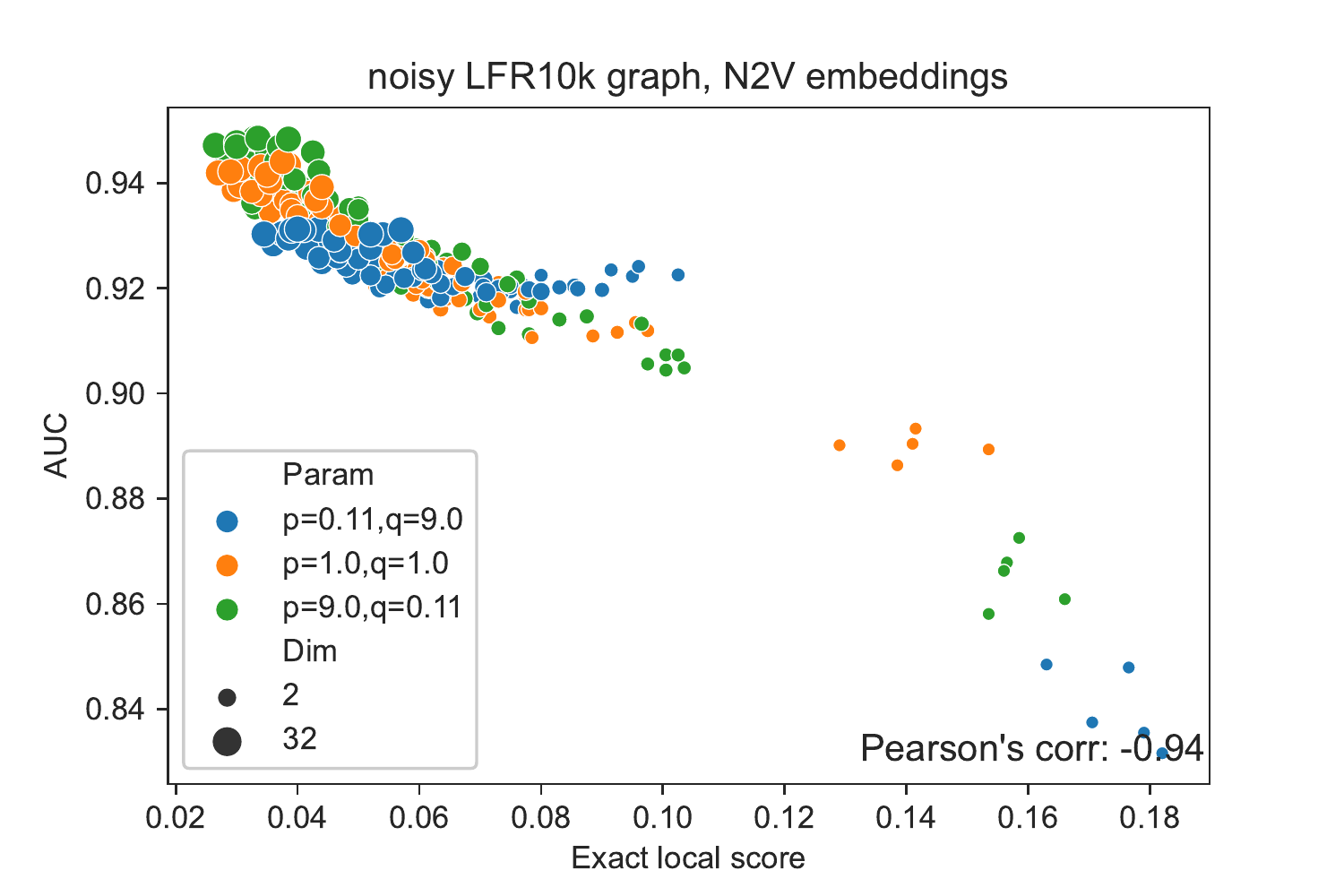}
    \vspace{.1cm}
    \includegraphics[width=0.35\textwidth]{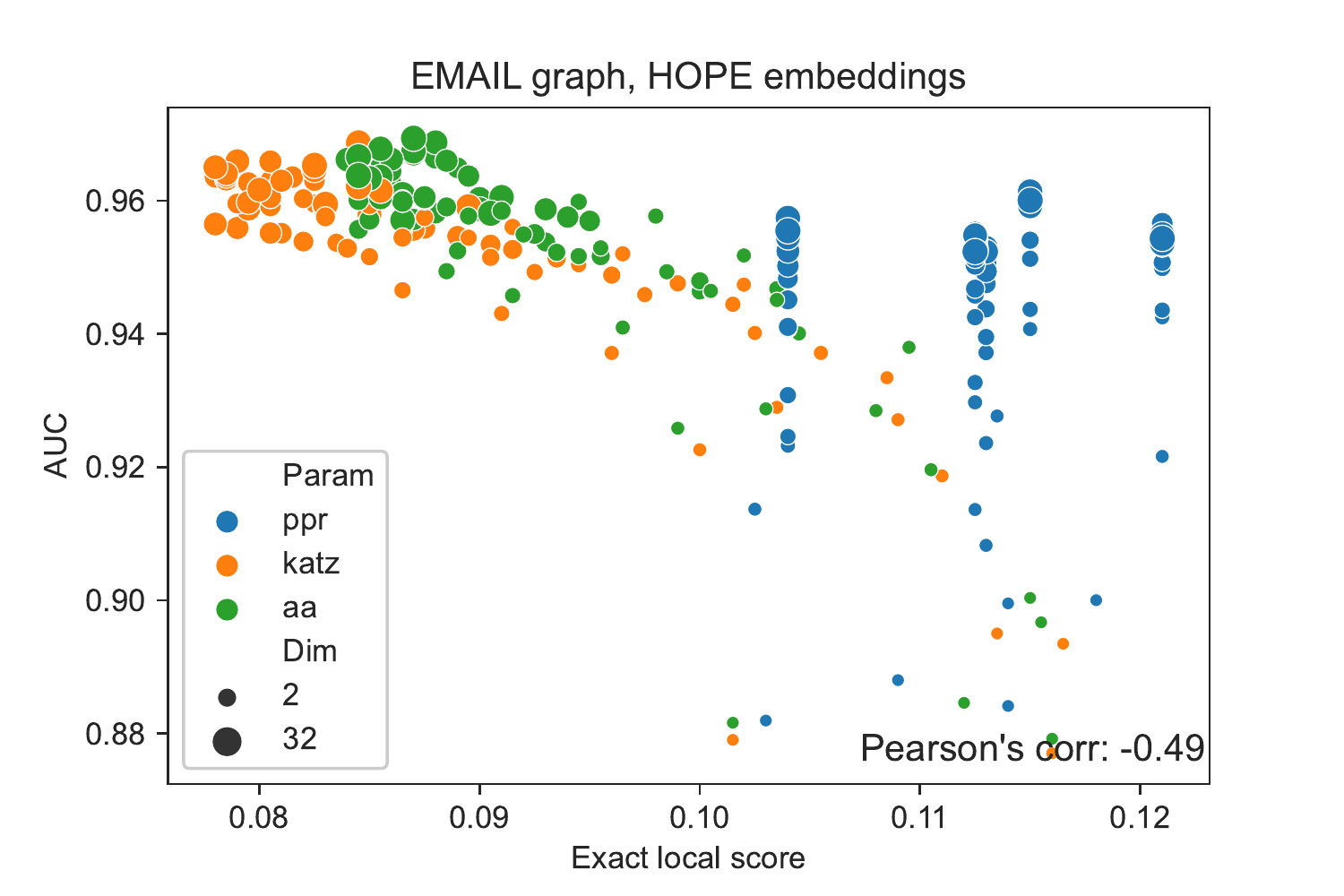}
        \hspace{.1cm}
    \includegraphics[width=0.35\textwidth]{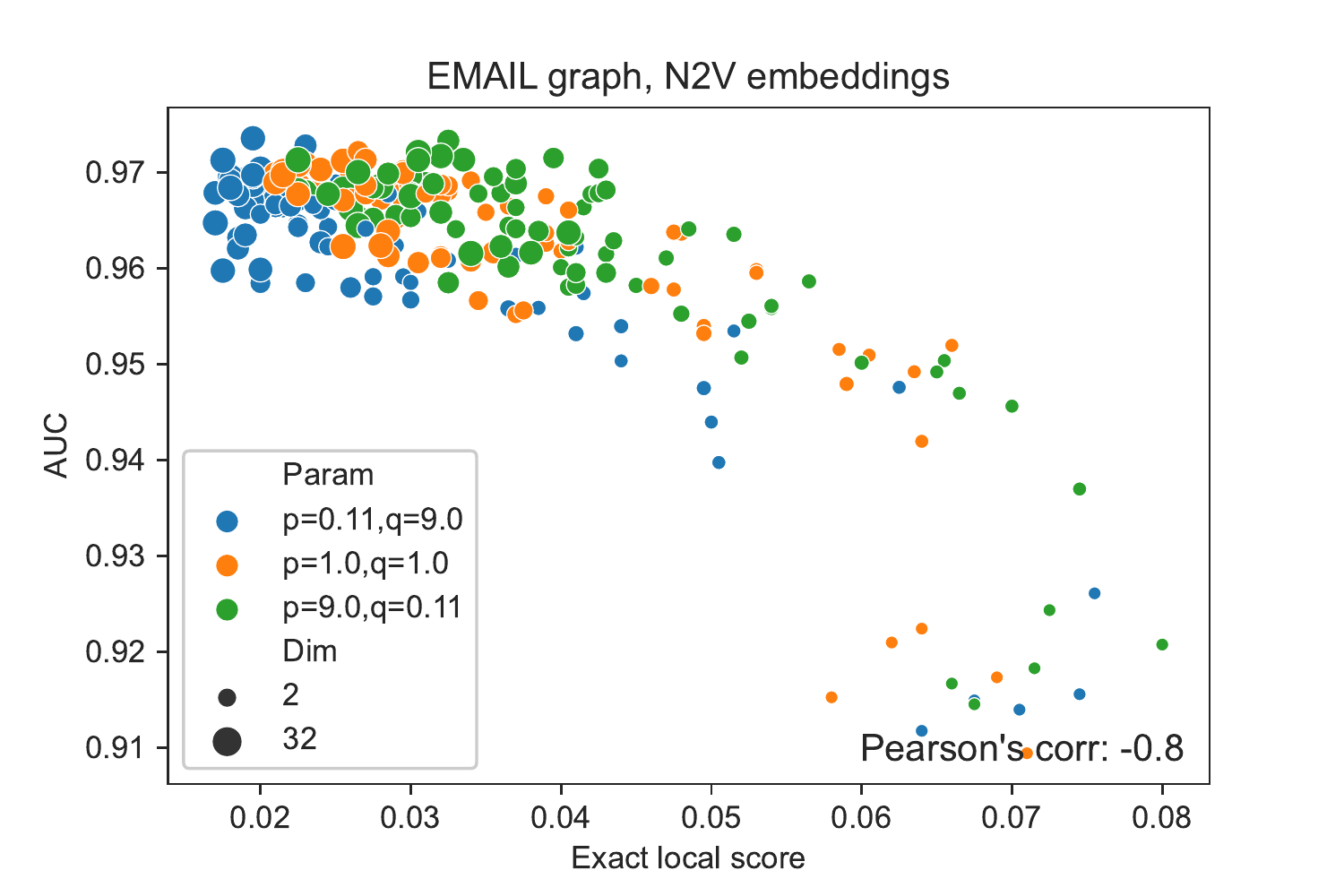}
    \caption{Link prediction: local score vs.\ AUC for \textbf{SBM}, \textbf{LFR}, \textbf{noisy-LFR}, \textbf{EMAIL} graphs and \textbf{HOPE} (left), \textbf{Node2Vec} (right) embeddings.}
    \label{fig:link_pred}
\end{figure}

In this section, we present a few more plots that are complementing the ones shown in the main paper. 
\begin{itemize}
    \item Figure~\ref{fig:10kexact_2measseed}: The global and local scores returned by the framework are presented for the four graphs considered earlier (\textbf{SMB}, \textbf{LFR}, \textbf{noisy-LFR}, and \textbf{EMAIL}). The results for 3 variants of \textbf{Node2Vec} and \textbf{HOPE} (shown in different colours) and a range of dimensions (size of the dots) are presented. One can see some correlation between the two scores, but not a perfect one. Higher dimension generally yields better results but with \textbf{Node2Vec} we observe that several choices of parameters give similarly good results. This can be important for some applications: for example, selecting a lower dimensional embedding (requiring less storage space) that still gives good results.
    \item Figure~\ref{fig:10kapprox_exact_divseed}: With the three synthetic graphs (\textbf{SMB}, \textbf{LFR}, and \textbf{noisy-LFR}) and the same algorithms as above, we compare the exact global score with the landmark-based approximate score when using the number of landmarks as described in Subsection~\ref{sec:landmarks}. The correlation coefficients are also reported. All plots show that the landmark-based approximations are very good. 
    \item Figure~\ref{fig:10kapprox_exact_aucseed}: The counterpart of Figure~\ref{fig:10kapprox_exact_divseed} but for the local score. Again, we see good results for the approximations with just a little more variability than what was observed for the global score.
    \item Figure~\ref{fig:10kxgboost}: This plot shows the relationship between the accuracy for the task of node classification and the framework's global score. The experiments are performed for the same graphs and embeddings as in Figure~\ref{fig:10kexact_2measseed}.
    \item Figure~\ref{fig:10kkmeans}: This plot shows the relationship between the AMI for the task of community detection and the framework's global score. Again, the experiments are performed for the same graphs and embeddings as in Figure~\ref{fig:10kexact_2measseed}.
    \item Figure~\ref{fig:link_pred}: This plot shows the relationship between the AUC for the task of link prediction and the framework's local score. One more time, the experiments are performed for the same graphs and embeddings as in Figure~\ref{fig:10kexact_2measseed}.
\end{itemize}

\clearpage

\subsection{{Illustration of the Framework}}


 \clearpage

\subsection{Approximating the Two Scores}

%
%

 \clearpage


\clearpage

\subsection{Node Classification and Community Detection vs.\ the Global Score}

\clearpage

%


\clearpage 

\clearpage 

\subsection{Link Prediction vs.\ the Local Score}

\end{document}